\documentclass[traditabstract]{aa} 
\bibpunct{(}{)}{;}{a}{}{,}

\newcommand{\ergps}{erg\thinspace s$^{-1}$}

\newcommand{\ergpspsqcm}{erg\thinspace s$^{-1}$\thinspace cm$^{-2}$}
\newcommand{\psqcm}{cm$^{-2}$}
\newcommand{\nH}{$N_{\rm H}$}

\usepackage[varg]{txfonts}
\usepackage{graphicx}
\usepackage{subfig}
\begin{document}

\title{The XMM deep survey in the CDFS}

   \subtitle{XI. X-ray spectral properties of 185 bright sources\thanks{Table 1 and Figure 2 are available in electronic form at the CDS via anonymous ftp to cdsarc.u-strasbg.fr (130.79.128.5) or via http://cdsweb.u-strasbg.fr/cgi-bin/qcat?J/A+A/}}

   \author{K. Iwasawa\inst{1,2}
          \and
          A. Comastri\inst{3}
          \and
          C. Vignali\inst{4,3}
          \and
          R. Gilli\inst{3}
          \and
          G. Lanzuisi\inst{3}
          \and
          W.~N. Brandt\inst{5,6,7}
          \and
          P. Tozzi\inst{8}
          \and
          M. Brusa\inst{4,3}
          \and
          F.~J.~Carrera\inst{9}
          \and
          P.~Ranalli\inst{10}
          \and
          V.~Mainieri\inst{11}
          \and
          I. Georgantopoulos\inst{12}
          \and
          S. Puccetti\inst{13}
          \and
          M. Paolillo\inst{14,15,16}
}

\institute{Institut de Ci\`encies del Cosmos (ICCUB), Universitat de Barcelona (IEEC-UB), Mart\'i i Franqu\`es, 1, 08028 Barcelona, Spain
         \and
ICREA, Pg. Llu\'is Companys 23, 08010 Barcelona, Spain
\and
INAF-Osservatorio di Astrofisica e Scienza dello Spazio di Bologna, via Gobetti 93/3, 40129 Bologna, Italy
\and
Dipartimento di Fisica e Astronomia, Universit\`a di Bologna, via Gobetti 93/2, 40129 Bologna, Italy
\and
Department of Astonomy and Astrophysics, 525 Davey Lab, The Pennsylvania State University, University Park, PA 16802, USA
\and
Institute of Gravitation and the Cosmos, The Pennsylvania State University, University Park, PA 16802, USA
\and
Department of Physics, 104 Davey Laboratory, The Pennsylvania State University, University Park, PA 16802, USA
\and
INAF-Osservatorio Astrofisico di Arcetri, Largo E. Fermi, 5, 50125 Firenze, Italy
\and
Instituto de F\'sica de Cantabria (CSIC-UC), Avenida de los Castros, 39005, Santander, Spain
\and
Combient MiX AB, PO Box 2150, 40313 Gothenburg, Sweden
\and
European Southern Observatory, Karl-Schwarzschild-Str. 2, 85748 Garching bei M\"unchen, Germany
\and
National Observatory of Athens, V. Paulou \& I. Metaxa, 15236, Greece
\and
Agenzia Spaziale Italiana-Unit\`a di Ricerca Scientifica, via del Politecnico, 00133 Roma, Italy
\and
Dipartimento di Fisica, Universit\`a degli studi di Napoli Federico II, via Cinthia, 80126 Napoli, Italy
\and
INAF - Osservatorio Astronomico di Capodimonte, Via Moiariello 16, 80131 Napoli, Italy
\and
INFN – Sezione di Napoli, Via Cinthia 9, 80126 Napoli, Italy
}


 
\abstract{We present the X-ray spectra of 185 bright sources detected in
  the XMM-Newton deep survey of the Chandra Deep Field South  with the three EPIC cameras combined. The 2-10 keV flux limit of the
  sample is $2\times 10^{-15}$ \ergpspsqcm. The sources are
  distributed over a redshift range of $z$ =0.1-3.8, with 11 new X-ray
  redshift measurements included. A spectral analysis was
  performed using a simple model to obtain absorbing column densities,
  rest-frame 2-10 keV luminosities, and Fe K line properties of 180
  sources at $z>0.4$. Obscured AGN are found to be more abundant
  toward higher redshifts. Using the XMM-Newton data alone, seven
  Compton-thick AGN candidates were identified, which set the
  Compton-thick AGN fraction at $\simeq 4$\%. An exploratory
  spectral inspection method with two rest-frame X-ray colours and an
  Fe line strength indicator was introduced and tested against the
  results from spectral fitting. This method works reasonably well to
  characterise a spectral shape and can be useful for a pre-selection
  of Compton-thick AGN candidates. We found six objects exhibiting
  broad Fe K lines out of 21 unobscured AGN of best data quality,
  implying a detection rate of $\sim 30$\%. Five redshift spikes, each
  with more than six sources, are identified in the redshift
  distribution of the X-ray sources. Contrary to the overall trend,
  the sources at the two higher redshift spikes, at $z=1.61$ and
  $z=2.57,$ exhibit a puzzlingly low obscuration.}

\titlerunning{XMM CDFS XI}
\authorrunning{K. Iwasawa et al.}
   \maketitle
%

\section{Introduction}

To understand the evolution of galaxies and black hole growth in the
Universe, sensitive observations of distant galaxies are essential, particularly of those that are heavily obscured,  and deep X-ray observations
play an important role for studying the population of active galactic
nuclei (AGN) with obscuration \citep{brandt05,brandt15}. The Chandra
Deep Field South \citep[hereafter CDFS,][]{giacconi01} is one of the
key fields that are subject to multiwavelength deep surveys. X-ray
observations with the Chandra X-ray Observatory (Chandra) now reach a
7 Ms exposure, which makes it the deepest X-ray survey field
\citep{luo17}. XMM-Newton observed the CDFS in the period of
2008-2010, in addition to the early observations in 2001-2002, and the
survey overview is presented in \citet{comastri11}. The resulting
total exposure time is about 2.5 Ms. The 2-10 keV bright source
catalogue of 339 sources, in which the flux limit is $\sim 1\times
10^{-15}$ \ergpspsqcm, was published by
\citet{ranalli13}. Various works on specific subjects or individual
sources from the survey data have been carried out
\citep{i12cdfs,georgantopoulos13,falocco13,castellmor13,antonucci15,i15brtwo,vignali15,falocco17}.
In this paper, we present X-ray spectra of the brightest 185 sources,
obtained from the EPIC cameras of XMM-Newton
\citep{struder01,turner01}, along with their basic spectral
properties. While the 7-Ms Chandra exposure of the CDFS reaches a
fainter flux limit and provides good quality spectra of individual
sources \citep{luo17,liu17}, this XMM dataset provides an independent
resource for investigating the X-ray source properties.

Between the XMM-Newton and Chandra datasets, comparable source counts
were acquired for a source located in the central part of the
survey field at intermediate energies, that is, $\sim 2$ keV, given the
respective effective areas and exposure times. However, the background
sets their overall data quality apart. The sharp point spread
function (PSF) of Chandra data not only helps push the detection limit of a
point-like source to the faint flux level, but it also improves the overall
spectral data quality  over that of XMM-Newton since the data
extraction aperture in the former is much smaller and, thus, the background is
proportionally low, even if the respective detector background levels
(per area) are comparable. Moreover, it is unfortunate for the
XMM-Newton observations that the quiescent level of the background of
the EPIC cameras during the 2008-2010 period was elevated to
approximately twice that of the earlier observations, likely due to 
variations in Solar activity \citep{ranalli13}.

However, the XMM-Newton data can still prove competitive when it comes to measurements
of Fe K lines, an important probe for AGN surrounding. The EPIC cameras
of XMM-Newton have spectral resolution that are better than the ACIS-I detector
used in the Chandra observations by a factor of 1.5-2. Apart from a
few Compton-thick AGN, Fe K lines in most of AGN spectra lie above
strong continuum emission. The higher spectral resolution gives a
better sensitivity to a narrow emission line in these
sources. \citet{liu17} detected Fe K lines in 50 sources from the
Chandra 7-Ms dataset. We detected Fe lines in a comparable number of
sources, namely, 71. Fe K emission is the most prominent spectral line in the
X-ray band but the study of its properties in AGN has been limited mostly
to objects in the local Universe \citep[e.g.][]{bianchi09}
because, for example, of insufficient photon counts of distant, faint sources. The
deep XMM-CDFS data allow us to extend it beyond the redshift of $z=1$.

This paper presents the X-ray spectral atlas of the XMM-CDFS sources
with which we try to facilitate a quick, visual inspection of overall
spectral shape and the iron line feature of individual sources, using
combined data from the three EPIC cameras. We examine the possible
evolution of AGN obscuration and Fe line properties and search for
Compton-thick AGN candidates, based on basic but homogeneous spectral
fitting. We also test an exploratory spectral inspection method
using two rest-frame X-ray colours and an Fe K line strength indicator
for selecting Compton-thick AGN candidates.

The paper is organised as follows. A brief description of 185 X-ray
sources is presented in Sect. 2 , followed by the spectral atlas of
these sources in Sect. 3. The X-ray properties obtained
from spectral fitting are reported in Sect. 4, followed by the
description of the exploratory analysis method and its results in
Sect. 5. Composite spectra of various subsamples are shown in
Sect. 6. Finally, in Sect. 7, a discussion of Compton-thick AGN, the
evolution of AGN obscuration, X-ray sources in redshift spikes, and Fe
K lines is provided. In the Appendix sections, supplemental
details on X-ray redshift measurements, the exploratory analysis,
spectral stacking methods, and the spectral atlas figures are given.

The cosmology adopted here is $H_0=70$ km s$^{-1}$ Mpc$^{-1}$,
$\Omega_{\Lambda}=0.72$, $\Omega_{\rm M}=0.28$ \citep{komatsu11}.

\section{XMM-CDFS sources}

We present X-ray spectra of the brightest 185 sources in the 2-10 keV band
from the XMM-CDFS survey, obtained from the EPIC cameras. The sources
were selected using a combined criteria of $8\sigma $ detection and the
observed 2-10 keV flux, $f_{2-10} > 1.8\times 10^{-15}$ \ergpspsqcm,
from the 339 sources in the 2-10 keV XMM-CDFS source catalogue of
\citet{ranalli13}. The source identification number listed in the
catalogue paper by \citet{ranalli13} are used with the prefix
'PID'. There is redshift information provided, either spectroscopic or
photometric, for all the sources.

\begin{figure}
\centerline{\includegraphics[width=0.4\textwidth,angle=0]{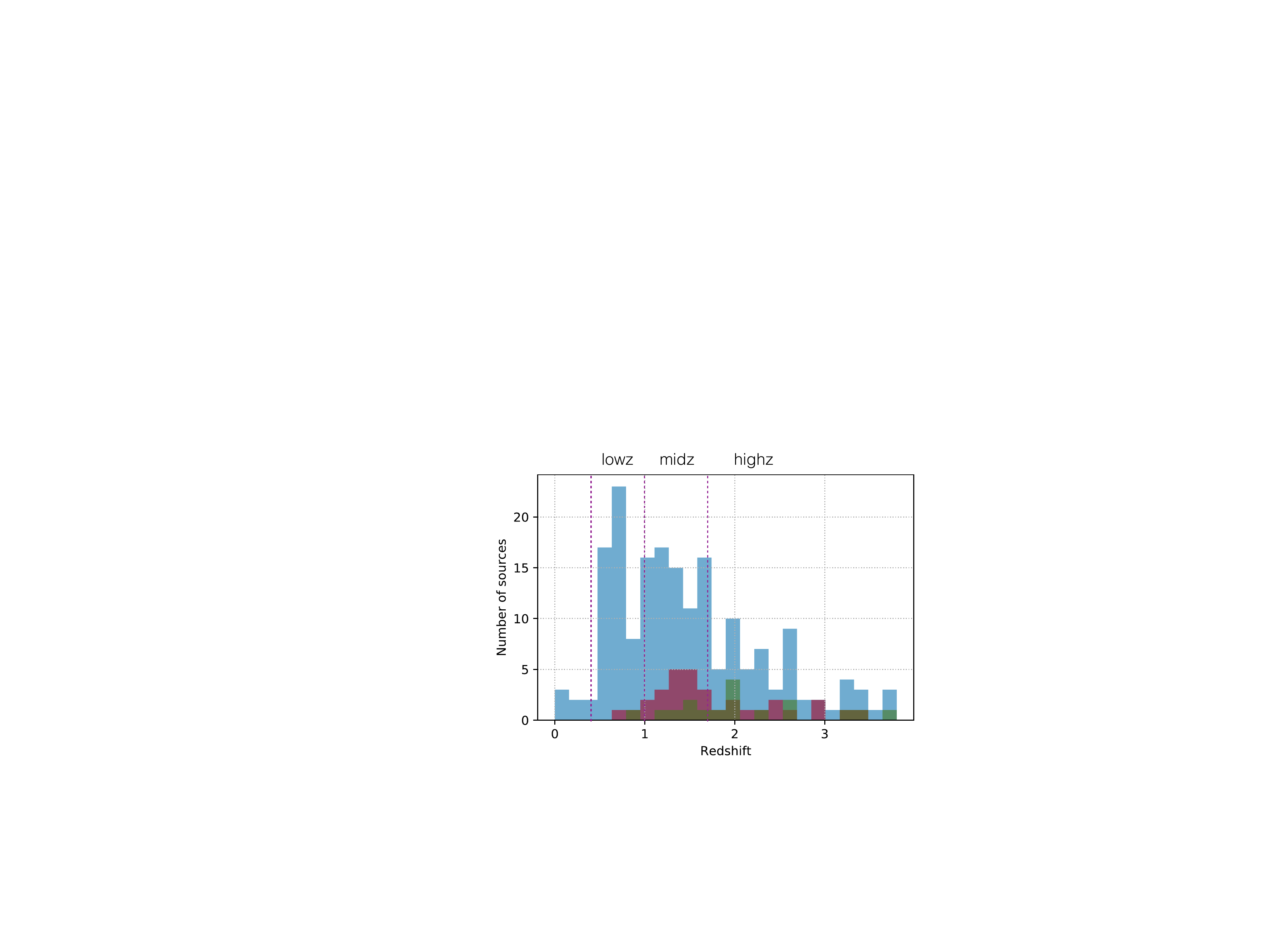}}
\caption{Redshift distribution of the 185 XMM-CDFS sources (blue). The
  majority are spectroscopic redshifts. Photometric redshifts and
  X-ray redshifts are overdrawn in red and green, respectively. The
  redshift ranges of lowz, midz and highz are labeled and their
  boundaries are marked by dashed lines.  }
\label{fig:zdist}
\end{figure}

The majority of the redshift information was taken from the CDFS
Chandra 7 Ms source catalogue by \citet{luo17}. When a secure
spectroscopic redshift was not available, we inspected the X-ray
spectra for a characteristic line or absorption feature in the Fe K
band to try to obtain an X-ray spectroscopic
redshift. X-ray redshift estimates have already been obtained for six
sources with $z\geq 1.6$ in \citet{i12cdfs} and \citet{vignali15}. We
have a further 11 sources for which X-ray spectroscopic redshifts are
adopted. The procedure of obtaining an X-ray redshift is described in
Sect. 4.3 and a brief summary of the new 11 sources with X-ray
redshifts is given in Appendix A. Consequently, among the total 185
sources, we have 135 sources with optical spectroscopic
redshifts, 33 sources with photometric redshifts and 17 sources with
X-ray spectroscopic redshifts. The redshift ranges from $z = 0.11$ to
$z = 3.74$. Their distribution is shown in Fig. \ref{fig:zdist}. A few
redshift spikes, which may not be clear in the histogram, are
identified and X-ray sources in the spikes are discussed in Sect. 7.3.

The source identification number, adopted redshift, redshift type
(spectroscopic, photometric or X-ray), along with the source
characteristics obtained in the spectral analysis, are presented for
the 185 sources in Table \ref{tab:big}. The electronic version
of this table available at the CDS has a few extra columns for Fe K line detection flag,
the spectral data quality indicator, the Chandra source identification
numbers of \citet{luo17} and \citet{xue16}, along with the classification of
radio-loud AGN.

\section{EPIC spectral atlas}

We constructed low-resolution spectra of all the 185 sources in the
0.5-7 keV band, as observed in a uniform manner for a visual
inspection of broad-band spectral shape. Figure 2 shows first three spectra in PID number and the remaining 182 spectra can be found in Appendix. All
the spectra are plotted in 24 identical, logarithmically equally
spaced energy intervals. The data are plotted in the flux density
units, $10^{-15}$ erg cm$^{-2}$ s$^{-1}$ keV$^{-1}$ , and the range
of flux density is always kept to 2 orders of magnitude to
facilitate a visual comparison between sources. Three dotted,
vertical lines drawn in each panel indicate a rest-frame of 3 keV, 6.4
keV and 10 keV, as expected from the adopted redshift. These energy lines
are intended to guide the three key rest-frame energies in the
observed spectra of sources at various redshifts. Each spectral panel
shows PID, adopted redshift, and its category (sp: spectroscopic; ph:
photometric; or x: X-ray) in parenthesis.

Since XMM-Newton carries three EPIC cameras, namely, pn, MOS1, and MOS2, each with different energy responses, we corrected a
count-rate spectrum from each camera for the detector response (the
effective area and the energy redistribution) and the Galactic
absorption \citep[\nH\ $= 9\times 10^{19}$ \psqcm,][]{dickey90} with
the same method used in \citet{i12cosmos} before averaging them using
the observed-frame 1-5 keV source net counts as a weight. Then these
averaged spectra were converted to flux-density units. Among the 185
sources, 169 have data from all the three cameras while the rest have
of two cameras of the three. All the spectral data in ASCII format are available from the CDS.

\begin{figure*}
\centerline{\includegraphics[width=0.75\textwidth,angle=0]{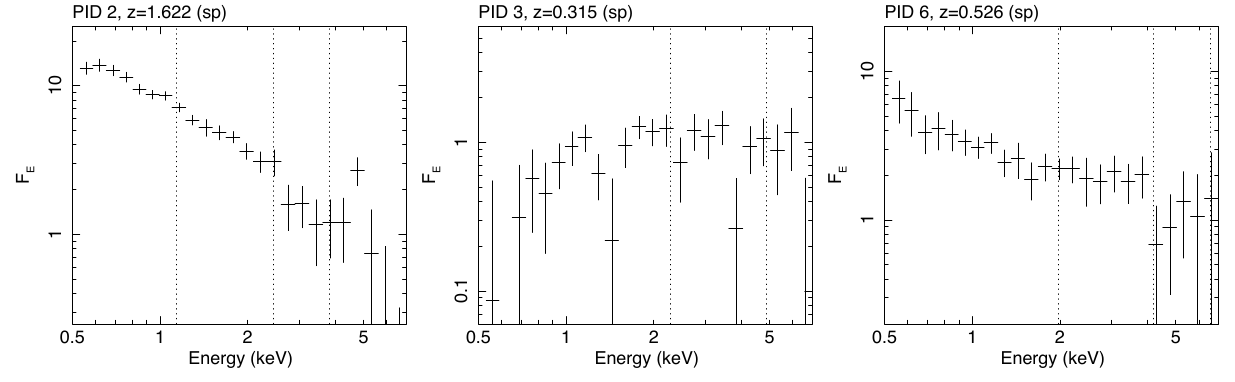}}
  \caption{Observed 0.5-7 keV spectra of 185 XMM-CDFS sources, shown in order of PID number. Only the first three spectra are shown here and the remaining 182 spectra can be found in Appendix D. Flux density is in units of $10^{-15}$ erg~s$^{-1}$~cm$^{-2}$~keV$^{-1}$. The energy scale is as observed and the three, vertical dotted lines indicate the rest-frame energies of 3 keV, 6.4 keV and 10 keV, expected from the adopted redshift.}
\end{figure*}

\section{Spectral fittings of individual sources}


\setcounter{table}{1}

\begin{table}
\caption{Three redshift groups.}
\label{tab:zgroups}
\centering
\begin{tabular}{lcccccc}
Group & $z$ & $\tilde z$ & $N$ & Range & Bins & Med cts\\[5pt]
lowz & 0.4-1.0 & 0.70 & 56 & 2-10 keV & 19 & 1253 \\
midz & 1.0-1.7 & 1.33 & 67 & 2-14 keV & 23 & 1885 \\
highz & 1.7-3.8 & 2.31 & 57 & 2-20 keV & 27 & 844 \\
\end{tabular}
\tablefoot{The table columns are redshift group, redshift range,
  median value of redshifts, number of sources, rest-frame energy
  range used, number of spectral bins, and median net counts of the
  spectra we use in the analysis in Sect. 4 and thereafter. Spectra of
  the three groups share the same energy intervals within the
  overlapping energy range. The column named `Med cts' is the median
  of the net counts, caculated as the sum of the three EPIC cameras in
  the respective rest-frame energy range.}
\end{table}

\subsection{Absorption and X-ray luminosity}

Spectral data were prepared for broad-band spectral fitting to extract
basic continuum spectral parameters, namely, intrinsic cold absorption
column density, \nH, spectral slope for some sources, and Fe K line
strength. We set 2-20 keV as the rest-frame energy range where each
spectrum is investigated and we defined 27 logarithmically equally-spaced
intervals to cover the energy range (note: they are different from the
24 intervals in the observed 0.5-7 keV band used for the spectral
atlas). Due to the limited XMM-Newton bandpass, the rest-frame
energies that the EPIC cameras cover differ depending on the source
redshift: data for lower redshift sources do not reach rest-frame 20
keV, conversely, data below the rest-frame 2 keV for highest redshift
sources are not covered. We divided all the sources, excluding 5
sources with $z\leq 0.4$, into three redshift groups, {\bf lowz}
($0.4<z\leq 1.0$); {\bf midz} ($1.0<z\leq 1.7$); and {\bf highz}
($1.7<z\leq 3.8$), and adopted reduced energy ranges for the lowz
(2-10 keV) and midz (2-14 keV) groups and the full 2-20 keV range to
the highz group. Then we rebinned individual spectra with energy
intervals corresponsing to the fixed rest-frame energy intervals
defined above (Table \ref{tab:zgroups}). The lower bound of the rest
energy of 2~keV is covered by the EPIC cameras for all the sources,
including the highest redshift object, and gives the data a
sensitivity to the absorbing column density of \nH $= 1\times 10^{22}$
\psqcm, which is conventionally used to divide obscured and unobscured
X-ray sources. The 'soft excess' emission is often present in
unobscured AGN and this component emerges below 2 keV. The 2 keV-limit
eliminates the soft excess emission affecting a continuum
measurement. This bandpass limit, however, makes the spectra
insensitive to absorption by a column density smaller than \nH $\simeq
1\times 10^{21}$ \psqcm.

The spectral model we use is a simple absorbed power law. The Galactic
absorption and a narrow Gaussian at the fixed rest-frame energy of 6.4
keV for an iron line are also included. For most sources, the
power-law slope\footnote{We use an energy index, $\alpha $, for a
  power-law slope throughout this paper. $f_\mathrm{E}\propto
  E^{-\alpha}$ and $\alpha $ is related to the conventionally used
  photon index $\Gamma = 1+\alpha $.} is assumed to be $\alpha = 0.8$
\citep[e.g.][]{nandrapounds94,ueda14} and the column density, \nH, in
the line of sight is fitted (Table \ref{tab:big}). There are a few
sources that have a clearly steeper continuum slope. These sources
were pre-selected by the rest-frame (2-5 keV) / (5-9 keV) colour
($=$S/M, see the following Sect. 5.1). As $\alpha = 0.8$ corresponds
to S/M $= 2.94$, when a source has S/M $>2.9$, it was considered to be
a steep-spectrum source candidate and both $\alpha $ and \nH\ were
fitted (Sect. 4.2).

All the spectral fitting was performed using count rate spectra with
background correction from the three EPIC cameras (or two cameras when
data from not all three cameras are available) jointly using the
model folded through the respective detector responses. A variable
constant factor is applied to each camera, primarily to accomodate the
inter-camera calibration errors, but sometimes also to take into account  larger variations
between the cameras, which are dominated by aperture differences
caused by the detector gap, etc. The rest-frame 2-10 keV
luminosity as observed ($L_{\rm 2-10}$) and that corrected for
absorption ($L^{\prime}_{\rm 2-10}$) are estimated from this absorbed
power-law (Table \ref{tab:big}). The median value of the luminosities from
the three cameras (because of the varying normalising constants) was
taken. The best-fitting model for each source was recorded and used
later as the continuum model to investigate the Fe K line further with
the higher resolution, narrow band data (Sect 4.3.1).

For sources in which a large absorbing column (\nH $>5\times 10^{23}$
\psqcm) is found, their \nH\ values and absorption-corrected
luminosities are re-evaluated, assuming a toroidal geometry for an
absorber, since the effects of Compton scattering become
non-negligible and alter the shape of their observed spectra. We use
the {\tt etorus} model \citep{ikeda09} in which the effects of Compton
scattering are included by Monte Carlo simulations assuming a torus
type of absorber. Since we do not know the geometry of the absorbing
torus (neither do the data have sufficient constraining power), we
assume an edge-on torus with a half opening angle of $30^{\circ}$. In
this particular torus geometry, the primary spectral alteration is due
to an addition of reflected light from the inner wall of the torus
which penetrates through the near side of the torus (and therefore
gets absorbed) before reaching an observer \citep[e.g. Reflection~1
  in][Fig. 2]{ikeda09}. The main effect is a deepened Fe K absorption
edge at a given \nH, resulting in the new \nH\ values being slightly
smaller than the original values. These refitted values of \nH\ and
absorption-corrected luminosities are reported in Table \ref{tab:big}.
We note that depending on the assumption of the torus geometry,
\nH\ values can also shift to higher values. In the case of a grazing
view of the torus edge of an inclined torus with a wide opening angle,
for example, $60^{\circ}$ (but with the central source still hidden), its inner
wall is well-exposed to an observer. Directly visible reflected light
from the inner wall \citep[e.g. Reflection~2 in][]{ikeda09} fills in
the softer energies below the Fe K band of the spectrum in this
\nH\ range where the primary continuum is suppressed by
absorption. This causes the line-of-sight absorption model to
underestimate the real value of \nH, that is, a larger \nH\ ,would be
obtained when replacing with this inclined torus. For example, log
$N_\mathrm{H}=23.82^{+0.11}_{-0.10}$ [\psqcm] given for PID 144
\citep{norman02, comastri11} could go up to log
$N_\mathrm{H}=23.92^{+0.18}_{-0.14}$ by changing the torus
configuration. Similar amounts of shift in log \nH\ are seen in other
five sources with log \nH $> 23.8$ (PID 66, 131, 245, 252, 316). In
the latter torus configuration, the 68\% error region of \nH\ for all
the six sources contains log \nH $= 24$. Thus, with no knowledge of the
torus geometry, $\sim 0.1$~dex of uncertainty in \nH\ always exists in
this \nH\ regime. Furthermore, the spectra are likely more complex,
due to scattered light and partial covering etc., than the simple
model we apply at high \nH.

\begin{figure}
  \centerline{\includegraphics[width=0.45\textwidth,angle=0]{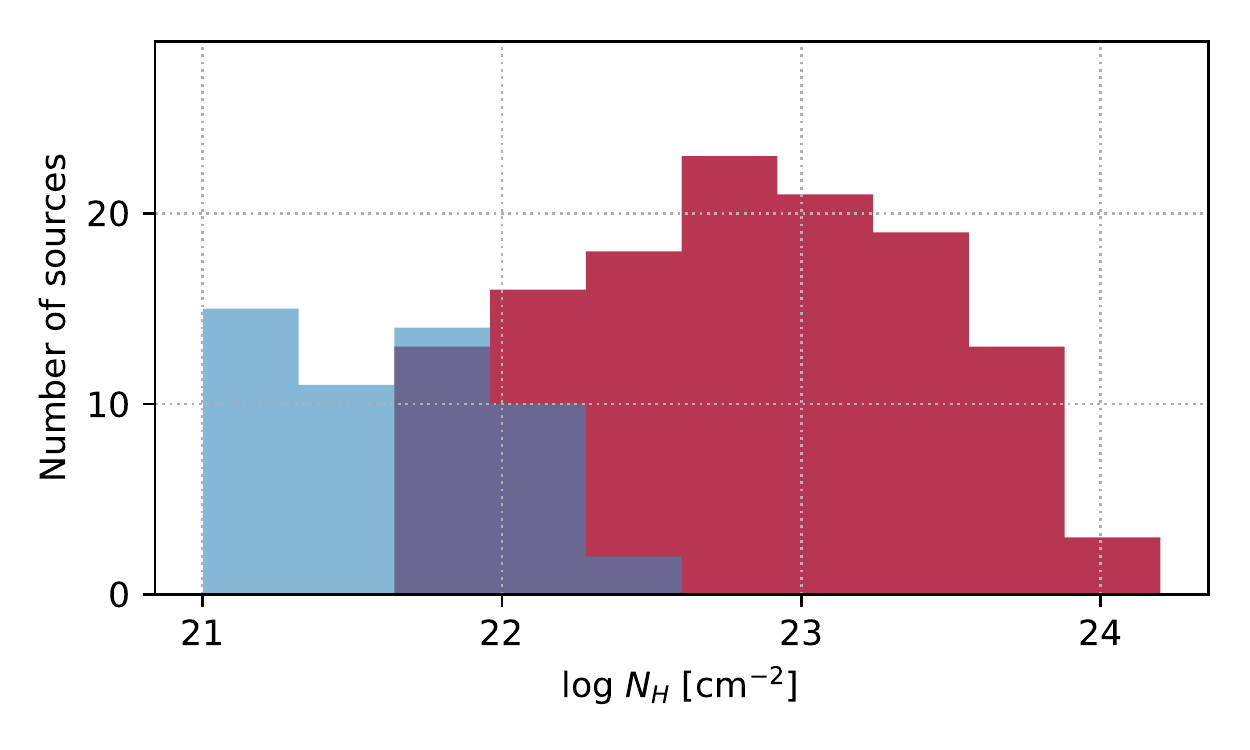}}
  \centerline{\includegraphics[width=0.38\textwidth,angle=0]{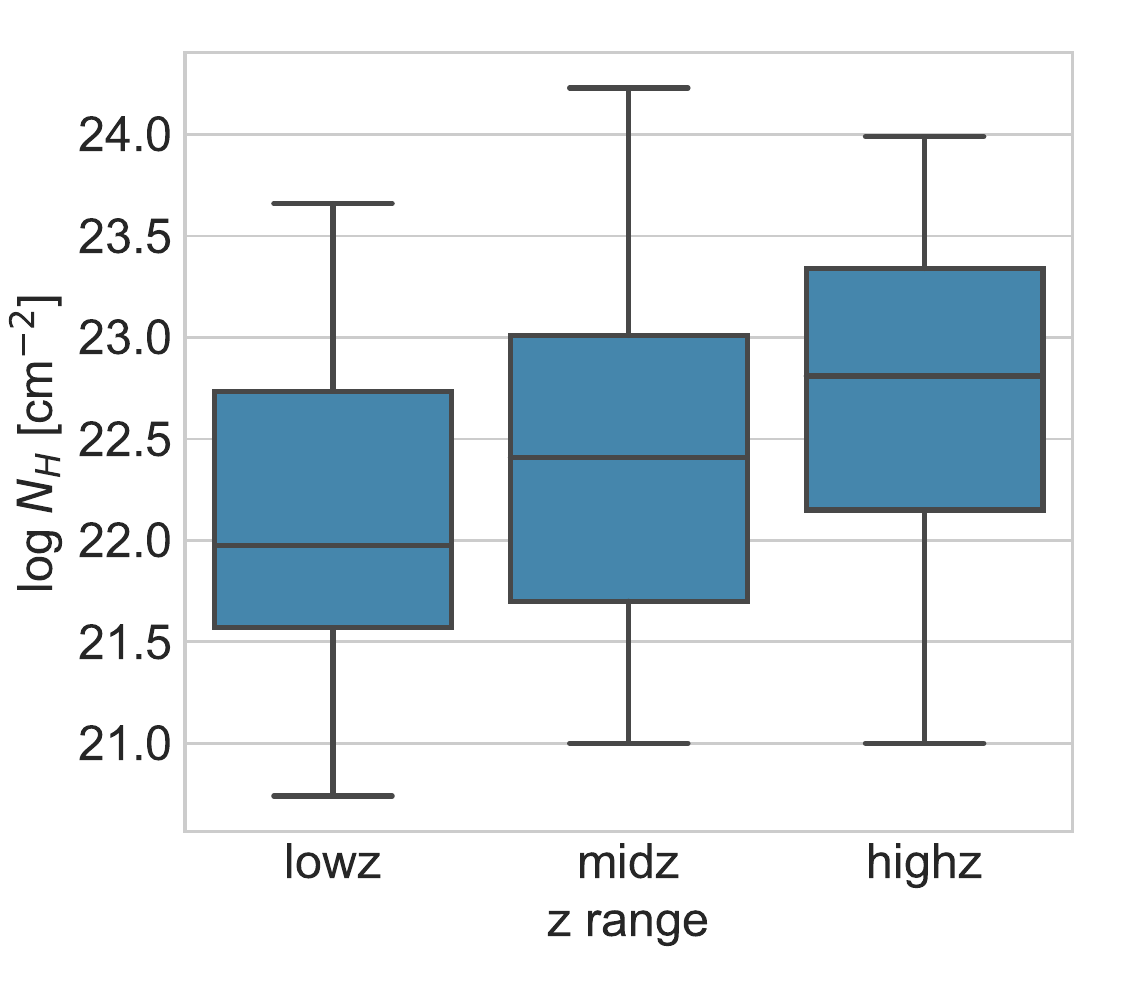}}
\caption{Upper panel: Distribution of best-fit $N_{\rm H}$ values
  obtained for all the 180 sources at $z>0.4$. The histogram in red
  shows the \nH\ distribution of absorption detected sources, while
  the blue histogram shows that of the upper limits. We note the limited
  but uniform sensitivity of our spectra to \nH\ due to the rest-frame
  bandpass (see text). Lower panel: Box-and-whisker plots of
  \nH\ measurements for the lowz, midz and highz groups. Following the
  normal convention, the box represents the inter quatile range (the
  middle 50\%), the bar in the middle represents the median value,and
  the whiskers show the range of the minimum and the maximum of the
  distribution.}
\label{fig:histo_nH}
\end{figure}

\begin{figure}
\centerline{\includegraphics[width=0.35\textwidth,angle=0]{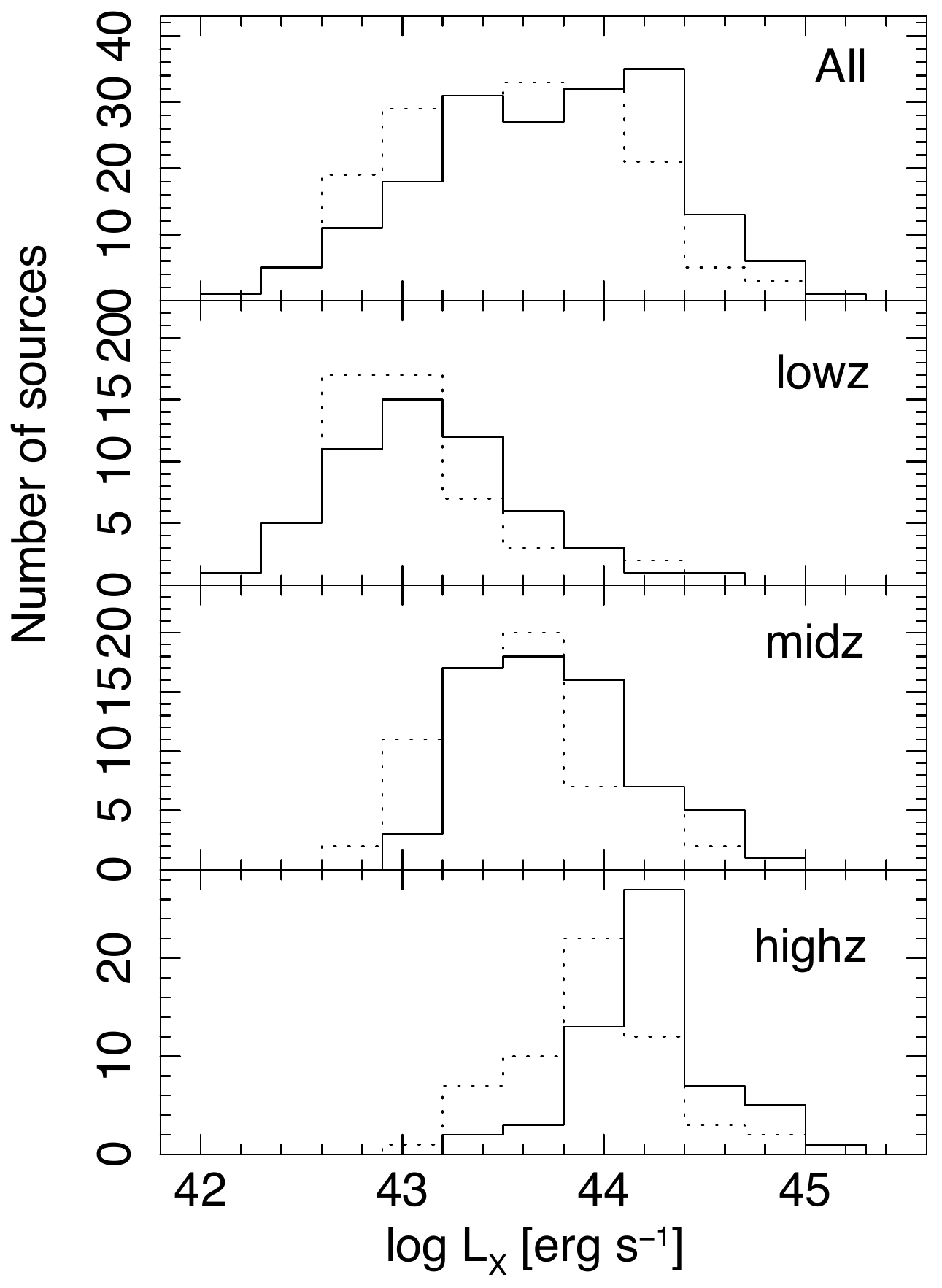}}
\caption{Rest-frame 2-10 keV luminosity distributions for all the
  180 sources, the lowz, midz and highz sources. The solid line
  histograms denote the distribution of absorption-corrected
  luminosity, while the dotted lines are for luminosity as observed. }
\label{fig:histo_lx}
\end{figure}

The \nH\ distribution is plotted in Fig. \ref{fig:histo_nH}. Median
values of log \nH, in units of cm$^{-2}$, are 22.48 for all the
sources, 21.98 for the lowz, 22.38 for midz, and 22.84 for highz
groups (see Table \ref{tab:median_zgrp}), suggesting an increasing trend
of absorption toward higher redshifts, as illustrated by the box-and-whisker plot
in Fig. \ref{fig:histo_nH}. This trend is translated to (on average)
larger absorption corrections of luminosity for sources at higher
redshift, as shown in Fig. \ref{fig:histo_lx}. The median values of
observed luminosity, absorption-corrected luminosity for each redshift
group are given in Table \ref{tab:median_zgrp}. As for the obscured AGN
fraction, when sources with \nH $> 1\times 10^{22}$ \psqcm\ (the
$1\sigma $ lower limits need to be larger than $1\times 10^{22}$
\psqcm) are counted as obscured AGN, $f_\mathrm{lowz} =
0.40^{+0.07}_{-0.06}$, $f_\mathrm{midz} = 0.52^{+0.06}_{-0.06}$, and
$f_\mathrm{highz} = 0.72^{+0.05}_{-0.05}$ for lowz, midz and highz
sources, respectively\footnote{Hereafter, uncertainty for a proportion
  value refers to the 68\% credible interval estimated assuming a
  binomial distribution with the uniform prior between 0 and 1, unless
  stated otherwise.}.

The median properties for sources with spectroscopic, photometric and
X-ray redshift estimates are also shown in Table
\ref{tab:median_zgrp}. A comparison between sources with spectroscopic
and photometric redshifts is discussed later (Sect. 6.2). X-ray
redshifts are derived from an Fe K absorption edge in many cases,
which naturally result in the large median absorbing column.

\begin{table*}
\begin{center}
  \caption{Median properties of sources in the three redshift ranges and three redshift estimate types.}
  \label{tab:median_zgrp}
\begin{tabular}{lcccccccc}
Group & $z$ & log \nH & log $L_{\rm X}$ & log $L^{\prime}_{\rm X}$ & EW & S/M & H/M & R(Fe)\\[5pt]
All & $1.34^{+0.08}_{-0.07}$ & $22.48^{+0.08}_{-0.16}$ & $43.53^{+0.03}_{-0.08}$ & $43.77^{+0.08}_{-0.11}$ & $0.15^{+0.02}_{-0.01}$ & $1.95^{+0.11}_{-0.20}$ & --- & $1.28^{+0.04}_{-0.02}$ \\[5pt]
lowz & $0.70^{+0.03}_{-0.03}$ & $21.98^{+0.17}_{-0.13}$ & $43.02^{+0.06}_{-0.09}$ & $43.11^{+0.07}_{-0.07}$ & $0.13^{+0.03}_{-0.02}$ & $2.26^{+0.09}_{-0.07}$ & --- & $1.30^{+0.05}_{-0.08}$ \\
midz & $1.33^{+0.04}_{-0.03}$ & $22.38^{+0.14}_{-0.30}$ & $43.54^{+0.05}_{-0.05}$ & $43.74^{+0.07}_{-0.09}$ & $0.17^{+0.03}_{-0.02}$ & $2.05^{+0.19}_{-0.33}$ & $0.54^{+0.02}_{-0.02}$ & $1.29^{+0.05}_{-0.05}$ \\
highz & $2.31^{+0.17}_{-0.03}$ & $22.84^{+0.09}_{-0.18}$ & $43.98^{+0.02}_{-0.08}$ & $44.25^{+0.02}_{-0.05}$ & $0.15^{+0.05}_{-0.02}$ & $1.34^{+0.34}_{-0.16}$ & $0.56^{+0.03}_{-0.01}$ & $1.27^{+0.09}_{-0.01}$ \\[5pt]
Specz & $1.21^{+0.01}_{-0.11}$ & $22.15^{+0.19}_{-0.15}$ & $43.51^{+0.07}_{-0.10}$ & $43.65^{+0.14}_{-0.06}$ & $0.14^{+0.01}_{-0.01}$ & $2.22^{+0.13}_{-0.12}$ & $0.52^{+0.02}_{-0.01}$ & $1.25^{+0.02}_{-0.02}$ \\
Photoz & $1.57^{+0.13}_{-0.13}$ & $22.61^{+0.20}_{-0.15}$ & $43.46^{+0.10}_{-0.10}$ & $43.71^{+0.15}_{-0.11}$ & $0.19^{+0.06}_{-0.07}$ & $1.64^{+0.29}_{-0.30}$ & $0.55^{+0.02}_{-0.01}$ & $1.31^{+0.04}_{-0.08}$ \\
Xz & $1.92^{+0.10}_{-0.09}$ & $23.34^{+0.18}_{-0.16}$ & $43.72^{+0.17}_{-0.24}$ & $44.18^{+0.08}_{-0.10}$ & $0.33^{+0.02}_{-0.07}$ & $0.64^{+0.09}_{-0.14}$ & $0.67^{+0.06}_{-0.04}$ & $1.64^{+0.11}_{-0.04}$ \\
\end{tabular}
\tablefoot{The luminosities are logarithmic values in units of \ergps,
  measured in the rest-frame 2-10 keV band. $L$ denotes the luminosity
  as estimated from the best-fitting absorbed power-law while
  $L^{\prime}$ denotes those corrected for absorption. The EW column
  shows Fe K line equivalent width in keV. S/M and H/M are the two
  rest-frame X-ray colours described in Sect. 5.1. R(Fe) is the Fe K
  line strength indicator defined in Sect. 5.2. The 68\% confidence
  interval obtained from bootstrap \citep{efron79} is quoted for each
  value. H/M values are available for sources with $z>1$. The median
  values of H/M are obtained for sources for which H/M are available
  in each category.}
\end{center}
\end{table*}

\subsection{Steep spectrum sources}

As mentioned above, sources that have a possibly steep continuum
slope, i.e., $\alpha > 0.8$, were selected by S/M $>2.94$ and the
absorbed power-law fit was performed by allowing both the slope and
absorption to vary. Five sources (PID 40, 86, 243, 290, 375 and 407)
with noisy data which do not have constraining power for both slope and
\nH\ were excluded here. Measured slope (energy index $\alpha$) for
these 22 steep-spectrum source candidates are reported in Table
\ref{tab:steep}.

\begin{table}
\begin{center}
  \caption{Steep spectrum sources.}
  \label{tab:steep}
\begin{tabular}{lc}
PID & $\alpha $ \\[5pt]
2 & $1.20^{+0.06}_{-0.06}$ \\
45 & $1.40^{+0.60}_{-0.40}$ \\
57 & $1.18^{+0.08}_{-0.08}$ \\
62 & $1.13^{+0.05}_{-0.05}$ \\
81 & $0.84^{+0.04}_{-0.04}$ \\
118 & $1.02^{+0.06}_{-0.05}$\\
130 & $0.88^{+0.04}_{-0.03}$\\
178 & $1.45^{+0.35}_{-0.35}$\\
200 & $0.89^{+0.03}_{-0.03}$\\
201 & $0.91^{+0.08}_{-0.07}$\\
219 & $1.03^{+0.32}_{-0.26}$\\
244 & $1.17^{+0.17}_{-0.14}$\\
249 & $0.99^{+0.06}_{-0.05}$\\
262 & $0.88^{+0.07}_{-0.05}$\\
277 & $1.02^{+0.08}_{-0.06}$\\
298 & $1.20^{+0.20}_{-0.30}$\\
308 & $0.88^{+0.05}_{-0.06}$\\
319 & $0.89^{+0.01}_{-0.01}$\\
328 & $1.33^{+0.04}_{-0.04}$\\
337 & $1.00^{+0.02}_{-0.03}$\\
341 & $1.26^{+0.04}_{-0.05}$\\
345 & $1.14^{+0.08}_{-0.08}$\\
\end{tabular}
\begin{list}{}{}
\item[] Note:\ The spectral slope is in the unit of energy index $\alpha$ of a power-law continuum in energy - flux density: $f_{\rm E} \propto E^{-\alpha}$.  
\end{list}
\end{center}
\end{table}

\subsection{Spectral slopes of radio-loud AGN}

Some fraction of our sources are expected to be radio-loud AGN. Their
X-ray spectrum may have a contribution of jet, resulting in a
harder spectrum than the standard slope of $\alpha = 0.8$ assumed for the
above spectral analysis. Among the 185 sources, 16 sources are found
to match radio sources classified as radio-loud AGN in
\citet{bonzini13}. This makes the fraction of radio-loud AGN to be
$\sim 9$\%. The radio source identification number (RID) of
\citet{bonzini13} for the 16 sources are listed in the online version
of Table \ref{tab:big}. We note that the radio-loud AGN classification
by \citet{bonzini13} follows \citet{padovani11} and is based on excess
radio emission above the radio (1.4 GHz) - mid-infrared (24 $\mu$m)
luminosity correlation for star-forming galaxies, instead of the
classical radio-loudness parameter $R$ \citep{kellermann89} defined by
the radio to optical ratio, which could pose a problem for obscured
AGN (we refer to e.g. \citet{lambrides20} for effects of obscuration
on the AGN classification schemes using multiwavelength
comparisons). These 16 radio-loud AGN include the two brightest X-ray
sources in the field, PID 203 and PID 319, and the optically faint,
X-ray bright obscured quasar at $z=1.6$, PID 352
\citep{vignali15}. PID 319 and PID 352 are reported to have an
extended radio morphology \citep{hales14,huynh15}: PID 352 is a powerful radio galaxy with log $P_{1.4}=27.1$ [W Hz$^{-1}$] with a
double radio-lobe \citep{vignali15}.

Out of the 16 radio-loud AGN, ten sources have X-ray absorption of \nH
$\geq 10^{22}$ \psqcm, which makes it diffiult to determine their
intrinsic X-ray slopes. We reexamined the remaining six unobscured
sources for their spectral slopes. Three sources, PID 182, PID 190,
and PID 359 are found to have slightly hard slopes, $\alpha =
0.55$-0.67 (Table \ref{tab:radioloud}), and therefore a jet
contribution to the X-ray emission is suspected. These three have the
1.4 GHz radio power of log $P_{1.4} = 24.6$-26.0 [W Hz$^{-1}$] for our
adopted redshits. The other three, PID 203, PID 319 and PID 407 have
X-ray slopes $\alpha = 0.8$-0.9, similar to the standard radio-quiet
AGN slope. The former two are the bright X-ray sources and exhibit
broad Fe K emission \citep{i15brtwo}, suggesting that the disc
emission dominates their X-ray spectra. The spectral slope of PID 407 is not well constrained but the source is one of the possible steep-spectrum sources selected in Sect. 4.2.

\begin{table}
  \caption{Unobscured radio-loud AGN with a hard spectrum}
  \label{tab:radioloud}
  \centering
    \begin{tabular}{ccccc}
PID & RID & log $P_{1.4}$ & log $L^{\prime }_{\rm X}$ & $\alpha $\\[5pt]
182 & 719 & 26.04 & 43.41 & $0.55^{+0.10}_{-0.08}$ \\  
190 & 674 & 24.58 & 44.20 & $0.59^{+0.10}_{-0.10}$ \\  
359 & 348 & 25.60 & 44.72 & $0.67^{+0.04}_{-0.03}$ \\  
    \end{tabular}
    \tablefoot{ The three unobscured sources with radio-loud AGN
      classification exhibiting hard X-ray slopes. RID is the radio
      source identification number of \citet{bonzini13}; $P_{1.4}$ is
      the 1.4 GHz radio power in units of W Hz$^{-1}$ for our adopted
      redshift; $L^{\prime }$ is absorption corrected, rest-frame 2-10
      keV luminosity in units of \ergps; $\alpha $ is fitted X-ray
      slope.}
\end{table}

\subsection{Fe K lines}

\subsubsection{Narrow Fe K emission}

The spectral binning used for the broad band spectral fits includes a
rest-frame 6.1-6.6 keV interval where a neutral Fe K line at 6.4 keV
would fall. In general, when a adopted source redshift is correct, the
6.4 keV line intensity derived from the broad-band fit is expected to
be correct even with this low-resolution setting. However, when, for
example, an adopted photometric redshift is slightly wrong, or the Fe
K emission comes from highly ionised iron, Fe {\sc xxv} or Fe {\sc
  xxvi}, some line photons fall out of the Fe K interval, leading to
an inaccurate line measurement. To assure more reliable line
measurements, we optimise the resolution of the spectrum in a limited
energy range. The source spectra were rebinned with 30-eV intervals in
the observed frame and data corresponding to the rest-frame 5-8 keV were
used. This choice of spectral resolution gives sufficient oversampling
for a narrow Fe line from high-redshift sources at $z\sim 3$ (for
which a Fe line would appear at an energy range where the spectral
resolution is $\sim 100$ eV in FWHM) as well as lower redshift
sources. A high-ionisation Fe K line can be distinguished from a 6.4
keV cold line. In case of sources with loosely constrained photometric
redshifts, fitting a line energy could give a more precise
redshift. When this happened, the broad-band fitting was repeated to
revise the absorbing column density and luminosity, using the updated
X-ray redshift information.

The best-fitting absorbed power-law obtained for each source in the
previous subsection was used as the continuum model. A narrow Gaussian
was fitted with one of the line energies of 6.4 keV, 6.7 keV and 6.97
keV appropriate for cold Fe, Fe {\sc xxv,} and Fe {\sc xxvi},
respectively, whichever describes the observed line best, with the
normalisation set as a free parameter. The adopted line centroid and
rest-frame equivalent width (EW) are reported in Table
\ref{tab:big}. For Fe line driven X-ray redshift measurements, for example,
for PID 352 \citep{vignali15} and PID 215, the Fe K line is assumed to
be a cold line at 6.4 keV, since most Fe lines in our sample are found
at that energy, as shown below, and the source redshift was fitted
using these fine resolution spectra.

\begin{figure}
\centerline{\includegraphics[width=0.5\textwidth,angle=0]{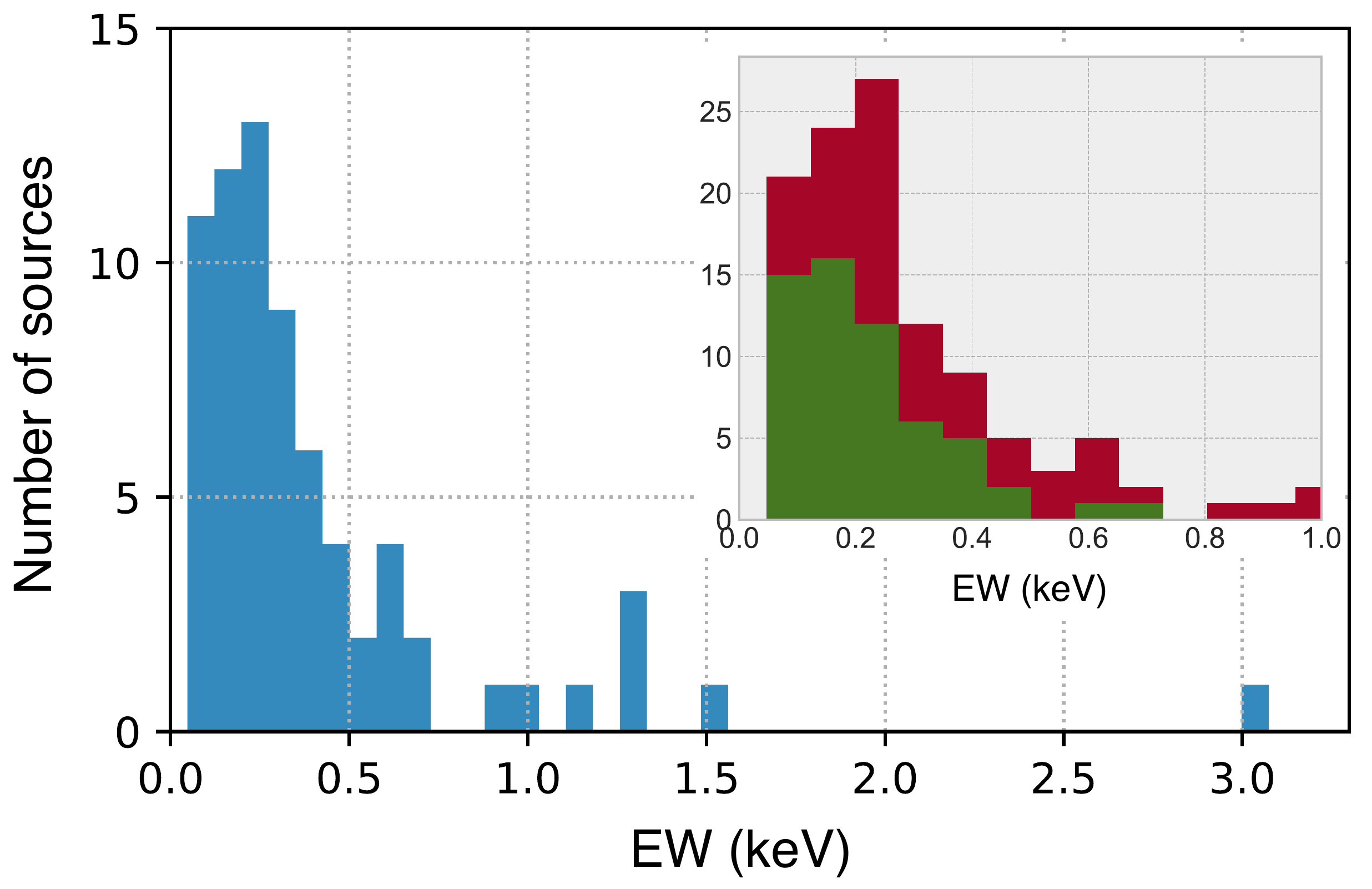}}
\caption{Main: Fe K line equivalent width distributions of the 71 sources with the line detection at the 90\% significance. Inset: the same distribution for the lines of the 68\% line detection (red) and that for the upper limits of no detection (green).}
\label{fig:ewdist}
\end{figure}

There are 71 sources with Fe K line detection at the 90\% confidence
level or higher. The majority (85\%) of the lines are found at 6.4 keV
and the rest are found at 6.7 keV and 6.97 keV with a comparable
share. The distribution of Fe K EW stretches up to 3 keV but most
lines have EW smaller than 0.6 keV (Fig. \ref{fig:ewdist}). The number
of Fe K line sources increases to 119 when the detection threshold is
relaxed to the 68\% confidence level. The composition of line energy
slightly changes and the fraction of the high-energy lines increases
to 25\%. The EW distribution of these sources is shown in the inset
of Fig. \ref{fig:ewdist} along with the upper limits for non-detected
sources. It suggests that the majority of sources with no line
detection have EW smaller than 0.2 keV and they would be populated
increasingly toward smaller EW if the spectral sensitivity were
sufficient.

The median line EW for all the 180 sources inspected is 0.15 keV and
the 68\% interval of the bootstrap error is (0.14-0.17) keV (Table
\ref{tab:median_zgrp}). No significant differences are found between
the median EW values for the three redshift ranges.

\subsubsection{Broad Fe K lines}

\begin{figure}
\centerline{\includegraphics[width=0.5\textwidth,angle=0]{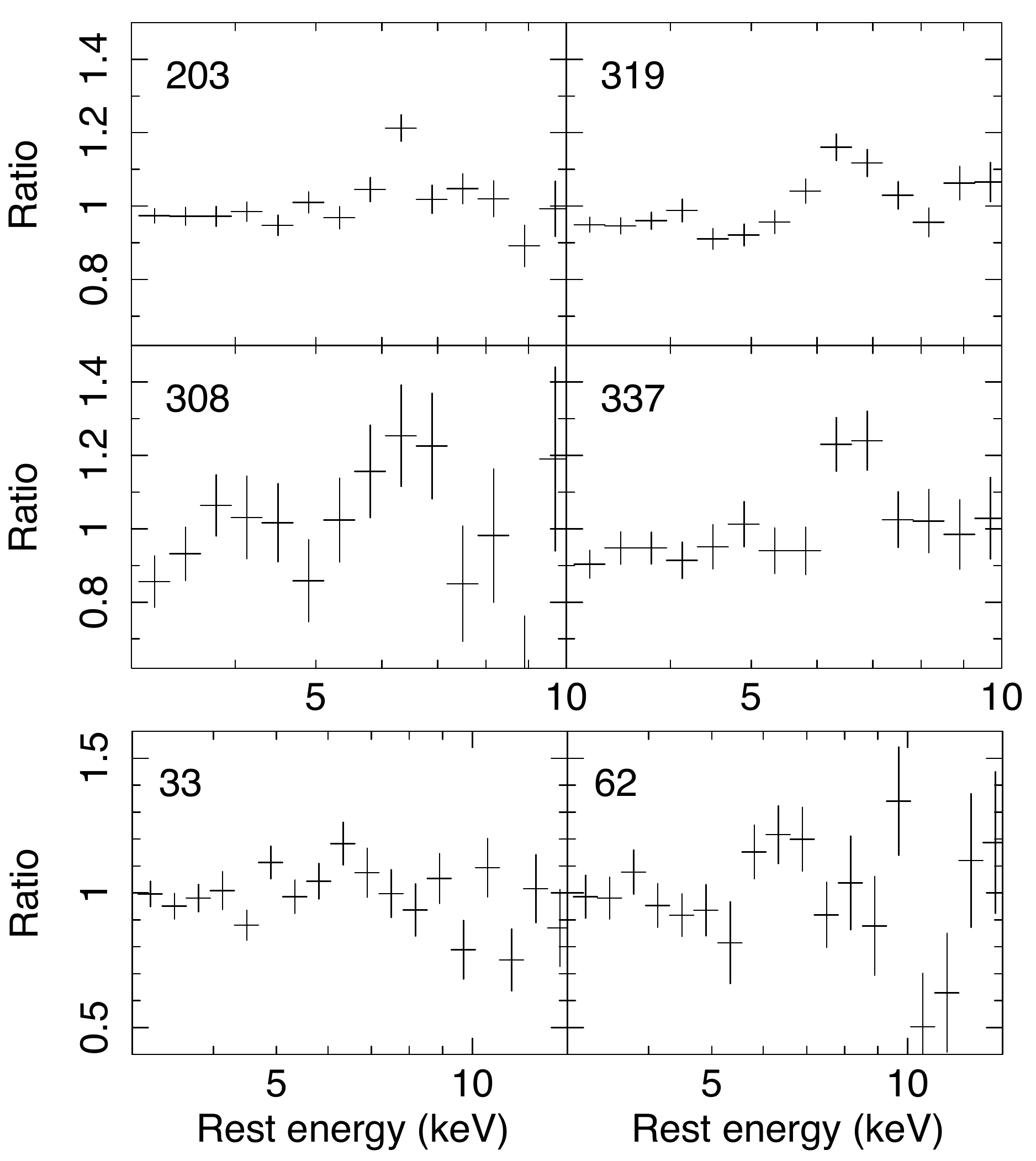}}
\caption{Plot of spectral data of six unobscured sources exhibiting
  broad Fe K features. The data are divided by the best-fit
  power-law. The source identification number is indicated. The upper
  four sources come from the lowz group while the lower two from the
  highz group (see Fig. \ref{fig:ecdf_bl})}.
\label{fig:rabl6}
\end{figure}

\begin{figure}
\centerline{\includegraphics[width=0.5\textwidth,angle=0]{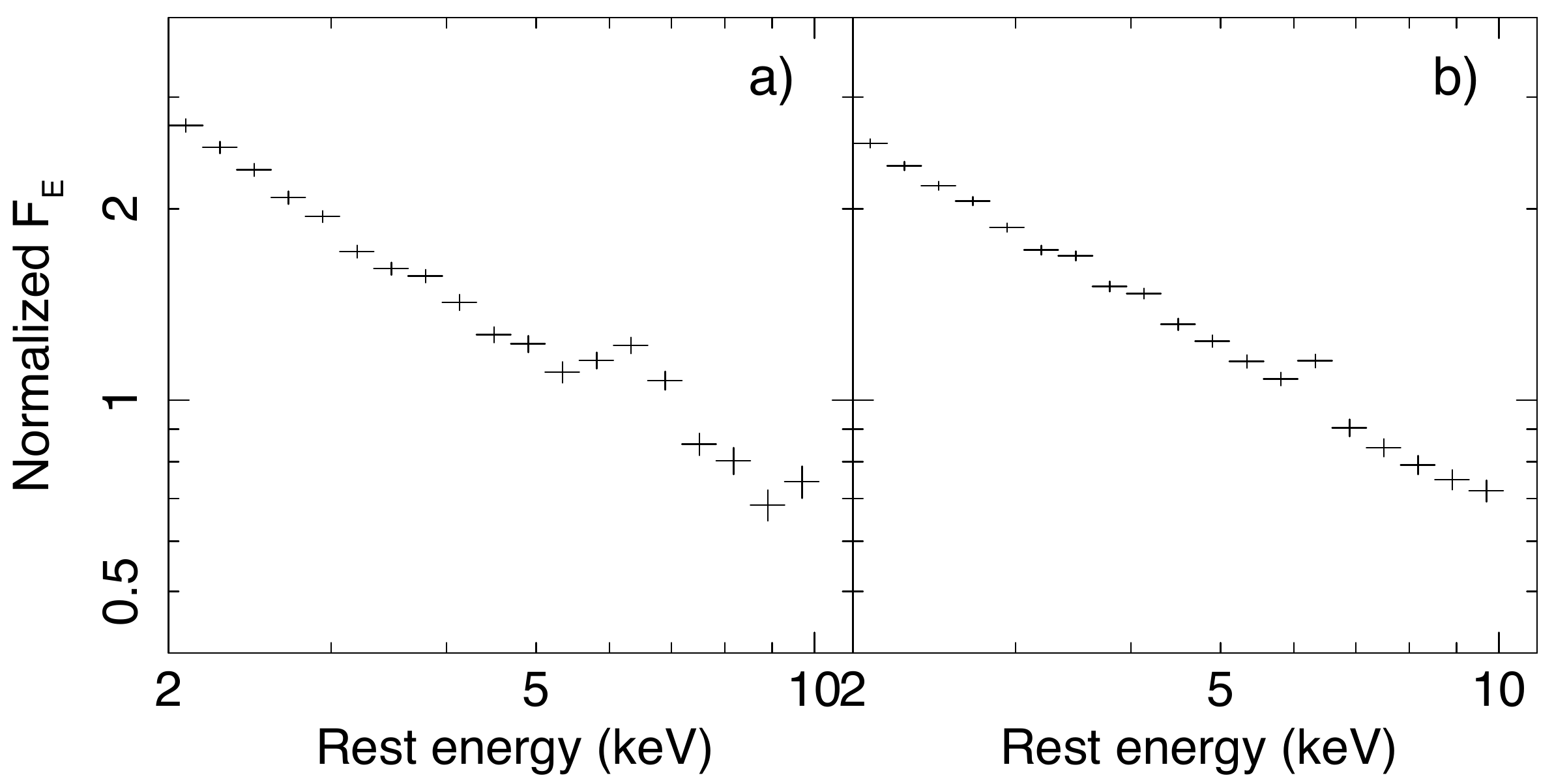}}
\caption{Stacked spectra of the six broad Fe-line AGN (a) and the
  other 15 AGN (b) of the 21 unobscured sources with highest quality data. }
\label{fig:blnlmeansp}
\end{figure}

\begin{figure}
\centerline{\includegraphics[width=0.35\textwidth,angle=0]{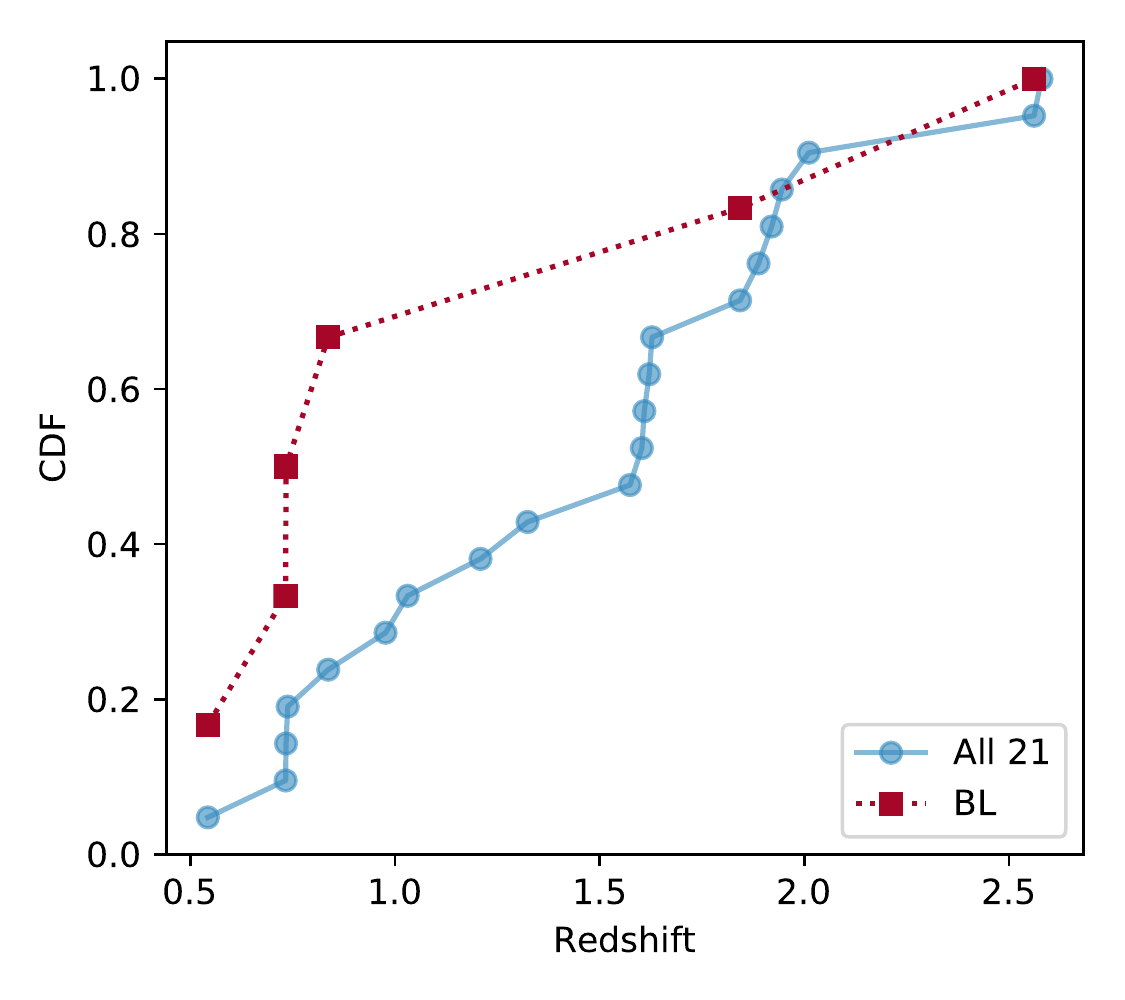}}
\caption{Cumulative distribution functions of the redshifts of the 21
  unobscured AGN with good data quality (blue circles) and those of
  the six in which broad Fe lines are detected (red squares). The
  redshifts of the sample AGN are spread over the range of 0.5-2.6 with
  some clustering in the $z = $1.6-2 range. In contrast, broad line
  detections may be frequent at low redshifts ($z<0.9$).}
\label{fig:ecdf_bl} 
\end{figure}

We investigated how common broad Fe K features are among the
unobscured AGN. Detecting a broad emission component requires high
quality data because of the low contrast against the continuum
\citep[e.g.][]{delacalle10}. The two brightest sources (PID 203 and PID 319)
in the field are unobscured AGN and indeed are broad Fe K line emitters
\citep{i15brtwo}. As a guide for data quality, we adopt the signal to
noise ratio of the rest-frame 3-10 keV ($sn_{e310}$: all the three
EPIC cameras combined together\footnote{$sn_{e310}$ can be found in
  the electronic form of Table \ref{tab:big}.}) and selected 21
unobscured sources with $sn_{e310}>30$ (including the above two
brightest sources). Their rest-frame 2-10 keV, low-resolution spectra
used in Sect. 4.1 were examined by modelling with a power-law
continuum with a Gaussian line. The line centroid is fixed at 6.4 keV
and the line width is fitted as well as line intensity. Because of our
rest-frame spectral resolution setting, fits are sensitive to line
broadening of $\sigma\ga 0.3$~keV. This degree of broadening is
appropriate for relativistically broadened line
\citep[e.g.][]{fabian00}. It also minimises the detection bias against high
redshift sources in which Fe emission appears in energies where
spectral resolution is lower than in lower redshift sources.

We detected line broadening (when the lower bound of the 90\%
confidence interval of the line width is positive) in six sources listed
in Table \ref{tab:bl6}, in which line widths, line EW, and the source
flux measured in the observed 1-5 keV band can be found. In the other
15 sources, Fe lines are either unresolved or not detected. The
broad-line detection rate is thus 29\% in this sample of 21
sources. \citet{delacalle10} obtained a detection rate of 36\% for
the 31-source flux-limited FERO sample of nearby bright Seyfert galaxies
observed with XMM-Newton. The 95\% credible interval of our broad-line
detection rate when the FERO result is used as a prior is 22-46\%. The
composite spectra of the six broad-line sources and the rest are shown in
Fig. \ref{fig:blnlmeansp} to illustrate their differences.

Four of the six broad-line AGN belong to the lowz group ($z<1$) while
the remaining two lie at $z=1.8$ and $z=2.6$. There is no broad line
detection in the nine sources in the intermediate redshifts between $z=
$0.9-1.8. The 21 sources we investigated are distributed over $z =
$0.5-2.6 (Fig. \ref{fig:ecdf_bl}) with no bias towards low
redshifts. Ten sources are clustered in the $z =$1.6-2 range where one
broad-line detection is recorded. The cumulative distribution
functions of the sample and the broad-line detected sources in
Fig. \ref{fig:ecdf_bl} seem largely deviated, suggesting a possible
change in broad-line detection rate with redshift, or deficit of
broad-line detection in the intermediate redshifts, although there is
no obvious reason why it should change. A two-sample K-S test on the
broad-line and non-broad-line samples give a p-value of 0.03,
suggesting marginally that they might come from a different
distribution. 

Excluding the two brightest sources (PID 203 and PID 319) that feature
exceptionally high-quality data, $sn_{e310}$ is comparable between the
sources in the three redshift groups, indicating that data quality is
unlikely to be the reason. If the data quality were all that mattered, the stacked
spectrum of the 15 sources without individual broad-line detections
would show a sign of line broadening, which is not seen
(Fig. \ref{fig:blnlmeansp}). Other properties such as median X-ray
luminosities of the sources with and without broad-line detections are
compatible (44.1 and 44.2 in log $L_\mathrm{X}$ [\ergps],
respectively). Assuming external conditions are equal among the sample
sources, the hypothesis that broad-line detection rate remains
constant ($\sim 29$\%) over the whole redshift range was examined by
random sampling of six of the 21 redshifts. If we set four redshifts
below $z =0.9$ being found among the chosen six as a test measure
given our observation, the p-value is estimated to be 0.01, indicating
that our result is a relatively rare case under the hypothesis of the
constant broad Fe K line detection rate. The complete lack of
broad-line detection in the nine sources at $z=$0.9-1.8 could occur at a
probability of 0.03 by chance. Given the small sample size and the
presence of two exceptionally bright sources with detected broad lines
\citep[PID 203 and 319, see][]{i15brtwo}, we withhold from drawing a
conclusion on a possible redshift variation of broad Fe lines.


\begin{table}
\caption{Six broad Fe K line sources}
\label{tab:bl6}
\centering
\begin{tabular}{rcccc}
PID & $z$ & $\sigma $ & EW & $F_\mathrm{1-5}$ \\
 & & keV & keV & $10^{-14}$ \ergpspsqcm \\[5pt]
33 & 1.843 & $0.46^{+0.50}_{-0.25}$ & $0.23^{+0.15}_{-0.10}$ & 3.0 \\
62 & 2.561 & $0.44^{+0.21}_{-0.11}$ & $0.46^{+0.18}_{-0.13}$ & 0.78 \\
203 & 0.543 & $0.30^{+0.15}_{-0.07}$ & $0.23^{+0.06}_{-0.05}$ & 7.6 \\
308 & 0.733 & $0.48^{+0.07}_{-0.05}$ & $0.57^{+0.09}_{-0.06}$ & 1.2 \\
319 & 0.734 & $0.43^{+0.11}_{-0.09}$ & $0.22^{+0.05}_{-0.05}$ & 7.5 \\
337 & 0.837 & $0.38^{+0.34}_{-0.12}$ & $0.37^{+0.18}_{-0.11}$ & 2.8 \\
\end{tabular}
\tablefoot{Six unobscured sources in which broad Fe K emission is
  detected, out of the 21 bright sources. When a Gaussian of the
  centroid energy of 6.4 keV is fitted, and the 90\% confidence
  interval of the line width is positive, the source made this
  table. The line EW is measured in the galaxy rest frame. The source
  flux in the observed 1-5 keV range, $F_\mathrm{1-5}$, is shown for
  reference.}
\end{table}

\section{Exploratory spectral inspection with two X-ray colours and Fe K line strength indicator}

We present a few quick-look spectral inspection methods 
and compare them with the results obtained from the conventional
spectral fitting. When there are many spectra, a quick
exploratory analysis using these methods may be useful for characterising
the source population or to select sources with particular properties,
such as Compton-thick AGN, if its accuracy turns out to be reasonable.

We used the same data that was prepared for the broad-band spectral fit. The
data from the three cameras were averaged in the same way as those in
the spectrum atlas (Sect. 3) and converted to flux density units and
the energy scale was replaced by the rest-frame energy. The 27 energy
intervals over the rest 2-20 keV range are denoted by indicies $i=$ 0,
1...26. Hereafter we use these rest-frame flux density spectra for the
analysis below.

\subsection{Broad band continuum}

\begin{figure}
\centerline{\includegraphics[width=0.4\textwidth,angle=0]{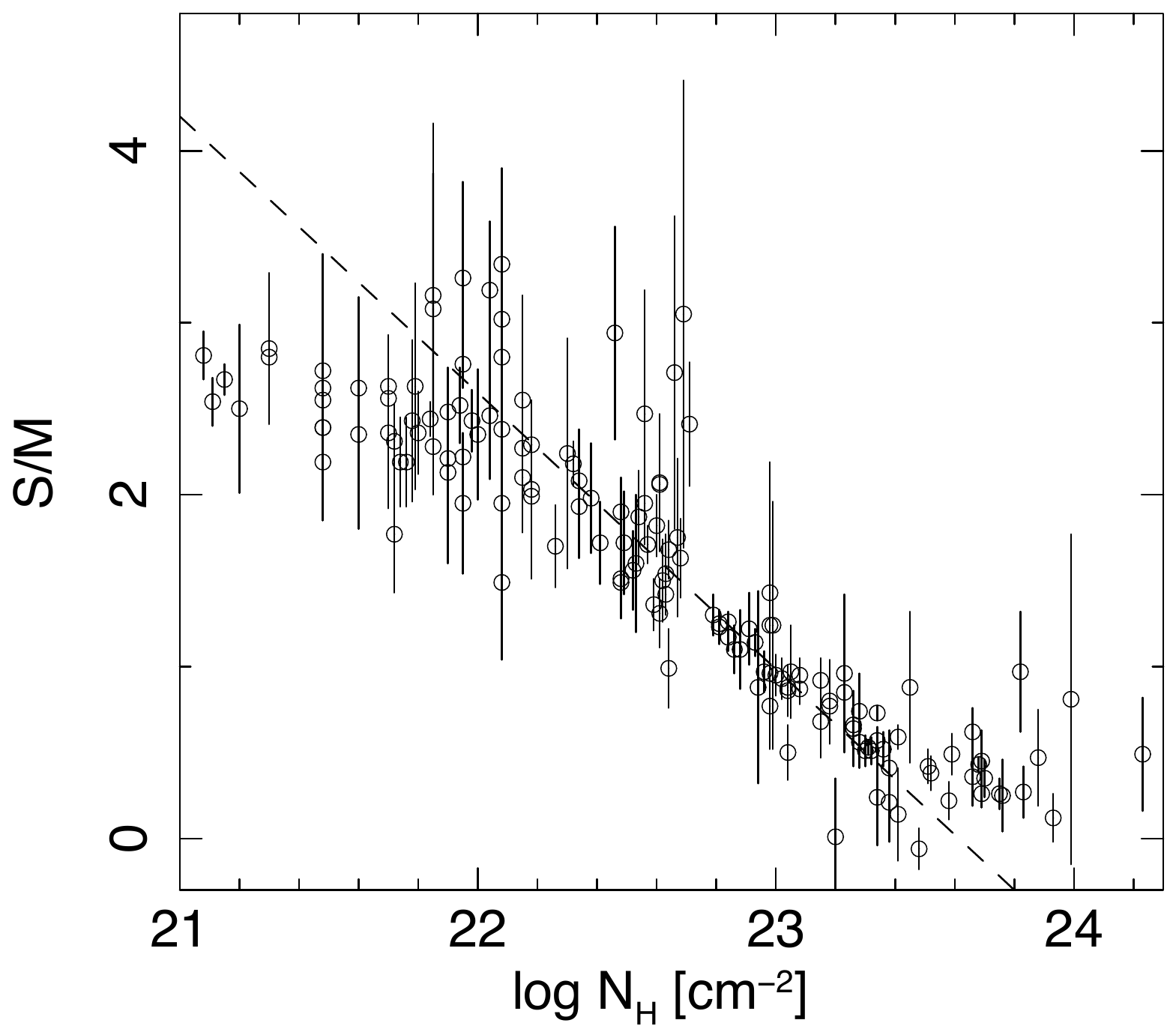}}
\caption{Plot of S/M colours against best-fitting \nH, obtained from
  the conventional spectral fit. The dashed line indicates the
  best-fitting relation in the range of 22-23.3 in log
  \nH\ [cm$^{-2}$] (see text).}
\label{fig:smnH}
\end{figure}

First, we define three rest-frame energy bands: S: 2-5 keV; M: 5-9
keV; and H: 9-14 keV, where S, M, and H are integrated flux densities
over the respective bands. Two X-ray colours, S/M and H/M, are then
obtained. We note that there is no information on H for the sources in
the lowz group because of the limited hard band data (see Table
\ref{tab:zgroups}), and thus only S/M colours are available. The S/M
colour is, in principle, a proxy for absorption if \nH\ lies in the
$10^{22}$ \psqcm\ - $10^{23}$ \psqcm\ range. Fig. \ref{fig:smnH} shows
a plot of S/M values against the best fitting \nH. On both ends of the
\nH\ range, S/M values are saturated because the sensitivity of S/M to
\nH\ becomes weak due to the limited band passes of S and M,
respectively. This results in a flatter correlation for the full range
of log \nH\ (21-24): the slope comes out as $a = -1.10\pm 0.02$ for
S/M $= a \times \mathrm{log}\thinspace N_\mathrm{H} + b$. The S/M
ratio is sensitive in the middle part, that is, in the range of 22-23.3 in
log \nH, the slope becomes steeper ($a\sim -1.6$) with a good fit (a
$\chi^2$ test gives a chance probability of 2\% with $\chi^2 =
110.5$ for 82 degrees of freedom). Thus, S/M can be considered as a
reasonably good \nH\ indicator in this range with the parameters, $a =
1.60\pm 0.05$ and $b = 38\pm 1$ in the above formula.

\begin{figure}
\centerline{\includegraphics[width=0.4\textwidth,angle=0]{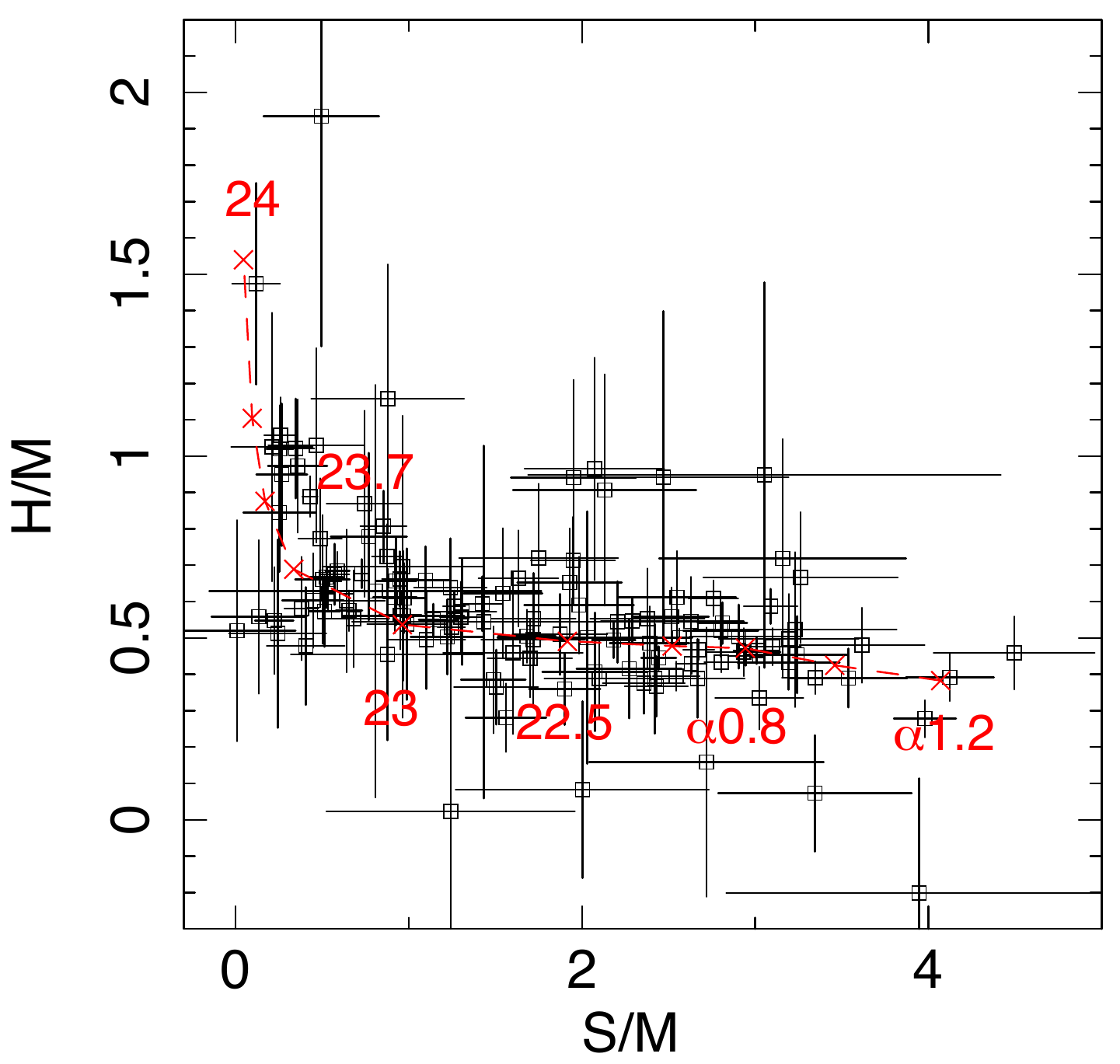}}
\caption{Colour-colour (H/M - S/M) diagram for the midz and highz
  sources. The locus in dashed red line indicates the evolution of
  (S/M, H/M) when a power-law continuum of energy index $\alpha = 0.8$
  is modified by cold absorption with various column densities: log
  \nH\ [cm$^{-2}$] of 22, 22.5, 23, 23.5, 23.7, 23.85 and 24, as
  marked by cross symbols. Colours for a simple power-law with no
  absorption with $\alpha $ of 0.8, 1.0, and 1.2 are also marked.}
\label{fig:colcol}
\end{figure}

On the other hand, the H/M colour is sensitive to absorption only when
\nH\ exceeds $10^{23}$ \psqcm, complementing S/M in the \nH\ estimate
at the higher value range. Figure \ref{fig:colcol} shows the
colour-colour diagram of S/M and H/M for the midz and highz sources
\citep[similar to that in][]{i12cdfs}. Predicted positions of sources
with a power-law spectrum with $\alpha = 0.8$ and various absorbing
columns, as well as those with steeper continuum slopes (up to $\alpha
= 1.2$) and no absorption, are indicated in the figure. Sources
located in the upper left corner are likely to be strongly absorbed sources,
such as \nH $> 5\times 10^{23}$ \psqcm.

\subsection{Estimate of Fe K band continuum and strong Fe line sources}

\begin{figure}
\centerline{\includegraphics[width=0.5\textwidth,angle=0]{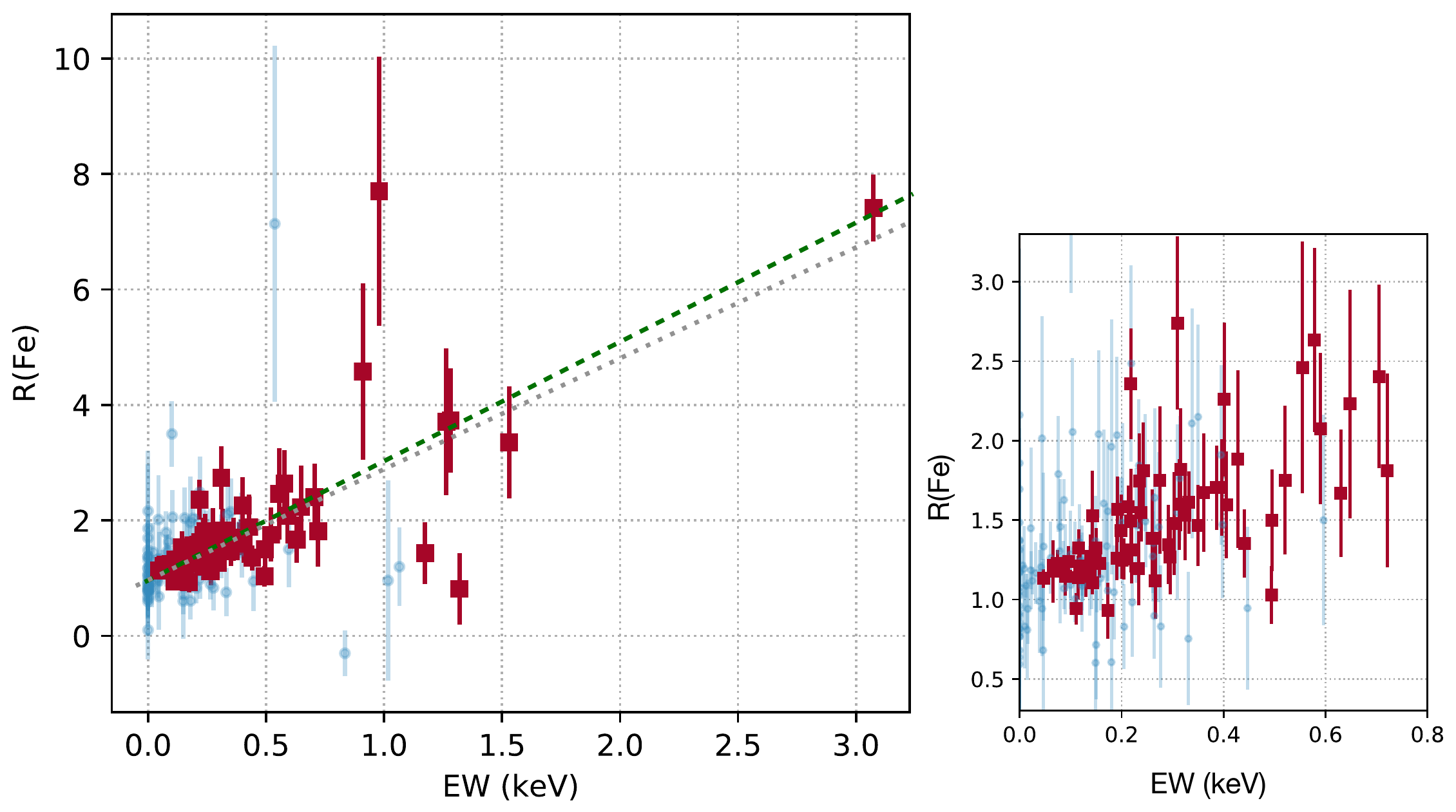}}
  \caption{Left: Plot of the Fe K line strength indicator, $R$(Fe),
    against the Fe K line equivalent width (EW) obtained from spectral
    fitting. The filled squares (red) show data for Fe lines with the
    $\geq 90$\% detection while the circles (light blue) show data for
    weak/no detection for the whole sample. The dashed line indicates
    the fitted linear correlation (see text). The dotted line
    indicates the expected relation when the continuum level estimate
    of the $i=13$ bin is accurate. Right: Same plot for the
    crowded region of EW $<0.8$ keV.}
  \label{fig:ew-rfe}
\end{figure}

Secondly, we consider a relatively crude method for noisy spectra to
select sources with strong Fe K line emission. Strong Fe line emission
with EW $\ga 1$ keV is an unique characteristic of a
reflection-dominated spectrum from heavily obscured AGN. This proposed
method may help quickly select such strong Fe line sources.

With the adopted spectral binning, the $i=13$ bin corresponds to the
6.0-6.6 keV range where a cold Fe K line (6.4 keV) would appear. The
first step is to estimate the continuum level in the $i=13$ bin using
the data of neighbouring intervals. We adopt as the flux density in
bin $i=10$ at 4.9 keV ($i=16$ at 8.2 keV) the median of the flux
densities in bins $i=$ 8-12 ($i=$ 14-18). A logarithmic mean of these
two values of flux density gives an estimate of the continuum level at
the $i=13$ bin. This operation is equivalent to obtaining a continuum
level at $i=13$ by applying a power-law running through the two median
points because of the logarithmically equal intervals by design. A few
examples of its application to real data can be found in Appendix B.

Thus we have the continuum level estimates in the Fe K band for all
the sources. Taking the ratio of $f_{13}/f^{\prime}_{13} \equiv
R(\mathrm{Fe})$, where $f_{13}$ is the flux density of the source
measured at $i=13$ and $f^{\prime}_{13}$ is the estimated continuum
level, gives a measure of excess emission in that interval, likely due
to cold Fe K line emission. This ratio obtained for each source is
given in Table \ref{tab:big}. Since the $i=13$ bin has an energy
interval of 0.54 keV, $R(\mathrm{Fe})$ and Fe K line EW would be
related by $R(\mathrm{Fe}) = 1 + (\mathrm{EW}/0.54)$, if the continuum
estimate for the $i=13$ bin is accurate. While the ratio being unity
means no Fe line, selecting sources with $R({\rm Fe}) > 2$, for
instance, results in their spectra having an Fe K line with EW
$>0.54$ keV.

We examine how $R$(Fe) is actually related to the EW measured by
spectral fitting. First, the 71 sources with Fe lines of $\geq 90$\%
detection (Sect. 4.3.1) were inspected (Fig. \ref{fig:ew-rfe}). The
correlation between $R$(Fe) and EW is reasonably strong
($r=0.76$). Fitting a linear relation of $R$(Fe) $= \alpha + \beta
\mathrm{EW}$ gives $\alpha = 0.97^{+0.06}_{-0.02}$ and $\beta =
2.0^{+0.1}_{-0.4}$. The fraction of outliers is about 5\%. The
intercept, $\alpha$, is consistent with being 1. The slope, $\beta$,
is slightly steeper than the expected value, $\beta_0 = 1/0.54 = 1.85$
but is compatible. The uncertainty of the slope extending
predominantly toward lower values indicates that the best-fit slope is
largely driven by the single data point of large EW ($\sim 3$ keV of
PID 215) which lies very close to the best-fit. Since the error
interval was obtained by bootstrap, this implies that, when this data
point is excluded, the relationship becomes flatter than the best-fit
case and actually has a slope closer to the theoretical value. When
the data range is limited to the crowded region on the bottom-left
corner (right panel of Fig. \ref{fig:ew-rfe}), where $63/71\approx
90$\% of the sources are located, the correlation, albeit not very
tight, still persists ($r=0.66$). Re-fitting the linear model yields
$\alpha = 1.04^{+0.11}_{-0.09}$ and $\beta = 1.8^{+0.2}_{-0.4}$. There
are not many catastrophic outliers among the sources with no or weak
detection of Fe lines, as shown in Fig. \ref{fig:ew-rfe}. Therefore,
statistically, $R$(Fe) performs as an approximate EW indicator as
expected, with a dispersion of $\sigma = 0.32\pm 0.02$ among the
detected lines.

Without spectral fitting, whether a given $R$(Fe) is associated with a
significantly detected Fe line is not immediately clear. One measure
is the error bars of $R$(Fe) itself, since it reflects the data quality
of the Fe K line interval. Another is the broad-band data quality such
as $sn_{e310}$ (for the rest-frame 3-10 keV). Although good quality
data do not always guarantee a line detection, significant line
detections tend to be more frequent as $sn_{e310}$ increases, for example, the
probability of an Fe line being $\geq 90$\% significance is $\geq
74$\% when $sn_{e310}$ exceeds 30.

A few caveats on $R$(Fe) are: firstly, $R({\rm Fe})$ may sometimes
overestimate EW when a source is heavily absorbed (a prime example is
PID 316). A strongly absorbed spectrum exhibits a strong continuum
curvature with a prominent continuum discontinuity around 7 keV due to
an Fe K edge. Our simple interpolation method would then tend to place
$f^{\prime}_{13}$ lower than actually is, leading to an overestimate
of $R({\rm Fe})$ (that is, the continuum discontinuity is viewed as
part of an Fe line excess). A prime example is PID 316 but such a
large deviation occurs only in two sources. Secondly, this method
works only when the line photons fall well within the $i=13$ bin
(6.0-6.6 keV). Some sources show higher ionisation lines of Fe {\sc
  xxv} and Fe {\sc xxvi} (Sect. 4.3.1) which peak outside $i=13$ and
R(Fe) would fail to capture those lines. However, since the majority
of the Fe K lines appear to be cold 6.4 keV lines, $R$(Fe) should work
most of the time.

\section{Composite spectra}

We use spectral stacking to investigate average properties of the
XMM-CDFS sources. Here we use normalised, individual rest-frame
spectra to avoid a bias for bright sources. The normalising point is
the $i=13$ bin and each spectrum was normalised to the estimated
continuum flux, $f^{\prime}_{13}$, obtained above for individual
sources. Then we apply median stacking (see Appendix C). The 68\% confidence interval
for the median is estimated by bootstrap.

\begin{figure}
\centerline{\includegraphics[width=0.35\textwidth,angle=0]{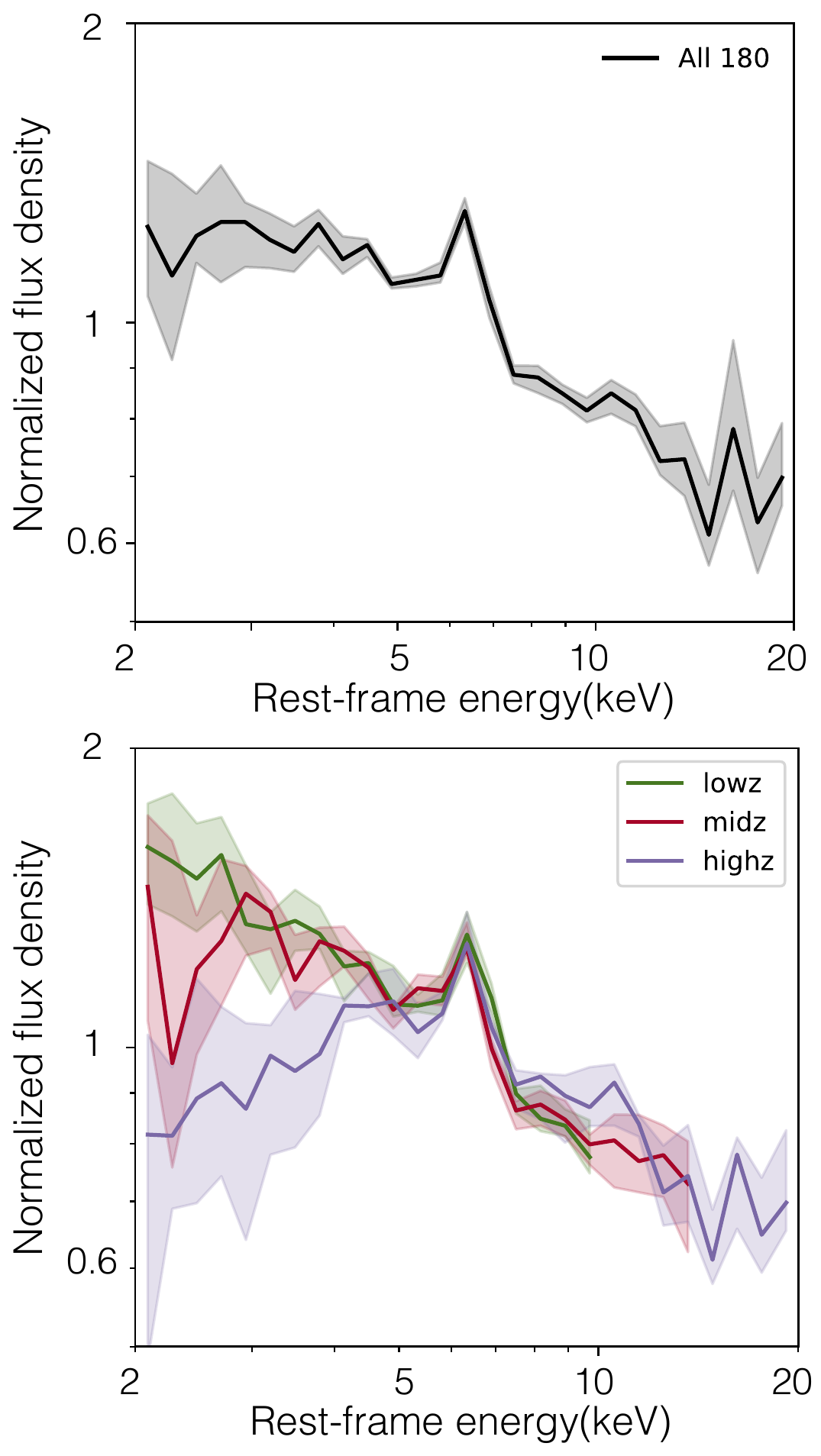}}
  \caption{Upper panel: Rest-frame 2-20 keV composite spectrum of
  all the 180 sources at $0.4<z<3.8$. The shaded area indicates 68\%
  confidence intervals from bootstrap. Lower panel: Same spectra
  for the lowz, midz and highz sources.}
\label{fig:medsp_all}
\end{figure}

Figure \ref{fig:medsp_all} shows the composite spectrum of all the 180
sources. Fitting a simple absorbed power-law for the continuum gives
$\alpha = 0.54^{+0.04}_{-0.05}$, \nH $=(1.3\pm 0.4)\times 10^{22}$
\psqcm. We note that the fitted continuum precisely matches the value
of unity at $i=13$, suggesting the continuum level estimates of
individual sources work well. A superposition of spectra with various
degrees of absorption makes this hard spectrum. The presence of the
absorption cut-off represented by the \nH\ value above in the
composite spectrum suggests that the \nH\ distribution is not
flat. Probably absorbed sources with \nH\ greater than $10^{22}$
\psqcm\ dominate at the typical XMM-CDFS source flux. The Fe line
peaks at $i=13$ with wings on both sides. Fe line emission in the
stacked spectrum is expected to be broadened to some degree since
lines from sources at high redshifts are observed at low energies
where the energy resolution, $\Delta E/E$, of the EPIC cameras
degrades (approximately $\propto E^{-0.5}$) and they result in the
artificial broadening \citep[see
  e.g.][]{i12cosmos,falocco13}. However, at $z\sim 2$, the
broadening is expected to be $\sigma = 0.15$ keV, a narrow line at 6.4
keV should be contained within the $i=13$ interval. The presence of
the wings hence means line broadening. The intrinsic line broadening
is estimated to be $\sigma =0.27\pm 0.1$ keV, when fitting a Gaussian
and taking into account the effect of de-redshifting, assuming the
median redshift of the sample $z=1.34$. However, this is more likely
due to the spread of the line energy than to many sources emitting
intrinsically broad Fe lines, since we observed that Fe K line
energies would be distributed around 6.4 keV with a dispersion of 0.26
keV when line energy is fitted instead of being fixed
(Sect. 4.3.1). The line equivalent width is found to be EW $=
0.25^{+0.06}_{-0.04}$ keV, when the line broadenig is taken into
account. When a narrow 6.4 keV line is assumed, the EW is reduced to
$0.14\pm 0.04$ keV.

\subsection{Redshift evolution of spectra}

\begin{figure}
\centerline{\includegraphics[width=0.35\textwidth,angle=0]{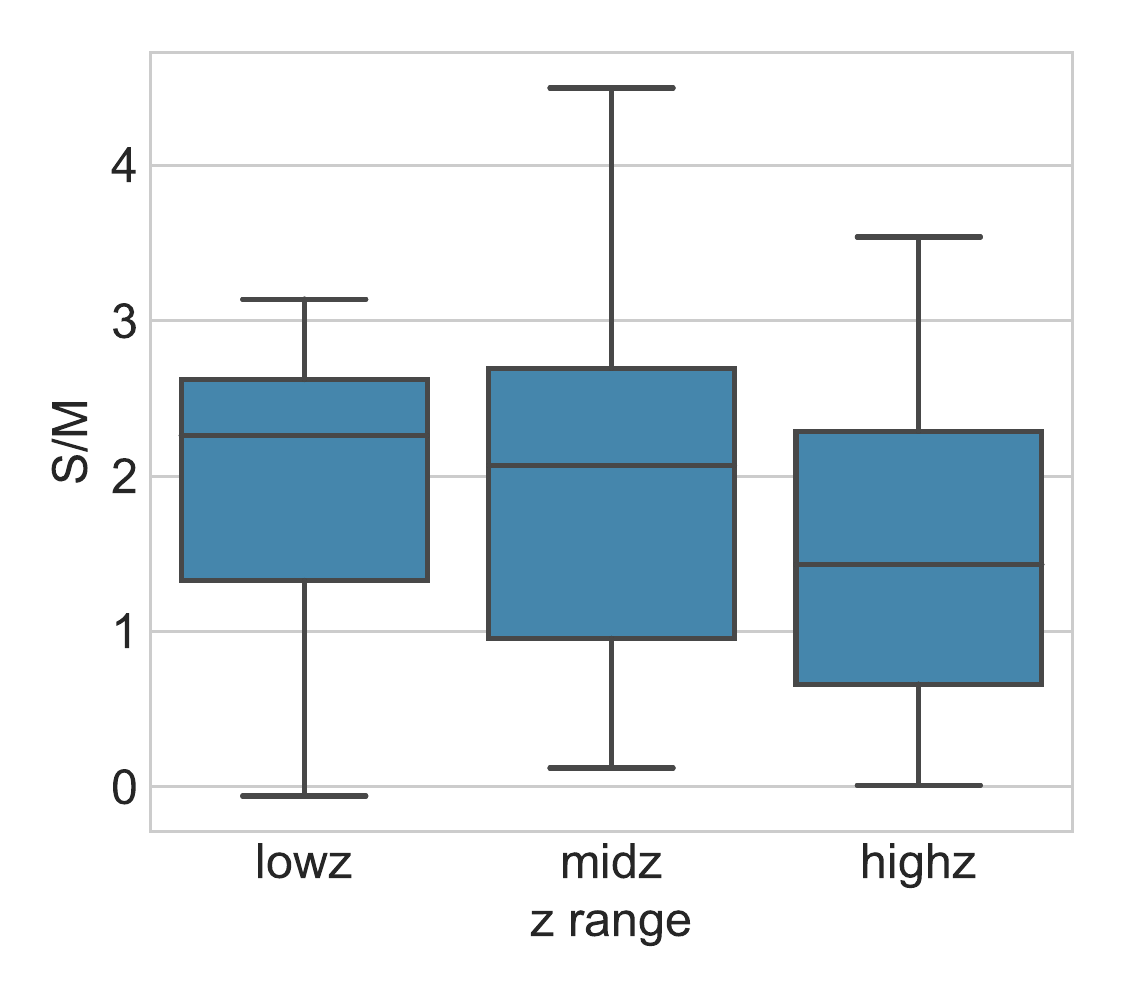}}
\caption{The S/M X-ray colour distributions of lowz, midz, and highz groups.}
\label{fig:smdist}
\end{figure}

We examine how the average spectral shape evolves between the three
redshift groups (Table \ref{tab:zgroups}). The composite spectra for
the lowz, midz and highz groups are shown in the lower panel of
Fig. \ref{fig:medsp_all}.
It is readily clear that the spectrum becomes more absorbed as
redshift increases. When an absorbed power-law, which has a common
slope but variable \nH\ between the three spectra, with a narrow
Gaussian Fe line at 6.4 keV is fitted, the slope is found to be
$\alpha = 0.52\pm 0.03$, comparable to that obtained for the whole
sample, as expected. \nH\ values found in the three spectra are given
in Table \ref{tab:fit_3zgrps}. The difference between \nH\ for lowz
and midz is marginal but that for highz is significantly
larger. Although the harder spectral slope than that of unobscured AGN
($\alpha \sim 0.8$) means that the obtained \nH\ values are not real
values of typical absorbing columns, they show that absorption becomes
progressively larger toward higher redshifts. This is in agreement
with the \nH\ distributions (Fig. \ref{fig:histo_nH}) and the X-ray
colour analysis (Fig. \ref{fig:smdist}). The EW of Fe K lines are
comparable between the three spectra: 0.14, 0.13 and 0.12 keV ($\pm
0.04$ keV), respectively, as illustrated by Fig. \ref{fig:medsp_all}.

\begin{table}
\begin{center}
\caption{Properties of composite spectra for sources in the three redshift ranges.}
\label{tab:fit_3zgrps}
\begin{tabular}{cccc}
$z$ range & $\alpha $ & \nH & EW \\
&& $10^{22}$ \psqcm & keV \\[5pt]
lowz & $0.52\pm 0.03$ & $0.4\pm 0.4$ & $0.14\pm 0.04$ \\
midz &  = & $1.0\pm 0.4$ & $0.13\pm 0.04$ \\
highz & = & $3.4\pm 0.5$ & $0.12\pm 0.04$ \\
\end{tabular}
\begin{list}{}{}
\item[] Note:\ The continuum slopes for the three spectra are
  assumed to be identical and tied together in the fit. The above
  results on the Fe K lines are obtained, assuming a narrow line at
  6.4 keV.
\end{list}
\end{center}
\end{table}

\subsection{Spectroscopic vs. photometric redshifts}

\begin{figure}
\centerline{\includegraphics[width=0.5\textwidth,angle=0]{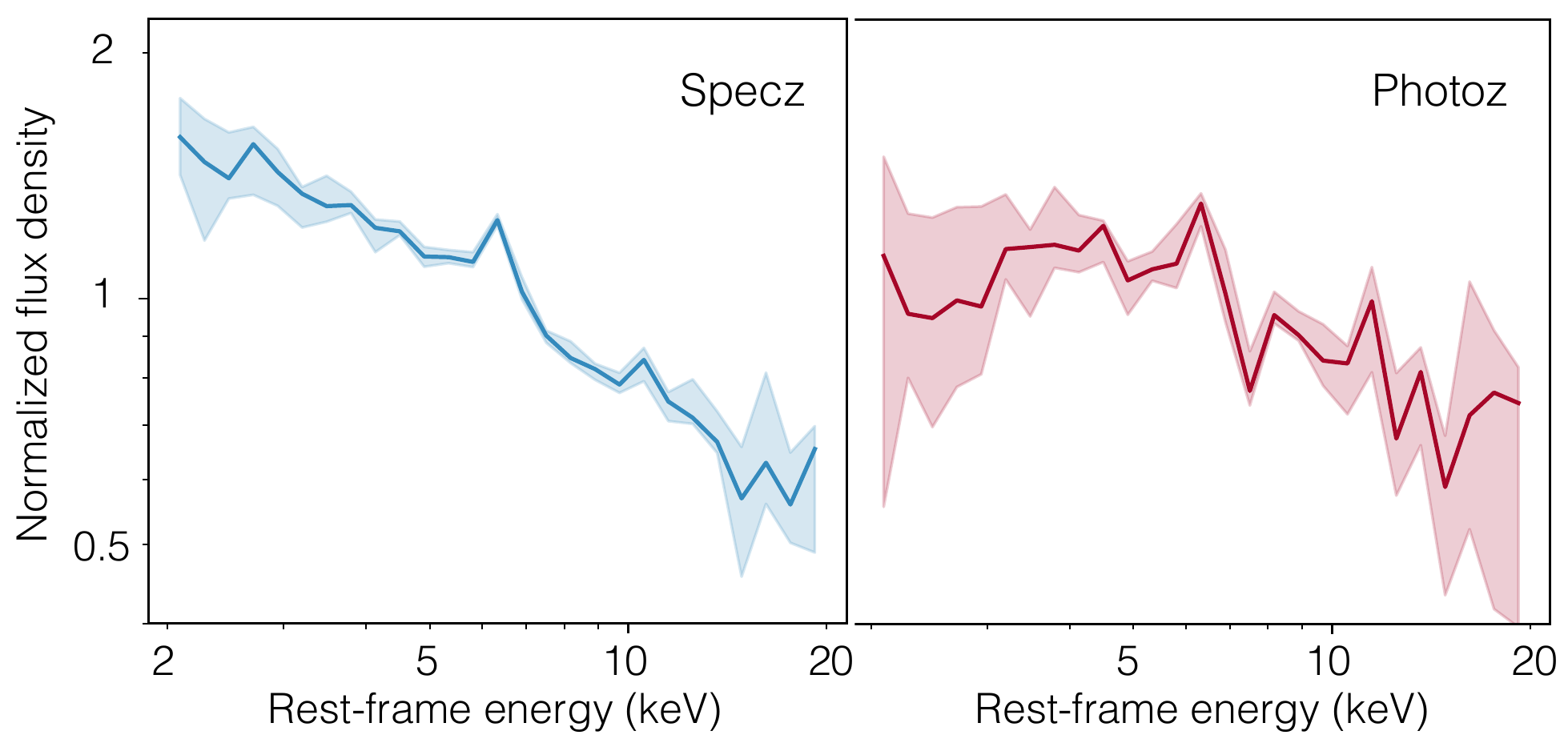}}
\caption{Composite spectra of sources with spectroscopic redshifts and
  and photometric redshift estimates.}
\label{fig:medsp_spph}
\end{figure}

Figure \ref{fig:medsp_spph} shows composite spectra of sources with
spectroscopic and photometric redshift estimates, respectively. The
median properties of these sources are given in Table
\ref{tab:median_zgrp}. The number of sources with spectroscopic
redshifts (131) is much larger than with photometric redshifts
(32). We are aware that not all the photometric redshifts are reliable
\citep[the typical photometric redshift accuracy and outlier rate for the
CDFS X-ray sources are $\sigma_\mathrm{NMAD} = 0.014$ and $\eta = 5.43$\%,
respectively for the work by][]{hsu14} but the presence of a
clear Fe K line in the composite spectrum suggests that the majority
of the photometric redshifts are of reasonable accuracy. An absorbed
power-law fit gives results shown in Table \ref{tab:fit_spph}. The
sources with photometric redshift are faint in optical and the
relatively hard composite spectrum agrees that their optical faintness
is related to nuclear obscuration.

\begin{table}
\begin{center}
\caption{Properties of composite spectra for sources with $z_{\rm sp}$
  and $z_{\rm ph}$.}
\label{tab:fit_spph}
\begin{tabular}{lccc}
$z$ type & $\alpha $ & \nH & EW \\
&& $10^{22}$ \psqcm & keV \\[5pt]
Specz & $0.57\pm 0.05$ & $0.9\pm 0.5$ & $0.12\pm 0.01$ \\
Photoz & $0.44\pm 0.09$ & $3.0\pm 0.9$ & $0.13\pm 0.03$ \\
\end{tabular}
\end{center}
\end{table}

\subsection{Unobscured AGN}

\begin{figure}
\hbox{\includegraphics[width=0.24\textwidth,angle=0]{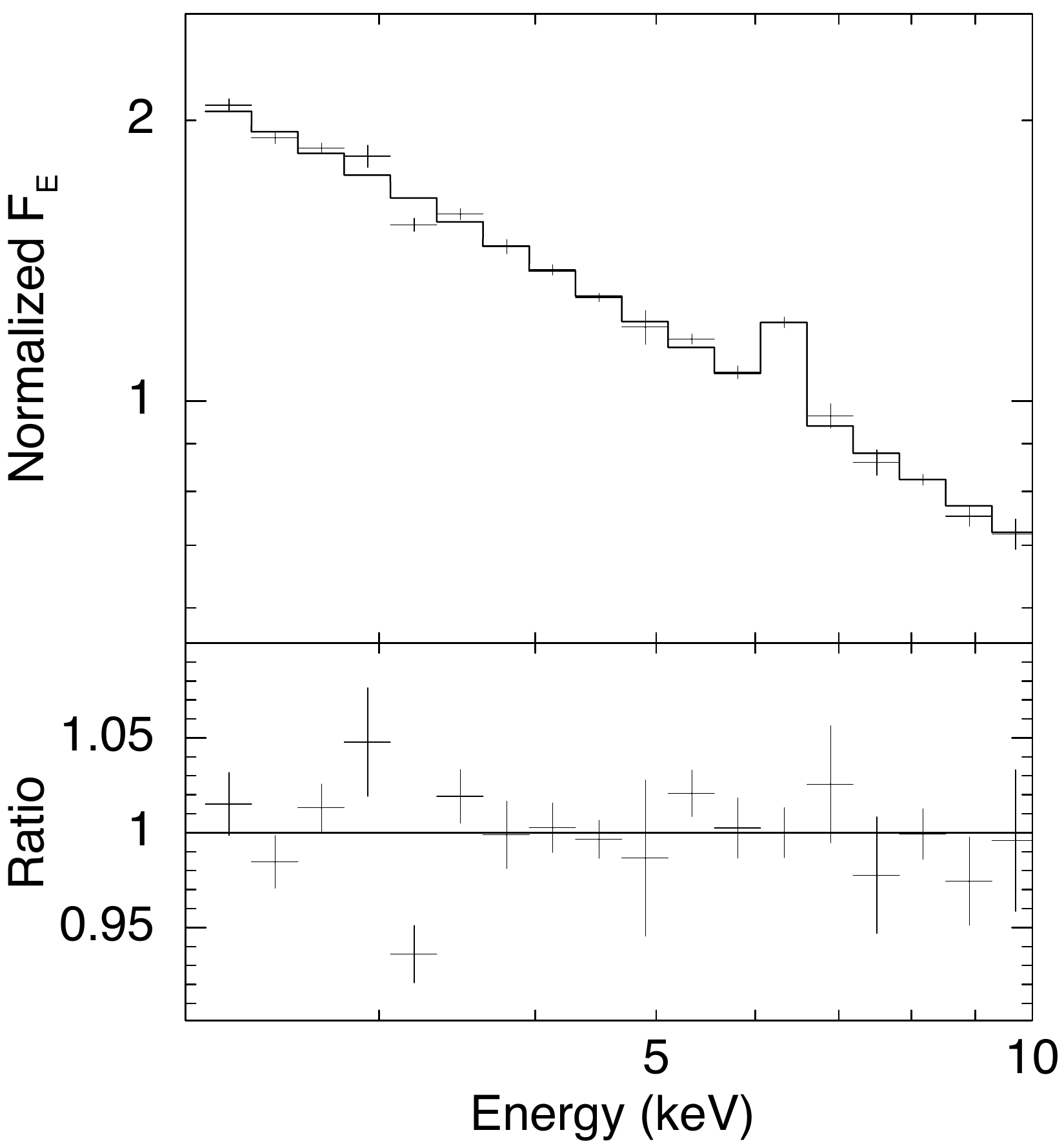}\includegraphics[width=0.253\textwidth,angle=0]{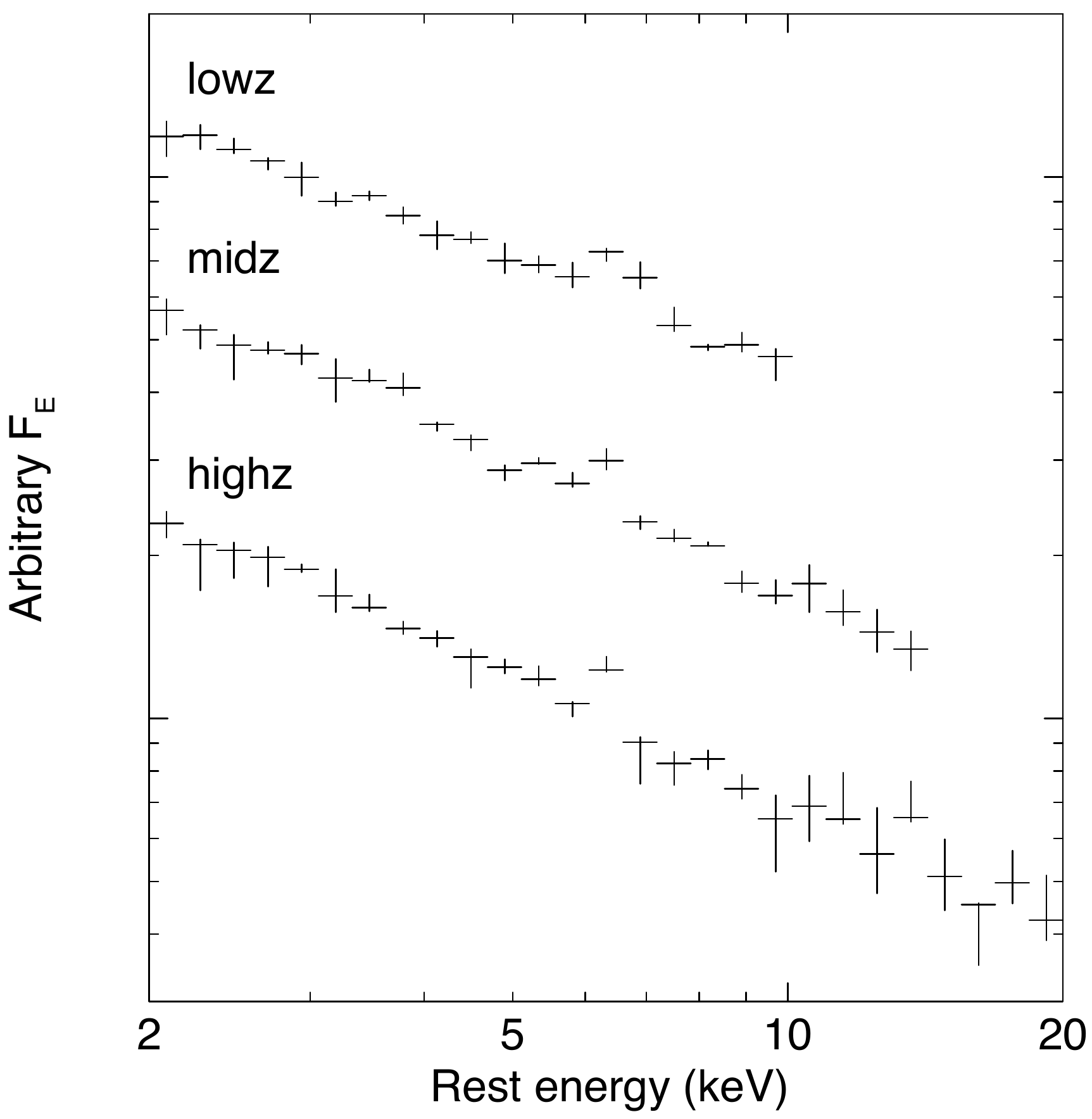}}
\caption{Left panel: Composite spectrum of all the 81 unobscured AGN
  with the best-fit power-law plus a narrow Gaussian for the Fe K
  emission and the residual in ratio of the data and the model. Right
  panel: Composite spectra of unobscured sources in the lowz, midz and
  highz groups. The spectra are plotted with arbitrary offsets for clarity.}
\label{fig:medsp_unobsc}
\end{figure}

X-ray obscured sources are defined as those with measured \nH\ larger
than $1\times 10^{22}$ \psqcm, excluding those for which the lower
bound of the $1 \sigma $ interval of \nH\ falling below the
threshold. Unobscured sources are thus those with absorbing
\nH\ undetected or smaller than $1\times 10^{22}$ \psqcm. There are 81
sources classified in this category. The median redshift and 2-10 keV
luminosity of these sources are \~z $= 1.13$ and log $L_{\rm X} =
43.60$ (log $L^{\prime}_{\rm X} = 43.64$). The composite spectrum
(Fig. \ref{fig:medsp_unobsc}) shows a power-law slope of $\alpha
=0.80\pm 0.02$, \nH $= (4.2\pm 0.7)\times 10^{21}$ \psqcm, and a Fe
K line EW of $0.11\pm 0.01$ keV.

Although the majority of these sources have spectroscopic redshifts (70
out of 81, most of the sources without spectroscopic redshifts are
from the midz group), we construct a composite spectrum by using only
sources with spectroscopic redshifts, in the interest of examining
line broadening. The spectral shape agrees with that including the
photometric redshift sources. No clear line broadening beyond the line
bin ($i=13$) is seen. Any redward excess emission above the continuum
is $<2$\% in the 5-6 keV band and $<1$\% level in the 4-5 keV band,
respectively. \citet{falocco17} performed a spectral stacking analysis
of the same XMM-CDFS dataset. While their spectral stacking method is
different from ours, the Fe line in their stacked spectrum for
unobscured sources shows no broadening, in agreement with our
result. The spectral analysis of the brightest individual sources
shows the broad-line detection rate to be $\sim 30$\%
(Sect. 4.3.2). The result of this composite spectrum suggests that no
significant increase of the broad-line frequency occurs among the
fainter sources.

Those 70 unobscured sources with spectroscopic redshifts are divided
into the previously defined three redshift groups and their composite
spectra are shown in the right panel of
Fig. \ref{fig:medsp_unobsc}. The Fe line in the lowz spectrum is
resolved and shows a blue wing. Fitting a Gaussian gives the line
centroid of $6.52^{+0.12}_{-0.15}$ keV and line width of $\sigma
=0.35\pm 0.19$ keV, EW $= 0.24^{+0.09}_{-0.07}$ keV. This is contrary to
a red wing expected from a relativistically broadened line and
probably due to distributed different line energies. The Fe lines in
the spectra of midz and highz are in agreement with that of the total
spectrum, confined within the $i=13$ bin and are consistent with a
narrow line at 6.4 keV with EW of $0.10\pm 0.03$ keV and $0.12\pm
0.04$ keV, respectively.

\begin{figure}
\centerline{\includegraphics[width=0.4\textwidth,angle=0]{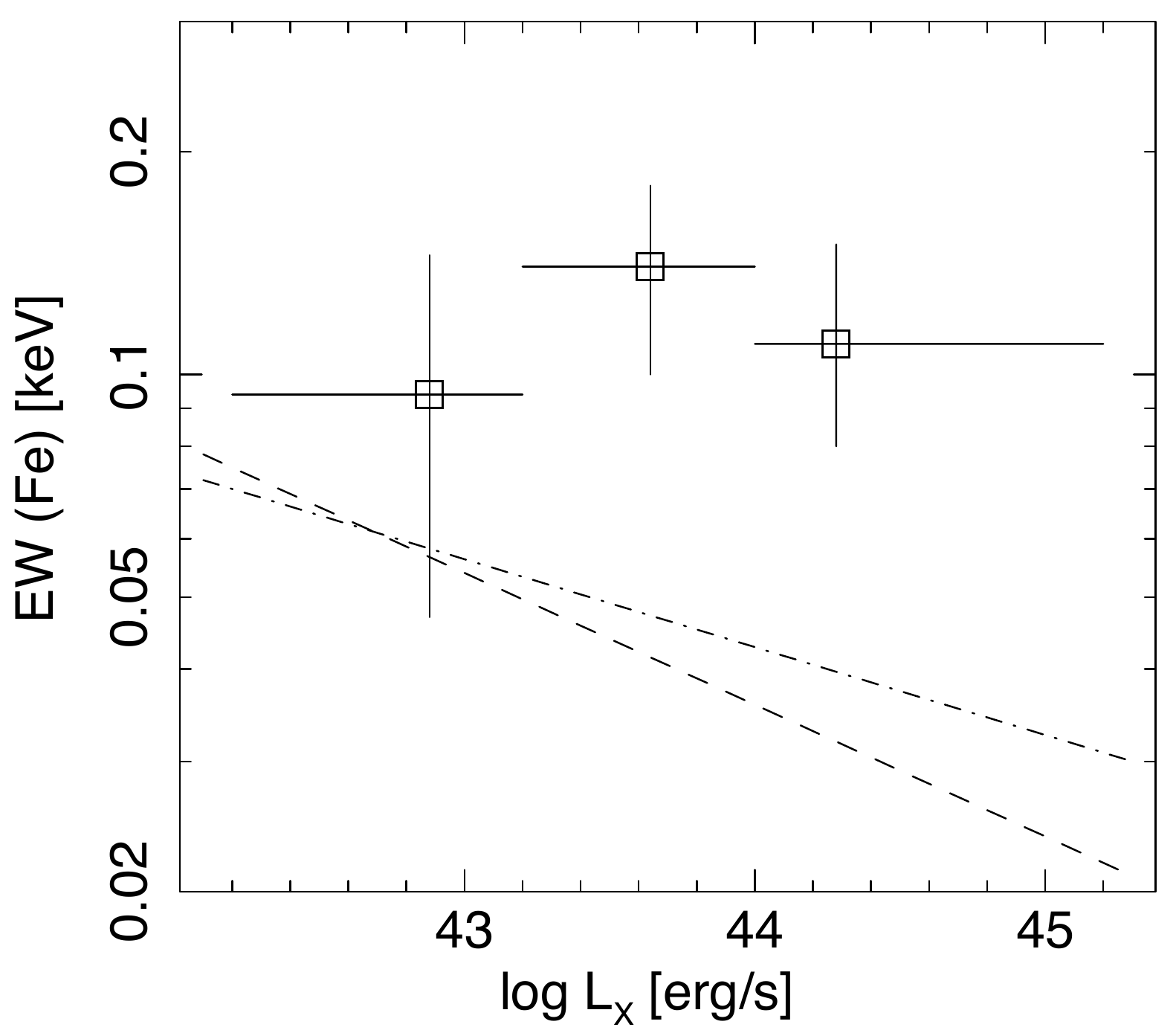}}
\caption{Plot of narrow Fe K line EW against X-ray luminosity obtained
  from the three luminosity slices of unobscured AGN (Table
  \ref{tab:it}). The dashed and dash-dotted lines indicate the I-T
  effect obtained for nearby AGN by \citet{bianchi07} and
  \citet{shu12}, respectively. }
\label{fig:iteffect}
\end{figure}

In order to see whether the anti-correlation between the EW of the
narrow Fe K line and X-ray luminosity, or Iwasawa-Taniguchi effect
\citep{it93}, seen in nearby unobscured AGNs \citep[see
  also][]{page04,bianchi07,shu12} holds, we constructed composite
spectra in three luminosity ranges lowLx: 42.2-43.2; midLx: 43.2-44.0;
and highLx: 44.0-45.2 in log $L_\mathrm{X}$ [\ergps]. The properties
of the sample of three luminosity slices and results of spectral fits
are given in Table \ref{tab:it}. The narrow line component show
comparable EW around 0.1 keV between the three luminosity ranges
(Fig. \ref{fig:iteffect}). There is no trend of decreasing EW as
increasing luminosity but also the EW are far above the relationship
found for nearby AGN.

\begin{table}
\caption{Properties of composite spectra of unobscured AGN in three luminosity slices.}
\label{tab:it}
\centering
\begin{tabular}{lccccc}
& $N$ & log$L^{\prime}$ & $\alpha$ & $N_\mathrm{H}$ & EW \\
  & & \ergps & & $10^{21}$ \psqcm & keV \\[5pt]
  lowLx & 24 & 42.88 & $0.64^{+0.07}_{-0.06}$ & $< 4.0$ & $0.09^{+0.05}_{-0.05}$ \\
  midLx & 22 & 43.68 & $0.85^{+0.03}_{-0.03}$ & $4.8^{+3.5}_{-2.4}$ & $0.14^{+0.02}_{-0.02}$ \\
  highLx & 24 & 44.28 & $0.82^{+0.04}_{-0.04}$ & $<3.1$ & $0.11^{+0.04}_{-0.03}$ \\
\end{tabular}
\tablefoot{Luminosty class, number of sources, median X-ray luminosity,
  and spectral parameters obtained from spectral fits by an absorbed
  power-law with a narrow Gaussian at 6.4 keV.}
\end{table}

\section{Discussion}

\subsection{Compton-thick AGN}

\begin{figure}
\centerline{\includegraphics[width=0.32\textwidth,angle=0]{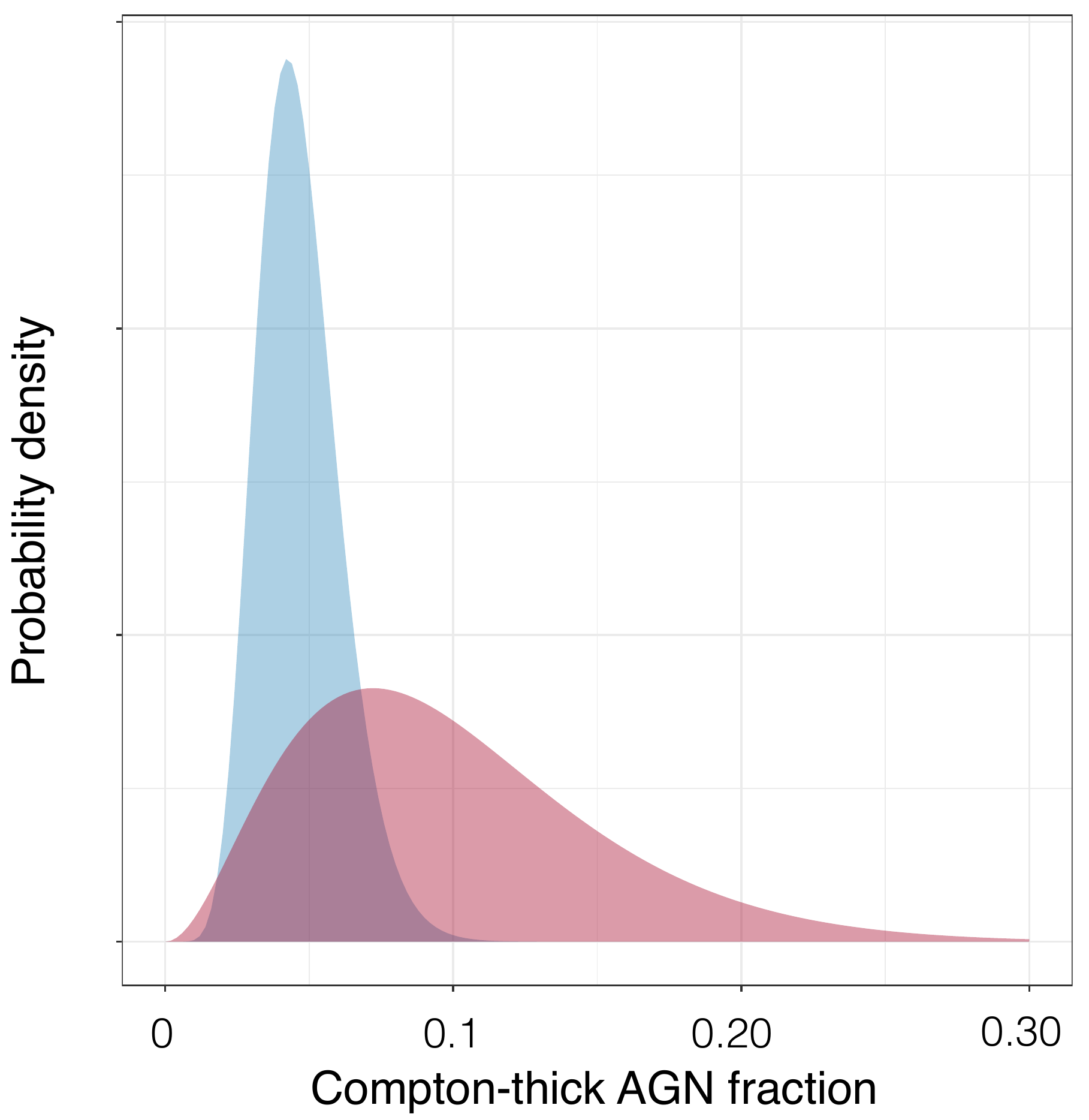}}
\caption{Probability distributions for the Compton-thick AGN
  fraction obtained from the XMM-CDFS (in blue) using the beta
  distribution prior (in red) which approximates the distribution of
  previous measurements for nearby AGN using hard X-ray instruments
  and Chandra CDFS measurements (references are given in text). The
  best estimate is found at 0.04 with the 95\% credible interval of
  (0.024-0.078). }
\label{fig:fct}
\end{figure}

Compton-thick AGN candidates were selected by two
characteristic spectral features: 1) large absorbing column density
\nH $\geq 10^{24}$ \psqcm; and 2) strong 6.4 keV Fe K line emission
arising from reflection fom thick, cold matter.

On taking face values of \nH\ measurements, the only source with
\nH\ exceeding $10^{24}$ \psqcm\ is PID 316. However, given the
uncertainty in the \nH\ measurements and the likely complexity of
spectra at large \nH\ discussed in Sect. 4.1, we select three more
candidates (PID 66, PID 131, PID 245) with log \nH $\geq 23.9$
[\psqcm], for which real values of their \nH\ could reach the
Compton-thick \nH\ threshold.

On the other hand, sources with a reflection-dominated spectrum are
often faint due to strong continuum suppression and their spectra are
generally noisy. Consequently, blindly applying the strong Fe K
line criterion could result in many false selection of Compton-thick
AGN. In order to make the selection more reliable, we
apply two complementary conditions in addition to the primary
condition for strong Fe K line (EW $\geq 1$ keV): 1) the line energy is
consistent with 6.4 keV and at above 90\% detection; and 2) hard X-ray
colours: S/M $< 1.5$ and H/M $> 0.5$ (where available), to comply with
the characteristics of a reflection-dominated spectrum in which Fe
line emission arises from cold matter and the underlying continuum has
a hard spectrum and extends well above 9 keV. This selection results
in five sources (PID 102, PID 114, PID 131, PID 215, and PID 398).
They are thus considered as reflection-dominated sources and their
true \nH\ would be in the Compton-thick regime. This alters the
\nH\ distribution in Fig. \ref{fig:histo_nH} slightly by stretching
the larger \nH\ end.

Among the above selected sources, we tentatively reserve the selection
of PID 131, mainly because of the poor data quality: although the
strong Fe K line seems to be real albeit lying close to the
instrumental Al line at 1.5 keV of the MOS cameras (see
Fig. \ref{fig:atlas}), the rest of the spectrum is very noisy.
Combining the two selections (and discounting PID 131), the number of
our Compton thick AGN candidates is seven out of 185
sources ($\sim 4$\%). Previously suggested Compton-thick AGN candidates, PID 144
\citep{norman02,comastri11}, PID 147
\citep{comastri11,georgantopoulos13}, and PID 252 \citep{i12cdfs}, for
which Chandra data were also used, fall just below the \nH\ threshold
on the use of the XMM-Newton data alone and may add up the number of
candidates to 10 (or 11 if PID 131 is included, $\sim$5-6\%).

There are a few measurements of Compton-thick AGN fraction for nearby
AGN using high-energy instruments, that is, Swift BAT and NuSTAR
\citep{burlon11,ricci15,koss16,masini18}, as well as the independent
measurements for CDFS sources
\citep{tozzi06,brightmanueda12,liu17}. They range from 3\% to 22\% and
their distribution can be approximated by $\sim $Beta(3,
28)\footnote{This is the beta distribution which is the conjugate
  prior for the binomial distribution we deal with here. The shape is
  controlled by the two shape parameters.}. Using this as a prior, our
seven detections yield the Compton-thick AGN fraction to be 4\% with
the 68\%/95\% credible intervals of (3.2-6.2)\%/(2.4-7.8)\%
(Fig. \ref{fig:fct}). The XMM-CDFS value lies on the smaller side of
the distribution of the previous measurements, largely coming from
hard X-ray surveys of nearby AGN, but is similar to those derived from
surveys with comparable depths: 5\% at the flux limit of log
$f_\mathrm{2-10}\simeq -15.0$ [\ergpspsqcm] \citep{tozzi06}; 5\% at
log $f_\mathrm{2-10}\simeq -15.1$ \citep{brightmanueda12}; 3\% at log
$f_\mathrm{2-10}\simeq -14.2$ \citep{lanzuisi18}. We note that these
values still have a selection bias against very Compton-thick
sources. The AGN population synthesis models of the X-ray background
give varying predictions of Compton-thick AGN fraction which depends
on observed flux
\citep[e.g.][]{ueda14,akylas12,ballantyne11,gilli07}. In general, the
Compton-thick AGN fraction starts to rise sharply below the flux limit
at $10^{-14}$ \ergpspsqcm. The typical 2-10 keV flux [\ergpspsqcm] of
our sources is log $f_\mathrm{X}\sim -14.4$ and the flux limit is
approximately $-14.7$. Our estimate of the Compton-thick AGN fraction
is compatible with the models by \citet{ueda14} and \citet{akylas12} (but lower
than those by \citep{ballantyne11,ananna19}). The deeper Chandra 7-Ms
dataset indeed gives a larger value of 8\% \citep{liu17}.

\begin{figure}
\centerline{\includegraphics[width=0.4\textwidth,angle=0]{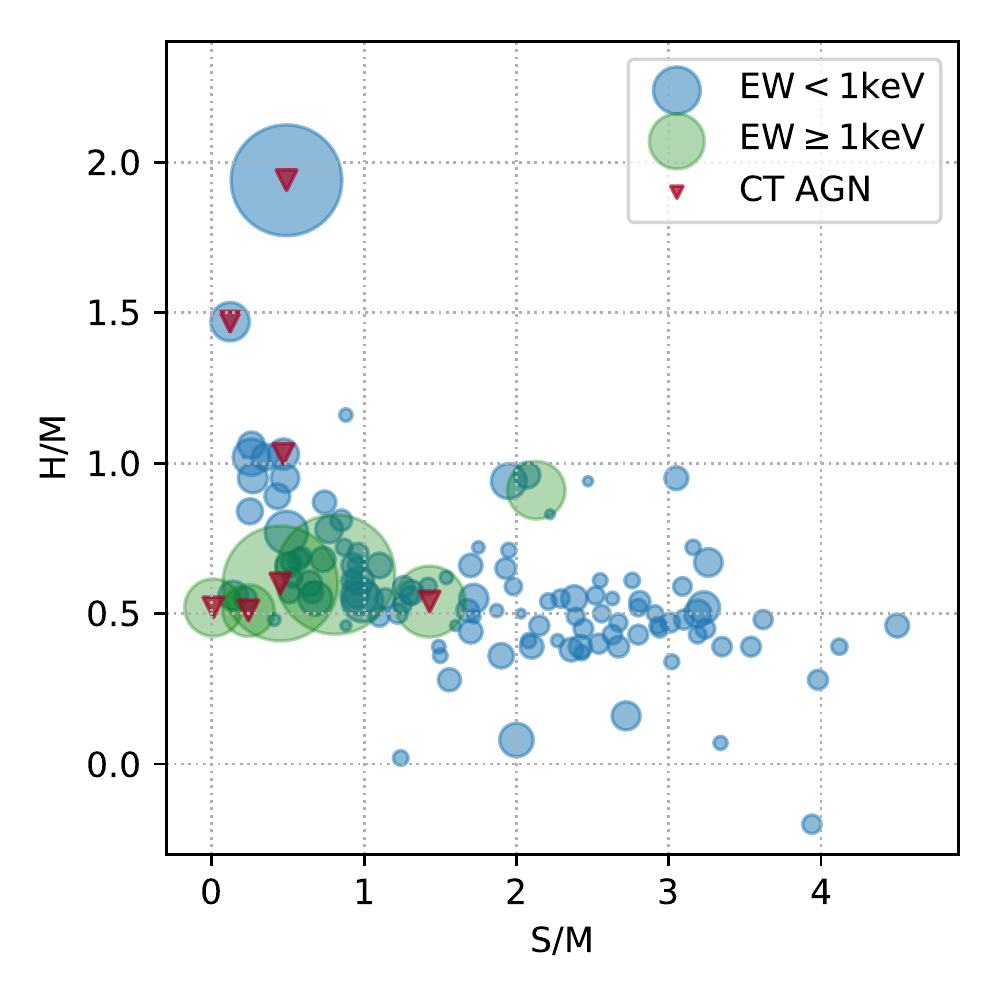}}
\caption{Same X-ray colour-colour diagram as Fig. \ref{fig:colcol}
  with information of R(Fe) added by the symbol size (the larger the
  size, the larger the R(Fe) value), to illustrate the Compton-thick
  AGN selection by the X-ray colours and R(Fe) alone (without spectral
  fitting). The error bars of the X-ray colours are omitted for
  clarity. Sources with strong Fe K lines (EW$\geq 1$ keV, measured by
  spectral fitting) are distinguished by colour. The seven Compton
  thick AGN candidates are marked by red triangles. Requiring S/M $<
  1.5$ and H/M $> 0.5$ ensures a hard spectrum of a
  reflection-dominated source.}
\label{fig:smhm_rfe}
\end{figure}

\begin{figure}
\centerline{\includegraphics[width=0.35\textwidth,angle=0]{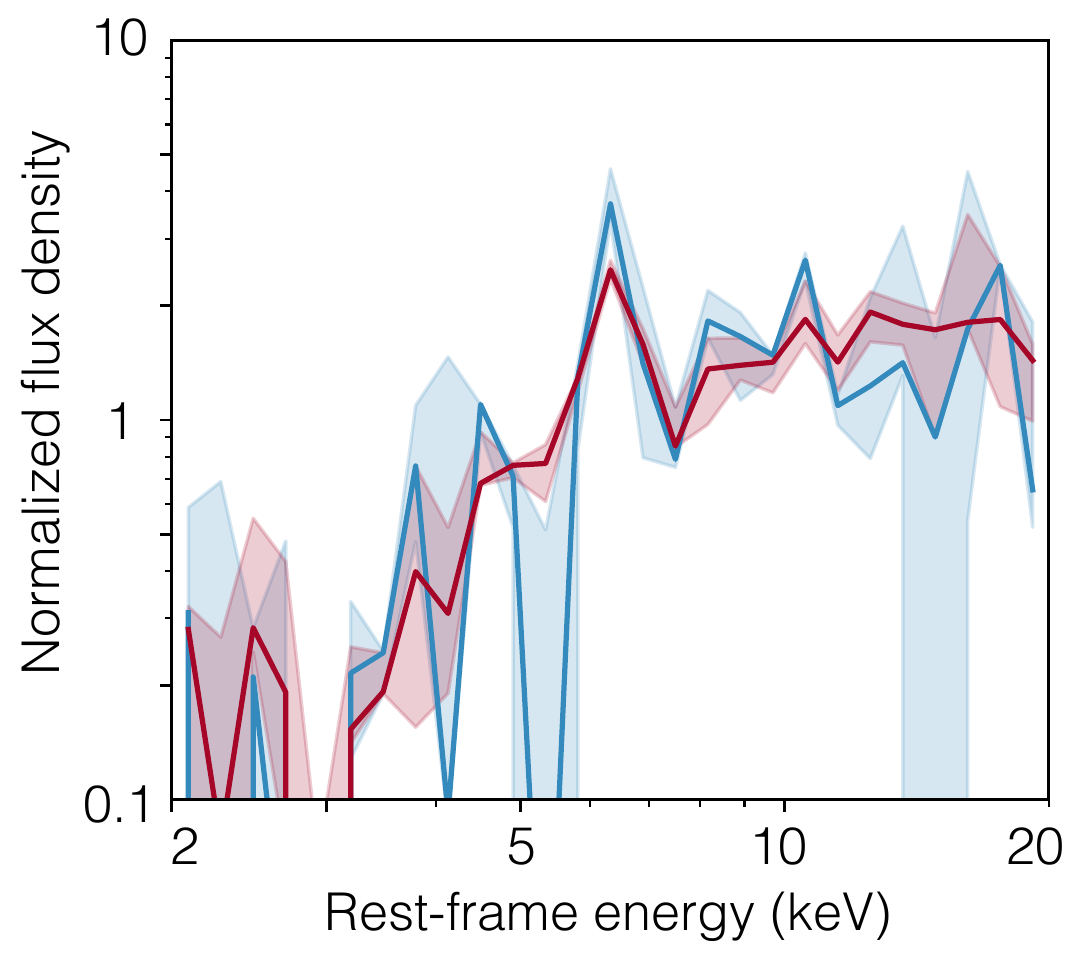}}
\caption{Composite spectra of the seven Compton-thick AGN candidates
  selected by spectral fitting (blue) and 19 candidates filtered by
  the two X-ray colours, S/M and H/M, and R(Fe) alone (red). }
\label{fig:composite_ctagn}
\end{figure}

The above selection was made with the spectral fitting results on
\nH\ and Fe line EW (with the supplemental filtering by the X-ray
colours). Then we investigate how those candidates can be selected
only by the simplified analysis, that is, the two X-ray colours and
R(Fe). Information on these values for each source is visualised in
Fig. \ref{fig:smhm_rfe}. The purpose is not to pin down Compton-thick
sources but to sift out promising candidates for a further inspection
by spectral fitting. By adjusting selection criteria for the above
candidates, we propose the following criteria.

\noindent 1) [S/M $< 1.5$ and H/M $>0.9$]: to select sources with
large \nH.

\noindent This colour selection select 10 sources, PID 24, 30, 59, 66,
142, 147, 180, 245, 252 and 316. Their measured log \nH\ ranges from
23.38 to 24.23 [\psqcm]. All the four Compton-thick AGN selected by
\nH\ are included, along with other previously suggested candidates
PID 147 \citep{comastri11,georgantopoulos13}, PID 252 \citep{i12cdfs}.

We note the very hard X-ray colours on PID 59 (S/M $=0.88$, H/M $=
1.16$). This source has a poorly constrained photometric redshift
2.881 (the 68\% uncertainty range 1.04-4.81) and its spectrum shows a
possible line-like feature of which energy does not match 6.4 keV for
the adopted redshift. If this was identified with a Fe K line, this
source would be another Compton-thick AGN candidate.

\noindent 2) [R(Fe) $>2.4$ and S/M $<1.5$ and H/M $>0.5$]: to select
reflection-dominated spectrum source with a strong Fe K line.

\noindent The selected sources are PID 66, 102, 114, 131, 144, 215, 221, 316, 324,
and 398. Without spectral fitting, we would not know whether the Fe K line
energy is 6.4 keV nor whether the line detection is significant, but all the six
candidates selected by large EW (Fe K) are included. A strong continuum
discontinuity due to an Fe K edge, when \nH\ is large, could lead to a
large R(Fe), as pointed out above. The rest of the sources
exhibiting `not sufficiently strong' Fe lines were selected for this
reason. The well-known Compton-thick AGN candidate PID 144
\citep{norman02,comastri11} and other previously suggested candidates,
PID 147 \citep{comastri11,georgantopoulos13} and PID 324
\citep{georgantopoulos13}, are present in the selected sources.

By design the above two criteria select all the spectral fitting-based, Compton-thick AGN candidates (supposing they are 'true positives'
here). The number of sources selected by these criteria is 19,
indicating that they do not pick up too many false positives (12). The
false positives include Compton-thick AGNs (PID 144, 147, 252, and
324), previously suggested by a combined analysis of Chandra or XMM-Newton and Chandra. Therefore, the X-ray colour + R(Fe) selection may
work as a pre-selection method. Figure \ref{fig:composite_ctagn} shows
the composite spectrum of the seven Compton-thick AGN candidates,
overplotted by that of the 19 sources of the colours + R(Fe)
selection. The two composite spectra appear very similar apart from
the Fe K line strength (although both are strong: EW $=2.2\pm 0.6$ keV
for the seven sources and EW $=1.0\pm 0.2$ keV for the 19
sources). The continuum is rising towards higher energies with the
slope of $\alpha\sim -2$.

\subsection{Evolution of the obscured AGN fraction}

\begin{figure}
\centerline{\includegraphics[width=0.5\textwidth,angle=0]{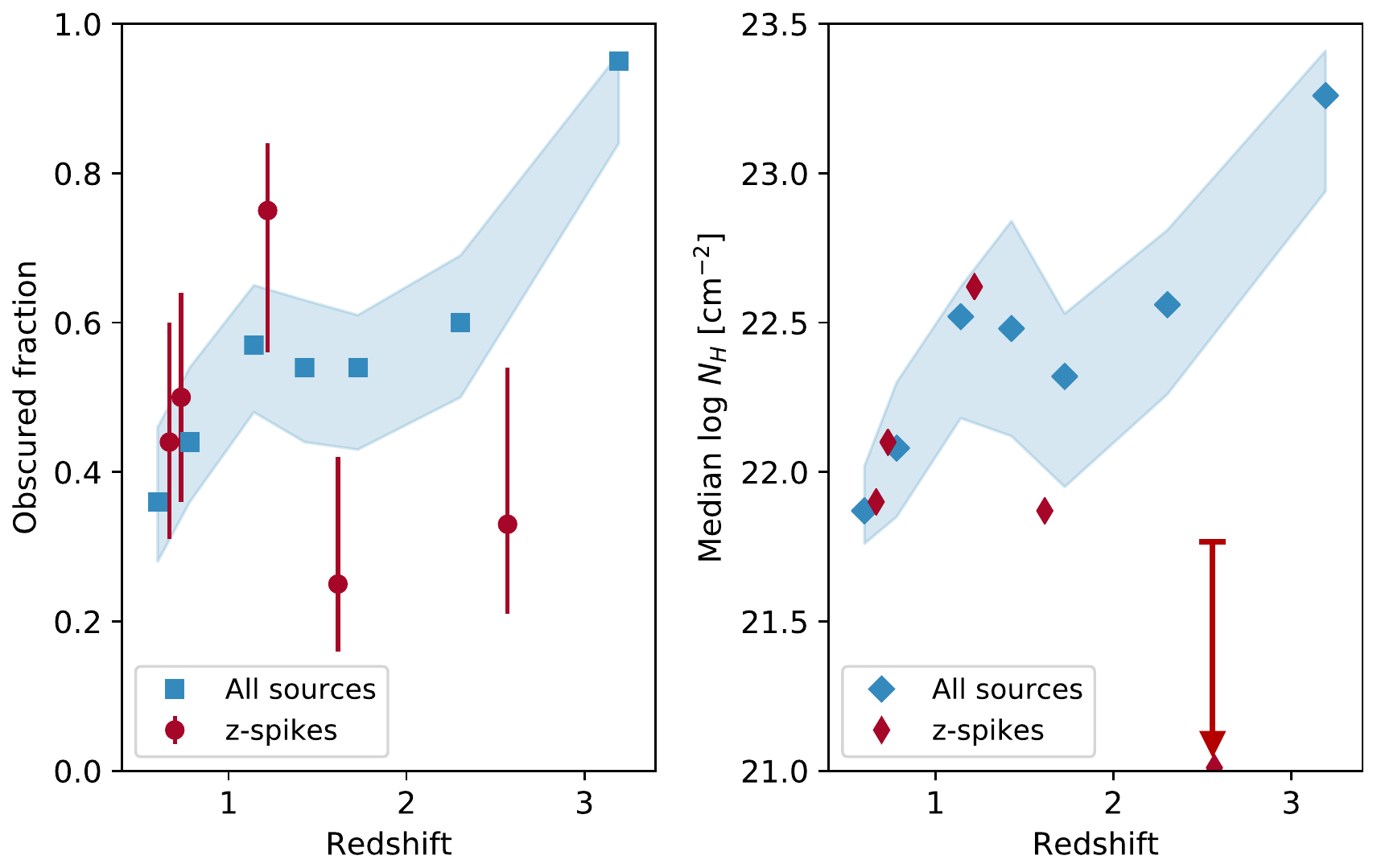}}
\caption{Left: Proportion of obscured AGN as a function of
  redshift. The blue squares show data for seven redshift ranges: 0.4-0.7,
  0.7-1, 1-1.3, 1.3-1.6, 1.6-2, 2-2.6, and 2.6-3.8 in $z$. The 68\%
  credible interval is indicated by the shaded area. The red filled
  circles are data for sources in the five redshift spikes. Right:
  Median absorbing column density for the whole sample as a function
  of redshift. The shaded area indicates the 68\% bootstrap error
  interval. The red symbols show the values for the five redshift spikes.}
\label{fig:fobscz}
\end{figure}

The distribution of absorbing column density, \nH, of the whole sample
has a median value of log \nH $= 22.5$ [\psqcm]
(Fig. \ref{fig:histo_nH}) and a shape similar to those obtained
previously \citep[e.g.][]{ueda14,liu17}. We note that this distribution
has incompleteness originating from the bias against very
Compton-thick AGN discussed in Sect. 4.1 and other various biases
discussed in detail by \citet[e.g.][]{liu17}. AGN obscuration becomes
higher at increasing redshifts: the \nH\ distribution, obscured AGN
fraction, X-ray colour analysis, and stacked spectra as a function of
redshift all agree. Here, instead of the three redshift groups in
Table \ref{tab:zgroups}, originally defined for the sake of spectral
analysis, we rebin the redshifts into seven ranges, each of which
contains a similar number of sources (24-30, except for 15 in the
highest redshift range) and examine the obscured AGN fraction as a
function of redshift (Fig. \ref{fig:fobscz}). The increasing trend of
the obscured AGN fraction with increasing redshift agree with the
study with the three redshift bins. The finer sampling of redshift
suggests that there is perhaps a plateau between $z=1$ and $z=2$ and a
steep increase at $z>2.5$. In general, there is a bias for detection
of absorption in higher redshift sources. However, in our spectral
analysis, this bias was minimised by setting the same lower-bound
rest-frame energy for all the spectra. Not only does the obscured AGN
fraction, but also the median absorbing column also shows a similar
increasing trend with redshift (Fig. \ref{fig:fobscz}).

The AGN obscured fraction is known to be luminosity dependent and
decreases towards higher luminosity \citep[e.g.][]{gilli07,ueda14}. For
our sample, sources in higher redshift groups have, on average, larger
luminosity as a result of the survey sensitivity limit. Therefore,
under a hypothesis of no evolution of obscured AGN fraction, a
decreasing trend of obscured fraction would be observed in
XMM-CDFS. This was verified against the mock AGN
catalogue\footnote{http://www.oabo.inaf.it/~gilli/mock.html} produced
by the AGN population model by \citet{gilli07} where a constant
obscured fraction and the luminosity dependence are assumed. Observing
the opposite trend over redshift supports the credibility of the
increasing obscured fraction towards high redshift. This evolution of
obscured AGN fraction has been suggested previously by a number of
works
\citep{lafranca05,treisterurry06,hasinger08,ebrero09,i12cdfs,ueda14,liu17,vito18}
and it appears to be luminosity
dependent \citep{buchner15,georgakakis15,aird15}.

\subsection{Sources in redshift spikes}

\begin{figure}
\centerline{\includegraphics[width=0.5\textwidth,angle=0]{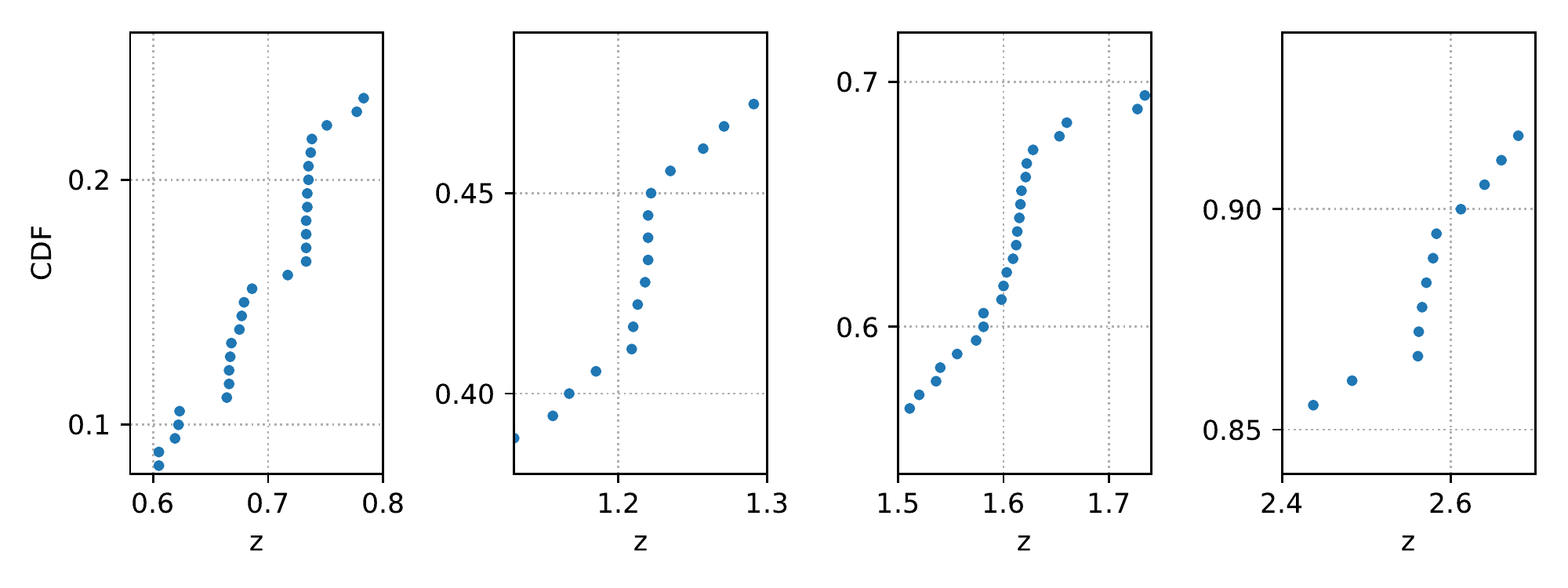}}
\caption{Cumulative distribution function (CDF) of the XMM-CDFS source
  redshifts in the four redshift ranges, where the five redshift
  spikes at $z\simeq 0.67$, $z\simeq 0.73$, $z=1.22$, $z=1.62$,
  $z\simeq 2.57$ are found. A steep rise in CDF represets a redshift
  clustering. The two spikes at $z\simeq 0.67$ and $z\simeq 0.73$ are
  well separated in the first figure.}
\label{fig:zcdf}
\end{figure}

\begin{figure}
\centerline{\includegraphics[width=0.35\textwidth,angle=0]{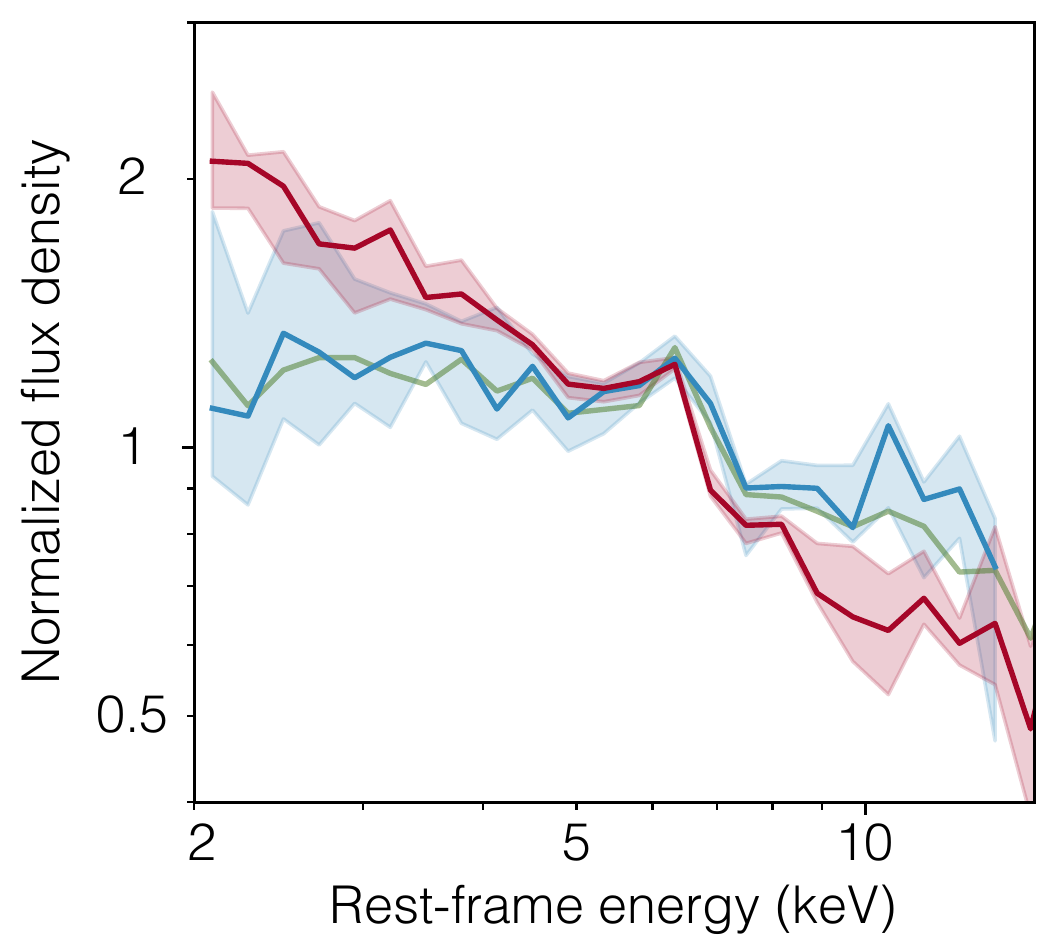}}
\caption{Composite spectra for the three $z$-spikes in lower redshifts
  (blue, $z\simeq 0.67$, $z\simeq 0.73$, $z\simeq 1.22$) and in higher
  redshifts (red, $z\simeq 1.61$ and $z\simeq 2.57$). The composite spectrum
  of the whole sample is also plotted in green for comparison. While
  the spectrum of the low-z spikes agrees with that of the whole
  sample, the spectrum of the high-z spikes is clearly softer.}
\label{fig:medsp_zspikes}
\end{figure}

A few redshift spikes in the CDFS have been observed through optical
spectroscopic surveys \citep[e.g.][]{gilli03,silverman10}. We
identified five redshift spikes composed of XMM-CDFS X-ray sources in
our sample. Since our sample size is not large, they may easily be
missed in a histogram like Fig. \ref{fig:zdist} due to the binning
bias. A cumulative distribution function (CDF) is instead free from the
binning bias. The CDF around the five redshift-spikes are shown in
Fig. \ref{fig:zcdf}, in which a steep rise of the CDF is identified
with each spike. These spikes have between six and 12 sources within $\Delta z\leq
0.03$, respectively (Table \ref{tab:zspikes}). Each of them has a
counterpart found by optical spectroscopy of galaxies in the field
\citep[e.g.][]{silverman10}. The total number of sources in these
spikes is 45, in which three sources have photometric redshifts. The
median properties of the sources belonging to each redshift spike are
presented in Table \ref{tab:zspikes}.

As noted above, obscured AGN seem to be more abundant towards higher
redshift and the sources in the three, lower redshift spikes ($z=$
0.67, 0.73 and 1.22) follow the same trend
(Fig. \ref{fig:fobscz}). However, those in the two, higher redshift
spikes ($z=$ 1.61 and 2.57) deviate from this trend and show lower obscured
AGN fractions and lower absorbing columns (Table
\ref{tab:zspikes}). This behaviour is illustrated by the composite
spectra of the sources in Fig. \ref{fig:medsp_zspikes}. The
composite spectrum of the lower-$z$ spikes is relatively hard ($\alpha
= 0.56\pm 0.06$, \nH $=(1.8\pm 0.5)\times 10^{22}$ \psqcm) and agrees
with that of the whole sample. In contrast, that of
the higher-$z$ spikes is clearly softer ($\alpha = 0.80\pm 0.05$, \nH
$=(5\pm 4)\times 10^{21}$ \psqcm). Both spectra show Fe K lines with
EW $\simeq 0.1$ keV. We note that the higher-$z$ spikes composite
spectrum shows an asymmetric Fe line profile with a red wing extending
down to 5 keV. If the {\tt diskline} model \citep{fabian89} is fitted,
the line EW would increase to $0.25\pm 0.02$ keV.

The deficit of obscured AGN in the $z=1.61$ and $z=2.57$ spikes is
puzzling. Since they are X-ray selected sources, there is little bias
for unobscured AGN, which would be likely to occur in an optical
spectroscopic selection. Indeed, the sources in the lower-z spikes
show no difference in obscuration from the sources in the similar
redshift range. It could be related to deficiency of gas in the
nuclear regions or in the host galaxies. Faster gas consumption by
star formation taking place earlier or faster transition of obscured
to unobscured AGN by expelling, for example, the surrounding gas
\citep{sanders88}, relative to the other sources, may provide an
explanation. Whether this can be attributed to an environmental effect
is uncertain. They are likely to be located in large-scale structures
\citep{gilli03} but their environments are not as dense as in a galaxy
cluster. Gas removal by ram pressure stripping (etc.) should not be as
efficient as in a more dense environment. A study of
general galaxy properties in these redshift spikes, for example, of post-starburst proportions, may provide a clue for understanding the
apparent deficit of obscuration.

\begin{table}
\caption{X-ray source properties of redshift spikes.}
\label{tab:zspikes}
\centering
\begin{tabular}{ccccccc}
  {\it \~z} & $N$ & $\Delta z$ & log\nH & log$L^{\prime}$ & EW & $f_{>22}$ \\
  & & & \psqcm & \ergps & keV & \\[5pt]
0.67 & 9 & 0.022 & 21.9 & 43.11 & 0.12 & 0.44 \\
0.73 & 10 & 0.005 & 22.1 & 43.08 & 0.17 & 0.50 \\
1.22 & 8 & 0.013 & 22.6 & 43.51 & 0.29 & 0.75 \\
1.61 & 12 & 0.030 & 21.9 & 44.13 & 0.15 & 0.25 \\
2.57 & 6 & 0.022 & 21.0 & 44.39 & 0.09 & 0.33 \\[5pt]
All & 45 & - & 21.9 & 43.59 & 0.15 & 0.44 \\
\end{tabular}
\tablefoot{Table columns are median redshift of each spike, number
  of sources, redshift range, median value of log \nH, median
  absorption-corrected 2-10 keV luminosity, median EW of the line,
  proportion of sources with \nH\ exceeding $10^{22}$ \psqcm.}
\end{table}

\subsection{Fe K emission lines}

Fe K lines are frequently found in the spectra. The line energy is
most frequently found at around 6.4 keV, indicating the majority of
AGN emit cold Fe K lines. For the narrow component of the Fe K line in
unobscured AGN, the Iwasawa-Taniguchi (IT) effect is well established
among the nearby AGN \citep{bianchi07,page04,shu12}. The correlation
is often interpreted as the consequence of decreasing torus covering
factor with increasing source luminosity, as in \citet{ricci13}. On the
contrary, as far as the composite spectra are concerned, the XMM-CDFS
sources show no luminosity dependence of the line EW. Furthermore, the
observed EW around 0.1 keV seen in high-luminosity sources is well
above the value predicted from the correlation
(Fig. \ref{fig:iteffect}). However, this may not be surprising given
that the XMM-CDFS sources are sampled over a wide redshift
range. Statistically, larger luminosity sources come from higher
redshifts. As discussed in the previous section for the whole sample,
there is an evolution of obscuration of AGN, that is, more obscuration at
higher redshifts. This could result from a larger gas mass fraction at
higher redshifts \citep[e.g.][]{carilliwalter13} as discussed in
\citet{i12cdfs}. The increased obscuration for higher redshift sources
(hence higher luminosity sources on average) then means more cold gas
surrounding the AGN regardless of their types (obscured or unobscured
in the line-of sight). It would offset the decline of line intensity
due to whatever mechanism produces the IT effect by providing more
cold gas to be illuminated by the central source for the line
production. The evolution of the gas fraction $f_\mathrm{gas}\propto
(1+z)^2$ \citep{geach11} means an increase of $\sim 0.5$ dex of cold
gas from $z = 0.7$ to $z=2$, which roughly matches the required
offset. \citet{chaudhary10} found the IT effect in the AGN sample of
the serendipitous 2XMM catalog \citep{watson09}, which includes
sources with redshifts up to $z\sim 5$, using spectral stacking. We
note that since their sample is dominated by low redshift objects, the
IT effect is probably driven by sources in the limited, low redshift
range.

Substantial line broadening of Fe K emission, most likely due to the
relativistic effects at inner radii of the accretion disc, in distant
AGN have been studied by many authors using spectral stacking
\citep{streblyanska05,brusa05,corral08,chaudhary12,i12cosmos,falocco13,falocco14}. Results
vary between the works but some evidence of line broadening has been
found. We observe a 30\% of detection rate of line broadening with
$\sigma\geq 0.3$ keV among the bright, individual unobscured sources
in the XMM-CDFS (sect. 4.3.2), which is similar to that for nearby AGN
\citep{delacalle10}. The lack of strong line broadening in our
composite spectrum of all the unobscured AGN (see also
\citet{falocco13}) suggests that broad Fe K lines in distant AGN are
not much more common than in the local AGN.

However, when we have a limited signal-to-noise ratio for spectral
data, a relativistically broadened line could escape from
detection. The archetypal broad Fe line shape \citep{tanaka95},
as with, for example, the disc inclination of $\sim 30^{\circ}$, the inner radius of
$6r_{\rm g}$, would be the easiest to detect with the spectral
resolution of a CCD detector. What we are looking to examine is specifically this type of
line broadening. Under more extreme relativistic effects, line
emission is altered to become so broad and its contrast against the
continuum is so low that its detection is deemed to be
difficult. Presence of a narrow line component originating from
distant matter makes it even more complicated.

\begin{acknowledgements}
This research made use of data obtained from XMM-Newton and software packages of HEASoft, R \citep{rcoreteam17}, jags \citep{Plummer03jags:a}, ggplot2 \citep{wickham09}, IPython \citep{PER-GRA:2007}, pandas \citep{mckinney-proc-scipy-2010}, seaborn and Matplotlib \citep{Hunter:2007}. We
acknowledge financial contribution from the agreement ASI-INAF
n.2017-14-H.O. KI acknowledges support by the Spanish MINECO under
grant AYA2016-76012-C3-1-P and MDM-2014-0369 of ICCUB (Unidad de
Excelencia 'Mar\'ia de Maeztu'). AC acknowledges the Caltech Kingsley
visitor program. FJC acknowledges financial support through grant
AYA2015-64346-C2-1P (MINECO/FEDER), by the Spanish Ministry MCIU under
project RTI2018-096686-B-C21 (MCIU/AEI/FEDER, UE), cofunded by FEDER
funds, and funded by the Agencia Estatal de Investigaci\'on, Unidad de
Excelencia Mar\'ia de Maeztu, MDM-2017-0765. WNB acknowledges support
from NASA grants 80NSSC19K0961 and 80NSSC18K0878.
\end{acknowledgements}

\bibliographystyle{aa} 
\bibliography{xmmcdfs}

\begin{thebibliography}{100}
\expandafter\ifx\csname natexlab\endcsname\relax\def\natexlab#1{#1}\fi

\bibitem[{{Aird} {et~al.}(2015){Aird}, {Coil}, {Georgakakis}, {Nandra},
  {Barro}, \& {P{\'e}rez-Gonz{\'a}lez}}]{aird15}
{Aird}, J., {Coil}, A.~L., {Georgakakis}, A., {et~al.} 2015, \mnras, 451, 1892

\bibitem[{{Akylas} {et~al.}(2012){Akylas}, {Georgakakis}, {Georgantopoulos},
  {Brightman}, \& {Nandra}}]{akylas12}
{Akylas}, A., {Georgakakis}, A., {Georgantopoulos}, I., {Brightman}, M., \&
  {Nandra}, K. 2012, \aap, 546, A98

\bibitem[{{Ananna} {et~al.}(2019){Ananna}, {Treister}, {Urry}, {Ricci},
  {Kirkpatrick}, {LaMassa}, {Buchner}, {Civano}, {Tremmel}, \&
  {Marchesi}}]{ananna19}
{Ananna}, T.~T., {Treister}, E., {Urry}, C.~M., {et~al.} 2019, \apj, 871, 240

\bibitem[{{Antonucci} {et~al.}(2015){Antonucci}, {Talavera}, {Vagnetti},
  {Trevese}, {Comastri}, {Paolillo}, {Ranalli}, \& {Vignali}}]{antonucci15}
{Antonucci}, M., {Talavera}, A., {Vagnetti}, F., {et~al.} 2015, \aap, 574, A49

\bibitem[{{Balestra} {et~al.}(2010){Balestra}, {Mainieri}, {Popesso},
  {Dickinson}, {Nonino}, {Rosati}, {Teimoorinia}, {Vanzella}, {Cristiani},
  {Cesarsky}, {Fosbury}, {Kuntschner}, \& {Rettura}}]{balestra10}
{Balestra}, I., {Mainieri}, V., {Popesso}, P., {et~al.} 2010, \aap, 512, A12

\bibitem[{{Ballantyne} {et~al.}(2011){Ballantyne}, {Draper}, {Madsen}, {Rigby},
  \& {Treister}}]{ballantyne11}
{Ballantyne}, D.~R., {Draper}, A.~R., {Madsen}, K.~K., {Rigby}, J.~R., \&
  {Treister}, E. 2011, \apj, 736, 56

\bibitem[{{Bianchi} {et~al.}(2007){Bianchi}, {Guainazzi}, {Matt}, \& {Fonseca
  Bonilla}}]{bianchi07}
{Bianchi}, S., {Guainazzi}, M., {Matt}, G., \& {Fonseca Bonilla}, N. 2007,
  \aap, 467, L19

\bibitem[{{Bianchi} {et~al.}(2009){Bianchi}, {Guainazzi}, {Matt}, {Fonseca
  Bonilla}, \& {Ponti}}]{bianchi09}
{Bianchi}, S., {Guainazzi}, M., {Matt}, G., {Fonseca Bonilla}, N., \& {Ponti},
  G. 2009, \aap, 495, 421

\bibitem[{{Bonzini} {et~al.}(2013){Bonzini}, {Padovani}, {Mainieri},
  {Kellermann}, {Miller}, {Rosati}, {Tozzi}, \& {Vattakunnel}}]{bonzini13}
{Bonzini}, M., {Padovani}, P., {Mainieri}, V., {et~al.} 2013, \mnras, 436, 3759

\bibitem[{{Brandt} \& {Alexander}(2015)}]{brandt15}
{Brandt}, W.~N. \& {Alexander}, D.~M. 2015, \aapr, 23, 1

\bibitem[{{Brandt} \& {Hasinger}(2005)}]{brandt05}
{Brandt}, W.~N. \& {Hasinger}, G. 2005, \araa, 43, 827

\bibitem[{{Brightman} \& {Ueda}(2012)}]{brightmanueda12}
{Brightman}, M. \& {Ueda}, Y. 2012, \mnras, 423, 702

\bibitem[{{Brusa} {et~al.}(2005){Brusa}, {Gilli}, \& {Comastri}}]{brusa05}
{Brusa}, M., {Gilli}, R., \& {Comastri}, A. 2005, \apjl, 621, L5

\bibitem[{{Buchner} {et~al.}(2015){Buchner}, {Georgakakis}, {Nandra},
  {Brightman}, {Menzel}, {Liu}, {Hsu}, {Salvato}, {Rangel}, {Aird}, {Merloni},
  \& {Ross}}]{buchner15}
{Buchner}, J., {Georgakakis}, A., {Nandra}, K., {et~al.} 2015, \apj, 802, 89

\bibitem[{{Burlon} {et~al.}(2011){Burlon}, {Ajello}, {Greiner}, {Comastri},
  {Merloni}, \& {Gehrels}}]{burlon11}
{Burlon}, D., {Ajello}, M., {Greiner}, J., {et~al.} 2011, \apj, 728, 58

\bibitem[{{Cardamone} {et~al.}(2010){Cardamone}, {van Dokkum}, {Urry},
  {Taniguchi}, {Gawiser}, {Brammer}, {Taylor}, {Damen}, {Treister}, {Cobb},
  {Bond}, {Schawinski}, {Lira}, {Murayama}, {Saito}, \&
  {Sumikawa}}]{cardamone10}
{Cardamone}, C.~N., {van Dokkum}, P.~G., {Urry}, C.~M., {et~al.} 2010, \apjs,
  189, 270

\bibitem[{{Carilli} \& {Walter}(2013)}]{carilliwalter13}
{Carilli}, C.~L. \& {Walter}, F. 2013, \araa, 51, 105

\bibitem[{{Casey} {et~al.}(2011){Casey}, {Chapman}, {Smail}, {Alaghband
  -Zadeh}, {Bothwell}, \& {Swinbank}}]{casey11}
{Casey}, C.~M., {Chapman}, S.~C., {Smail}, I., {et~al.} 2011, \mnras, 411, 2739

\bibitem[{{Castell{\'o}-Mor} {et~al.}(2013){Castell{\'o}-Mor}, {Carrera},
  {Alonso-Herrero}, {Mateos}, {Barcons}, {Ranalli}, {P{\'e}rez-Gonz{\'a}lez},
  {Comastri}, {Vignali}, \& {Georgantopoulos}}]{castellmor13}
{Castell{\'o}-Mor}, N., {Carrera}, F.~J., {Alonso-Herrero}, A., {et~al.} 2013,
  \aap, 556, A114

\bibitem[{{Chaudhary} {et~al.}(2010){Chaudhary}, {Brusa}, {Hasinger},
  {Merloni}, \& {Comastri}}]{chaudhary10}
{Chaudhary}, P., {Brusa}, M., {Hasinger}, G., {Merloni}, A., \& {Comastri}, A.
  2010, \aap, 518, A58

\bibitem[{{Chaudhary} {et~al.}(2012){Chaudhary}, {Brusa}, {Hasinger},
  {Merloni}, {Comastri}, \& {Nandra}}]{chaudhary12}
{Chaudhary}, P., {Brusa}, M., {Hasinger}, G., {et~al.} 2012, \aap, 537, A6

\bibitem[{{Comastri} {et~al.}(2011){Comastri}, {Ranalli}, {Iwasawa}, {Vignali},
  {Gilli}, {Georgantopoulos}, {Barcons}, {Brandt}, {Brunner}, {Brusa},
  {Cappelluti}, {Carrera}, {Civano}, {Fiore}, {Hasinger}, {Mainieri},
  {Merloni}, {Nicastro}, {Paolillo}, {Puccetti}, {Rosati}, {Silverman},
  {Tozzi}, {Zamorani}, {Balestra}, {Bauer}, {Luo}, \& {Xue}}]{comastri11}
{Comastri}, A., {Ranalli}, P., {Iwasawa}, K., {et~al.} 2011, \aap, 526, L9

\bibitem[{{Cooper} {et~al.}(2012){Cooper}, {Yan}, {Dickinson}, {Juneau},
  {Lotz}, {Newman}, {Papovich}, {Salim}, {Walth}, {Weiner}, \&
  {Willmer}}]{cooper12}
{Cooper}, M.~C., {Yan}, R., {Dickinson}, M., {et~al.} 2012, \mnras, 425, 2116

\bibitem[{{Corral} {et~al.}(2008){Corral}, {Page}, {Carrera}, {Barcons},
  {Mateos}, {Ebrero}, {Krumpe}, {Schwope}, {Tedds}, \& {Watson}}]{corral08}
{Corral}, A., {Page}, M.~J., {Carrera}, F.~J., {et~al.} 2008, \aap, 492, 71

\bibitem[{{de La Calle P{\'e}rez} {et~al.}(2010){de La Calle P{\'e}rez},
  {Longinotti}, {Guainazzi}, {Bianchi}, {Dov{\v c}iak}, {Cappi}, {Matt},
  {Miniutti}, {Petrucci}, {Piconcelli}, {Ponti}, {Porquet}, \&
  {Santos-Lle{\'o}}}]{delacalle10}
{de La Calle P{\'e}rez}, I., {Longinotti}, A.~L., {Guainazzi}, M., {et~al.}
  2010, \aap, 524, A50

\bibitem[{{Dickey} \& {Lockman}(1990)}]{dickey90}
{Dickey}, J.~M. \& {Lockman}, F.~J. 1990, \araa, 28, 215

\bibitem[{{Ebrero} {et~al.}(2009){Ebrero}, {Carrera}, {Page}, {Silverman},
  {Barcons}, {Ceballos}, {Corral}, {Della Ceca}, \& {Watson}}]{ebrero09}
{Ebrero}, J., {Carrera}, F.~J., {Page}, M.~J., {et~al.} 2009, \aap, 493, 55

\bibitem[{Efron(1979)}]{efron79}
Efron, B. 1979, Ann. Statist., 7, 1

\bibitem[{{Fabian} {et~al.}(2000){Fabian}, {Iwasawa}, {Reynolds}, \&
  {Young}}]{fabian00}
{Fabian}, A.~C., {Iwasawa}, K., {Reynolds}, C.~S., \& {Young}, A.~J. 2000,
  \pasp, 112, 1145

\bibitem[{{Fabian} {et~al.}(1989){Fabian}, {Rees}, {Stella}, \&
  {White}}]{fabian89}
{Fabian}, A.~C., {Rees}, M.~J., {Stella}, L., \& {White}, N.~E. 1989, \mnras,
  238, 729

\bibitem[{{Falocco} {et~al.}(2014){Falocco}, {Carrera}, {Barcons}, {Miniutti},
  \& {Corral}}]{falocco14}
{Falocco}, S., {Carrera}, F.~J., {Barcons}, X., {Miniutti}, G., \& {Corral}, A.
  2014, \aap, 568, A15

\bibitem[{{Falocco} {et~al.}(2013){Falocco}, {Carrera}, {Corral}, {Barcons},
  {Comastri}, {Gilli}, {Ranalli}, {Vignali}, {Iwasawa}, {Cappelluti},
  {Rovilos}, {Georgantopoulos}, {Brusa}, \& {Vito}}]{falocco13}
{Falocco}, S., {Carrera}, F.~J., {Corral}, A., {et~al.} 2013, \aap, 555, A79

\bibitem[{{Falocco} {et~al.}(2017){Falocco}, {Paolillo}, {Comastri}, {Carrera},
  {Ranalli}, {Iwasawa}, {Georgantopoulos}, {Vignali}, \& {Gilli}}]{falocco17}
{Falocco}, S., {Paolillo}, M., {Comastri}, A., {et~al.} 2017, \aap, 608, A32

\bibitem[{{Geach} {et~al.}(2011){Geach}, {Smail}, {Moran}, {MacArthur},
  {Lagos}, \& {Edge}}]{geach11}
{Geach}, J.~E., {Smail}, I., {Moran}, S.~M., {et~al.} 2011, \apjl, 730, L19

\bibitem[{{Georgakakis} {et~al.}(2015){Georgakakis}, {Aird}, {Buchner},
  {Salvato}, {Menzel}, {Brandt}, {McGreer}, {Dwelly}, {Mountrichas}, {Koki},
  {Georgantopoulos}, {Hsu}, {Merloni}, {Liu}, {Nandra}, \&
  {Ross}}]{georgakakis15}
{Georgakakis}, A., {Aird}, J., {Buchner}, J., {et~al.} 2015, \mnras, 453, 1946

\bibitem[{{Georgantopoulos} {et~al.}(2013){Georgantopoulos}, {Comastri},
  {Vignali}, {Ranalli}, {Rovilos}, {Iwasawa}, {Gilli}, {Cappelluti}, {Carrera},
  {Fritz}, {Brusa}, {Elbaz}, {Mullaney}, {Castello-Mor}, {Barcons}, {Tozzi},
  {Balestra}, \& {Falocco}}]{georgantopoulos13}
{Georgantopoulos}, I., {Comastri}, A., {Vignali}, C., {et~al.} 2013, \aap, 555,
  A43

\bibitem[{{Giacconi} {et~al.}(2001){Giacconi}, {Rosati}, {Tozzi}, {Nonino},
  {Hasinger}, {Norman}, {Bergeron}, {Borgani}, {Gilli}, {Gilmozzi}, \&
  {Zheng}}]{giacconi01}
{Giacconi}, R., {Rosati}, P., {Tozzi}, P., {et~al.} 2001, \apj, 551, 624

\bibitem[{{Gilli} {et~al.}(2003){Gilli}, {Cimatti}, {Daddi}, {Hasinger},
  {Rosati}, {Szokoly}, {Tozzi}, {Bergeron}, {Borgani}, {Giacconi}, {Kewley},
  {Mainieri}, {Mignoli}, {Nonino}, {Norman}, {Wang}, {Zamorani}, {Zheng}, \&
  {Zirm}}]{gilli03}
{Gilli}, R., {Cimatti}, A., {Daddi}, E., {et~al.} 2003, \apj, 592, 721

\bibitem[{{Gilli} {et~al.}(2007){Gilli}, {Comastri}, \& {Hasinger}}]{gilli07}
{Gilli}, R., {Comastri}, A., \& {Hasinger}, G. 2007, \aap, 463, 79

\bibitem[{{Hales} {et~al.}(2014){Hales}, {Norris}, {Gaensler}, {Middelberg},
  {Chow}, {Hopkins}, {Huynh}, {Lenc}, \& {Mao}}]{hales14}
{Hales}, C.~A., {Norris}, R.~P., {Gaensler}, B.~M., {et~al.} 2014, \mnras, 441,
  2555

\bibitem[{{Hasinger}(2008)}]{hasinger08}
{Hasinger}, G. 2008, \aap, 490, 905

\bibitem[{{Hsu} {et~al.}(2014){Hsu}, {Salvato}, {Nandra}, {Brusa}, {Bender},
  {Buchner}, {Donley}, {Kocevski}, {Guo}, {Hathi}, {Rangel}, {Willner},
  {Brightman}, {Georgakakis}, {Budav{\'a}ri}, {Szalay}, {Ashby}, {Barro},
  {Dahlen}, {Faber}, {Ferguson}, {Galametz}, {Grazian}, {Grogin}, {Huang},
  {Koekemoer}, {Lucas}, {McGrath}, {Mobasher}, {Peth}, {Rosario}, \&
  {Trump}}]{hsu14}
{Hsu}, L.-T., {Salvato}, M., {Nandra}, K., {et~al.} 2014, \apj, 796, 60

\bibitem[{Hunter(2007)}]{Hunter:2007}
Hunter, J.~D. 2007, Computing in Science \& Engineering, 9, 90

\bibitem[{{Huynh} {et~al.}(2015){Huynh}, {Bell}, {Hopkins}, {Norris}, \&
  {Seymour}}]{huynh15}
{Huynh}, M.~T., {Bell}, M.~E., {Hopkins}, A.~M., {Norris}, R.~P., \& {Seymour},
  N. 2015, \mnras, 454, 952

\bibitem[{{Ikeda} {et~al.}(2009){Ikeda}, {Awaki}, \& {Terashima}}]{ikeda09}
{Ikeda}, S., {Awaki}, H., \& {Terashima}, Y. 2009, \apj, 692, 608

\bibitem[{{Iwasawa} {et~al.}(2012{\natexlab{a}}){Iwasawa}, {Gilli}, {Vignali},
  {Comastri}, {Brandt}, {Ranalli}, {Vito}, {Cappelluti}, {Carrera}, {Falocco},
  {Georgantopoulos}, {Mainieri}, \& {Paolillo}}]{i12cdfs}
{Iwasawa}, K., {Gilli}, R., {Vignali}, C., {et~al.} 2012{\natexlab{a}}, \aap,
  546, A84

\bibitem[{{Iwasawa} {et~al.}(2012{\natexlab{b}}){Iwasawa}, {Mainieri}, {Brusa},
  {Comastri}, {Gilli}, {Vignali}, {Hasinger}, {Sanders}, {Cappelluti}, {Impey},
  {Koekemoer}, {Lanzuisi}, {Lusso}, {Merloni}, {Salvato}, {Taniguchi}, \&
  {Trump}}]{i12cosmos}
{Iwasawa}, K., {Mainieri}, V., {Brusa}, M., {et~al.} 2012{\natexlab{b}}, \aap,
  537, A86

\bibitem[{{Iwasawa} \& {Taniguchi}(1993)}]{it93}
{Iwasawa}, K. \& {Taniguchi}, Y. 1993, \apjl, 413, L15

\bibitem[{{Iwasawa} {et~al.}(2015){Iwasawa}, {Vignali}, {Comastri}, {Gilli},
  {Vito}, {Brandt}, {Carrera}, {Lanzuisi}, {Falocco}, \& {Vagnetti}}]{i15brtwo}
{Iwasawa}, K., {Vignali}, C., {Comastri}, A., {et~al.} 2015, \aap, 574, A144

\bibitem[{{Kellermann} {et~al.}(1989){Kellermann}, {Sramek}, {Schmidt},
  {Shaffer}, \& {Green}}]{kellermann89}
{Kellermann}, K.~I., {Sramek}, R., {Schmidt}, M., {Shaffer}, D.~B., \& {Green},
  R. 1989, \aj, 98, 1195

\bibitem[{{Komatsu} {et~al.}(2011){Komatsu}, {Smith}, {Dunkley}, {Bennett},
  {Gold}, {Hinshaw}, {Jarosik}, {Larson}, {Nolta}, {Page}, {Spergel},
  {Halpern}, {Hill}, {Kogut}, {Limon}, {Meyer}, {Odegard}, {Tucker}, {Weiland},
  {Wollack}, \& {Wright}}]{komatsu11}
{Komatsu}, E., {Smith}, K.~M., {Dunkley}, J., {et~al.} 2011, \apjs, 192, 18

\bibitem[{{Koss} {et~al.}(2016){Koss}, {Assef}, {Balokovi{\'c}}, {Stern},
  {Gandhi}, {Lamperti}, {Alexander}, {Ballantyne}, {Bauer}, {Berney}, {Brandt},
  {Comastri}, {Gehrels}, {Harrison}, {Lansbury}, {Markwardt}, {Ricci},
  {Rivers}, {Schawinski}, {Trakhtenbrot}, {Treister}, \& {Urry}}]{koss16}
{Koss}, M.~J., {Assef}, R., {Balokovi{\'c}}, M., {et~al.} 2016, \apj, 825, 85

\bibitem[{{Kurk} {et~al.}(2013){Kurk}, {Cimatti}, {Daddi}, {Mignoli},
  {Pozzetti}, {Dickinson}, {Bolzonella}, {Zamorani}, {Cassata}, {Rodighiero},
  {Franceschini}, {Renzini}, {Rosati}, {Halliday}, \& {Berta}}]{kurk13}
{Kurk}, J., {Cimatti}, A., {Daddi}, E., {et~al.} 2013, \aap, 549, A63

\bibitem[{{La Franca} {et~al.}(2005){La Franca}, {Fiore}, {Comastri}, {Perola},
  {Sacchi}, {Brusa}, {Cocchia}, {Feruglio}, {Matt}, {Vignali}, {Carangelo},
  {Ciliegi}, {Lamastra}, {Maiolino}, {Mignoli}, {Molendi}, \&
  {Puccetti}}]{lafranca05}
{La Franca}, F., {Fiore}, F., {Comastri}, A., {et~al.} 2005, \apj, 635, 864

\bibitem[{{Lambrides} {et~al.}(2020){Lambrides}, {Chiaberge}, {Heckman},
  {Gilli}, {Vito}, \& {Norman}}]{lambrides20}
{Lambrides}, E., {Chiaberge}, M., {Heckman}, T., {et~al.} 2020, arXiv e-prints,
  arXiv:2002.00955

\bibitem[{{Lanzuisi} {et~al.}(2018){Lanzuisi}, {Civano}, {Marchesi},
  {Comastri}, {Brusa}, {Gilli}, {Vignali}, {Zamorani}, {Brightman},
  {Griffiths}, \& {Koekemoer}}]{lanzuisi18}
{Lanzuisi}, G., {Civano}, F., {Marchesi}, S., {et~al.} 2018, \mnras, 480, 2578

\bibitem[{{Le F{\`e}vre} {et~al.}(2013){Le F{\`e}vre}, {Cassata}, {Cucciati},
  {Garilli}, {Ilbert}, {Le Brun}, {Maccagni}, {Moreau}, {Scodeggio}, {Tresse},
  {Zamorani}, {Adami}, {Arnouts}, {Bardelli}, {Bolzonella}, {Bondi},
  {Bongiorno}, {Bottini}, {Cappi}, {Charlot}, {Ciliegi}, {Contini}, {de la
  Torre}, {Foucaud}, {Franzetti}, {Gavignaud}, {Guzzo}, {Iovino}, {Lemaux},
  {L{\'o}pez-Sanjuan}, {McCracken}, {Marano}, {Marinoni}, {Mazure}, {Mellier},
  {Merighi}, {Merluzzi}, {Paltani}, {Pell{\`o}}, {Pollo}, {Pozzetti},
  {Scaramella}, {Tasca}, {Vergani}, {Vettolani}, {Zanichelli}, \&
  {Zucca}}]{lefvre13}
{Le F{\`e}vre}, O., {Cassata}, P., {Cucciati}, O., {et~al.} 2013, \aap, 559,
  A14

\bibitem[{{Liu} {et~al.}(2017){Liu}, {Tozzi}, {Wang}, {Brandt}, {Vignali},
  {Xue}, {Schneider}, {Comastri}, {Yang}, {Bauer}, {Paolillo}, {Luo}, {Gilli},
  {Wang}, {Giavalisco}, {Ji}, {Alexander}, {Mainieri}, {Shemmer}, {Koekemoer},
  \& {Risaliti}}]{liu17}
{Liu}, T., {Tozzi}, P., {Wang}, J.-X., {et~al.} 2017, \apjs, 232, 8

\bibitem[{{Luo} {et~al.}(2010){Luo}, {Brandt}, {Xue}, {Brusa}, {Alexander},
  {Bauer}, {Comastri}, {Koekemoer}, {Lehmer}, {Mainieri}, {Rafferty},
  {Schneider}, {Silverman}, \& {Vignali}}]{luo10}
{Luo}, B., {Brandt}, W.~N., {Xue}, Y.~Q., {et~al.} 2010, \apjs, 187, 560

\bibitem[{{Luo} {et~al.}(2017){Luo}, {Brandt}, {Xue}, {Lehmer}, {Alexander},
  {Bauer}, {Vito}, {Yang}, {Basu-Zych}, {Comastri}, {Gilli}, {Gu},
  {Hornschemeier}, {Koekemoer}, {Liu}, {Mainieri}, {Paolillo}, {Ranalli},
  {Rosati}, {Schneider}, {Shemmer}, {Smail}, {Sun}, {Tozzi}, {Vignali}, \&
  {Wang}}]{luo17}
{Luo}, B., {Brandt}, W.~N., {Xue}, Y.~Q., {et~al.} 2017, \apjs, 228, 2

\bibitem[{{Masini} {et~al.}(2018){Masini}, {Civano}, {Comastri}, {Fornasini},
  {Ballantyne}, {Lansbury}, {Treister}, {Alexander}, {Boorman}, {Brandt},
  {Farrah}, {Gandhi}, {Harrison}, {Hickox}, {Kocevski}, {Lanz}, {Marchesi},
  {Puccetti}, {Ricci}, {Saez}, {Stern}, \& {Zappacosta}}]{masini18}
{Masini}, A., {Civano}, F., {Comastri}, A., {et~al.} 2018, \apjs, 235, 17

\bibitem[{{Mignoli} {et~al.}(2005){Mignoli}, {Cimatti}, {Zamorani}, {Pozzetti},
  {Daddi}, {Renzini}, {Broadhurst}, {Cristiani}, {D'Odorico}, {Fontana},
  {Giallongo}, {Gilmozzi}, {Menci}, \& {Saracco}}]{mignoli05}
{Mignoli}, M., {Cimatti}, A., {Zamorani}, G., {et~al.} 2005, \aap, 437, 883

\bibitem[{{Mignoli} {et~al.}(2004){Mignoli}, {Pozzetti}, {Comastri}, {Brusa},
  {Ciliegi}, {Cocchia}, {Fiore}, {La Franca}, {Maiolino}, {Matt}, {Molendi},
  {Perola}, {Puccetti}, {Severgnini}, \& {Vignali}}]{mignoli04}
{Mignoli}, M., {Pozzetti}, L., {Comastri}, A., {et~al.} 2004, \aap, 418, 827

\bibitem[{{Morris} {et~al.}(2015){Morris}, {Kocevski}, {Trump}, {Weiner},
  {Hathi}, {Barro}, {Dahlen}, {Faber}, {Finkelstein}, {Fontana}, {Ferguson},
  {Grogin}, {Gr{\"u}tzbauch}, {Guo}, {Hsu}, {Koekemoer}, {Koo}, {Mobasher},
  {Pforr}, {Salvato}, {Wiklind}, \& {Wuyts}}]{morris05}
{Morris}, A.~M., {Kocevski}, D.~D., {Trump}, J.~R., {et~al.} 2015, \aj, 149,
  178

\bibitem[{{Nandra} \& {Pounds}(1994)}]{nandrapounds94}
{Nandra}, K. \& {Pounds}, K.~A. 1994, \mnras, 268, 405

\bibitem[{{Norman} {et~al.}(2002){Norman}, {Hasinger}, {Giacconi}, {Gilli},
  {Kewley}, {Nonino}, {Rosati}, {Szokoly}, {Tozzi}, {Wang}, {Zheng}, {Zirm},
  {Bergeron}, {Gilmozzi}, {Grogin}, {Koekemoer}, \& {Schreier}}]{norman02}
{Norman}, C., {Hasinger}, G., {Giacconi}, R., {et~al.} 2002, \apj, 571, 218

\bibitem[{{Padovani} {et~al.}(2011){Padovani}, {Miller}, {Kellermann},
  {Mainieri}, {Rosati}, \& {Tozzi}}]{padovani11}
{Padovani}, P., {Miller}, N., {Kellermann}, K.~I., {et~al.} 2011, \apj, 740, 20

\bibitem[{{Page} {et~al.}(2004){Page}, {O'Brien}, {Reeves}, \&
  {Turner}}]{page04}
{Page}, K.~L., {O'Brien}, P.~T., {Reeves}, J.~N., \& {Turner}, M.~J.~L. 2004,
  \mnras, 347, 316

\bibitem[{{Pentericci} {et~al.}(2018){Pentericci}, {McLure}, {Franzetti},
  {Garilli}, \& {the VANDELS team}}]{pentericci18}
{Pentericci}, L., {McLure}, R.~J., {Franzetti}, P., {Garilli}, B., \& {the
  VANDELS team}. 2018, arXiv e-prints, arXiv:1811.05298

\bibitem[{P\'erez \& Granger(2007)}]{PER-GRA:2007}
P\'erez, F. \& Granger, B.~E. 2007, Computing in Science and Engineering, 9, 21

\bibitem[{Plummer(2003)}]{Plummer03jags:a}
Plummer, M. 2003, JAGS: A program for analysis of Bayesian graphical models
  using Gibbs sampling

\bibitem[{{Popesso} {et~al.}(2009){Popesso}, {Dickinson}, {Nonino}, {Vanzella},
  {Daddi}, {Fosbury}, {Kuntschner}, {Mainieri}, {Cristiani}, {Cesarsky},
  {Giavalisco}, {Renzini}, \& {GOODS Team}}]{popesso09}
{Popesso}, P., {Dickinson}, M., {Nonino}, M., {et~al.} 2009, \aap, 494, 443

\bibitem[{{R Core Team}(2017)}]{rcoreteam17}
{R Core Team}. 2017, R: A Language and Environment for Statistical Computing, R
  Foundation for Statistical Computing, Vienna, Austria

\bibitem[{{Rafferty} {et~al.}(2011){Rafferty}, {Brandt}, {Alexander}, {Xue},
  {Bauer}, {Lehmer}, {Luo}, \& {Papovich}}]{rafferty10}
{Rafferty}, D.~A., {Brandt}, W.~N., {Alexander}, D.~M., {et~al.} 2011, \apj,
  742, 3

\bibitem[{{Ranalli} {et~al.}(2013){Ranalli}, {Comastri}, {Vignali}, {Carrera},
  {Cappelluti}, {Gilli}, {Puccetti}, {Brandt}, {Brunner}, {Brusa},
  {Georgantopoulos}, {Iwasawa}, \& {Mainieri}}]{ranalli13}
{Ranalli}, P., {Comastri}, A., {Vignali}, C., {et~al.} 2013, \aap, 555, A42

\bibitem[{{Ravikumar} {et~al.}(2007){Ravikumar}, {Puech}, {Flores}, {Proust},
  {Hammer}, {Lehnert}, {Rawat}, {Amram}, {Balkowski}, {Burgarella}, {Cassata},
  {Cesarsky}, {Cimatti}, {Combes}, {Daddi}, {Dannerbauer}, {di Serego
  Alighieri}, {Elbaz}, {Guiderdoni}, {Kembhavi}, {Liang}, {Pozzetti},
  {Vergani}, {Vernet}, {Wozniak}, \& {Zheng}}]{ravikumar07}
{Ravikumar}, C.~D., {Puech}, M., {Flores}, H., {et~al.} 2007, \aap, 465, 1099

\bibitem[{{Ricci} {et~al.}(2013){Ricci}, {Paltani}, {Awaki}, {Petrucci},
  {Ueda}, \& {Brightman}}]{ricci13}
{Ricci}, C., {Paltani}, S., {Awaki}, H., {et~al.} 2013, \aap, 553, A29

\bibitem[{{Ricci} {et~al.}(2015){Ricci}, {Ueda}, {Koss}, {Trakhtenbrot},
  {Bauer}, \& {Gandhi}}]{ricci15}
{Ricci}, C., {Ueda}, Y., {Koss}, M.~J., {et~al.} 2015, \apjl, 815, L13

\bibitem[{{Sanders} {et~al.}(1988){Sanders}, {Soifer}, {Elias}, {Madore},
  {Matthews}, {Neugebauer}, \& {Scoville}}]{sanders88}
{Sanders}, D.~B., {Soifer}, B.~T., {Elias}, J.~H., {et~al.} 1988, \apj, 325, 74

\bibitem[{{Shu} {et~al.}(2012){Shu}, {Wang}, {Yaqoob}, {Jiang}, \&
  {Zhou}}]{shu12}
{Shu}, X.~W., {Wang}, J.~X., {Yaqoob}, T., {Jiang}, P., \& {Zhou}, Y.~Y. 2012,
  \apjl, 744, L21

\bibitem[{{Silverman} {et~al.}(2010){Silverman}, {Mainieri}, {Salvato},
  {Hasinger}, {Bergeron}, {Capak}, {Szokoly}, {Finoguenov}, {Gilli}, {Rosati},
  {Tozzi}, {Vignali}, {Alexander}, {Brandt}, {Lehmer}, {Luo}, {Rafferty},
  {Xue}, {Balestra}, {Bauer}, {Brusa}, {Comastri}, {Kartaltepe}, {Koekemoer},
  {Miyaji}, {Schneider}, {Treister}, {Wisotski}, \& {Schramm}}]{silverman10}
{Silverman}, J.~D., {Mainieri}, V., {Salvato}, M., {et~al.} 2010, \apjs, 191,
  124

\bibitem[{{Straatman} {et~al.}(2016){Straatman}, {Spitler}, {Quadri},
  {Labb{\'e}}, {Glazebrook}, {Persson}, {Papovich}, {Tran}, {Brammer},
  {Cowley}, {Tomczak}, {Nanayakkara}, {Alcorn}, {Allen}, {Broussard}, {van
  Dokkum}, {Forrest}, {van Houdt}, {Kacprzak}, {Kawinwanichakij}, {Kelson},
  {Lee}, {McCarthy}, {Mehrtens}, {Monson}, {Murphy}, {Rees}, {Tilvi}, \&
  {Whitaker}}]{straatman16}
{Straatman}, C.~M.~S., {Spitler}, L.~R., {Quadri}, R.~F., {et~al.} 2016, \apj,
  830, 51

\bibitem[{{Streblyanska} {et~al.}(2005){Streblyanska}, {Hasinger},
  {Finoguenov}, {Barcons}, {Mateos}, \& {Fabian}}]{streblyanska05}
{Streblyanska}, A., {Hasinger}, G., {Finoguenov}, A., {et~al.} 2005, \aap, 432,
  395

\bibitem[{{Str{\"u}der} {et~al.}(2001){Str{\"u}der}, {Briel}, {Dennerl},
  {Hartmann}, {Kendziorra}, {Meidinger}, {Pfeffermann}, {Reppin}, {Aschenbach},
  {Bornemann}, {Br{\"a}uninger}, {Burkert}, {Elender}, {Freyberg}, {Haberl},
  {Hartner}, {Heuschmann}, {Hippmann}, {Kastelic}, {Kemmer}, {Kettenring},
  {Kink}, {Krause}, {M{\"u}ller}, {Oppitz}, {Pietsch}, {Popp}, {Predehl},
  {Read}, {Stephan}, {St{\"o}tter}, {Tr{\"u}mper}, {Holl}, {Kemmer}, {Soltau},
  {St{\"o}tter}, {Weber}, {Weichert}, {von Zanthier}, {Carathanassis}, {Lutz},
  {Richter}, {Solc}, {B{\"o}ttcher}, {Kuster}, {Staubert}, {Abbey}, {Holland},
  {Turner}, {Balasini}, {Bignami}, {La Palombara}, {Villa}, {Buttler},
  {Gianini}, {Lain{\'e}}, {Lumb}, \& {Dhez}}]{struder01}
{Str{\"u}der}, L., {Briel}, U., {Dennerl}, K., {et~al.} 2001, \aap, 365, L18

\bibitem[{{Szokoly} {et~al.}(2004){Szokoly}, {Bergeron}, {Hasinger}, {Lehmann},
  {Kewley}, {Mainieri}, {Nonino}, {Rosati}, {Giacconi}, {Gilli}, {Gilmozzi},
  {Norman}, {Romaniello}, {Schreier}, {Tozzi}, {Wang}, {Zheng}, \&
  {Zirm}}]{szokoly04}
{Szokoly}, G.~P., {Bergeron}, J., {Hasinger}, G., {et~al.} 2004, \apjs, 155,
  271

\bibitem[{{Tanaka} {et~al.}(1995){Tanaka}, {Nandra}, {Fabian}, {Inoue},
  {Otani}, {Dotani}, {Hayashida}, {Iwasawa}, {Kii}, {Kunieda}, {Makino}, \&
  {Matsuoka}}]{tanaka95}
{Tanaka}, Y., {Nandra}, K., {Fabian}, A.~C., {et~al.} 1995, \nat, 375, 659

\bibitem[{{Taylor} {et~al.}(2009){Taylor}, {Franx}, {van Dokkum}, {Quadri},
  {Gawiser}, {Bell}, {Barrientos}, {Blanc}, {Castander}, {Damen},
  {Gonzalez-Perez}, {Hall}, {Herrera}, {Hildebrandt}, {Kriek}, {Labb{\'e}},
  {Lira}, {Maza}, {Rudnick}, {Treister}, {Urry}, {Willis}, \&
  {Wuyts}}]{taylor09}
{Taylor}, E.~N., {Franx}, M., {van Dokkum}, P.~G., {et~al.} 2009, \apjs, 183,
  295

\bibitem[{{Tozzi} {et~al.}(2006){Tozzi}, {Gilli}, {Mainieri}, {Norman},
  {Risaliti}, {Rosati}, {Bergeron}, {Borgani}, {Giacconi}, {Hasinger},
  {Nonino}, {Streblyanska}, {Szokoly}, {Wang}, \& {Zheng}}]{tozzi06}
{Tozzi}, P., {Gilli}, R., {Mainieri}, V., {et~al.} 2006, \aap, 451, 457

\bibitem[{{Treister} \& {Urry}(2006)}]{treisterurry06}
{Treister}, E. \& {Urry}, C.~M. 2006, \apjl, 652, L79

\bibitem[{{Treister} {et~al.}(2009){Treister}, {Virani}, {Gawiser}, {Urry},
  {Lira}, {Francke}, {Blanc}, {Cardamone}, {Damen}, {Taylor}, \&
  {Schawinski}}]{treister09}
{Treister}, E., {Virani}, S., {Gawiser}, E., {et~al.} 2009, \apj, 693, 1713

\bibitem[{{Turner} {et~al.}(2001){Turner}, {Abbey}, {Arnaud}, {Balasini},
  {Barbera}, {Belsole}, {Bennie}, {Bernard}, {Bignami}, {Boer}, {Briel},
  {Butler}, {Cara}, {Chabaud}, {Cole}, {Collura}, {Conte}, {Cros}, {Denby},
  {Dhez}, {Di Coco}, {Dowson}, {Ferrando}, {Ghizzardi}, {Gianotti}, {Goodall},
  {Gretton}, {Griffiths}, {Hainaut}, {Hochedez}, {Holland}, {Jourdain},
  {Kendziorra}, {Lagostina}, {Laine}, {La Palombara}, {Lortholary}, {Lumb},
  {Marty}, {Molendi}, {Pigot}, {Poindron}, {Pounds}, {Reeves}, {Reppin},
  {Rothenflug}, {Salvetat}, {Sauvageot}, {Schmitt}, {Sembay}, {Short},
  {Spragg}, {Stephen}, {Str{\"u}der}, {Tiengo}, {Trifoglio}, {Tr{\"u}mper},
  {Vercellone}, {Vigroux}, {Villa}, {Ward}, {Whitehead}, \& {Zonca}}]{turner01}
{Turner}, M.~J.~L., {Abbey}, A., {Arnaud}, M., {et~al.} 2001, \aap, 365, L27

\bibitem[{{Ueda} {et~al.}(2014){Ueda}, {Akiyama}, {Hasinger}, {Miyaji}, \&
  {Watson}}]{ueda14}
{Ueda}, Y., {Akiyama}, M., {Hasinger}, G., {Miyaji}, T., \& {Watson}, M.~G.
  2014, \apj, 786, 104

\bibitem[{{Urrutia} {et~al.}(2019){Urrutia}, {Wisotzki}, {Kerutt}, {Schmidt},
  {Herenz}, {Klar}, {Saust}, {Werhahn}, {Diener}, {Caruana}, {Krajnovi{\'c}},
  {Bacon}, {Boogaard}, {Brinchmann}, {Enke}, {Maseda}, {Nanayakkara},
  {Richard}, {Steinmetz}, \& {Weilbacher}}]{urrutia19}
{Urrutia}, T., {Wisotzki}, L., {Kerutt}, J., {et~al.} 2019, \aap, 624, A141

\bibitem[{{Vanzella} {et~al.}(2008){Vanzella}, {Cristiani}, {Dickinson},
  {Giavalisco}, {Kuntschner}, {Haase}, {Nonino}, {Rosati}, {Cesarsky},
  {Ferguson}, {Fosbury}, {Grazian}, {Moustakas}, {Rettura}, {Popesso},
  {Renzini}, {Stern}, \& {GOODS Team}}]{vanzella08}
{Vanzella}, E., {Cristiani}, S., {Dickinson}, M., {et~al.} 2008, \aap, 478, 83

\bibitem[{{Vignali} {et~al.}(2015){Vignali}, {Iwasawa}, {Comastri}, {Gilli},
  {Lanzuisi}, {Ranalli}, {Cappelluti}, {Mainieri}, {Georgantopoulos},
  {Carrera}, {Fritz}, {Brusa}, {Brandt}, {Bauer}, {Fiore}, \&
  {Tombesi}}]{vignali15}
{Vignali}, C., {Iwasawa}, K., {Comastri}, A., {et~al.} 2015, \aap, 583, A141

\bibitem[{{Vito} {et~al.}(2018){Vito}, {Brandt}, {Yang}, {Gilli}, {Luo},
  {Vignali}, {Xue}, {Comastri}, {Koekemoer}, {Lehmer}, {Liu}, {Paolillo},
  {Ranalli}, {Schneider}, {Shemmer}, {Volonteri}, \& {Wang}}]{vito18}
{Vito}, F., {Brandt}, W.~N., {Yang}, G., {et~al.} 2018, \mnras, 473, 2378

\bibitem[{{Watson} {et~al.}(2009){Watson}, {Schr{\"o}der}, {Fyfe}, {Page},
  {Lamer}, {Mateos}, {Pye}, {Sakano}, {Rosen}, {Ballet}, {Barcons}, {Barret},
  {Boller}, {Brunner}, {Brusa}, {Caccianiga}, {Carrera}, {Ceballos}, {Della
  Ceca}, {Denby}, {Denkinson}, {Dupuy}, {Farrell}, {Fraschetti}, {Freyberg},
  {Guillout}, {Hambaryan}, {Maccacaro}, {Mathiesen}, {McMahon}, {Michel},
  {Motch}, {Osborne}, {Page}, {Pakull}, {Pietsch}, {Saxton}, {Schwope},
  {Severgnini}, {Simpson}, {Sironi}, {Stewart}, {Stewart}, {Stobbart}, {Tedds},
  {Warwick}, {Webb}, {West}, {Worrall}, \& {Yuan}}]{watson09}
{Watson}, M.~G., {Schr{\"o}der}, A.~C., {Fyfe}, D., {et~al.} 2009, \aap, 493,
  339

\bibitem[{{W}es {M}c{K}inney(2010)}]{mckinney-proc-scipy-2010}
{W}es {M}c{K}inney. 2010, in {P}roceedings of the 9th {P}ython in {S}cience
  {C}onference, ed. {S}t\'efan van~der {W}alt \& {J}arrod {M}illman, 56 -- 61

\bibitem[{Wickham(2009)}]{wickham09}
Wickham, H. 2009, ggplot2: Elegant Graphics for Data Analysis (Springer-Verlag
  New York)

\bibitem[{{Xue} {et~al.}(2016){Xue}, {Luo}, {Brandt}, {Alexander}, {Bauer},
  {Lehmer}, \& {Yang}}]{xue16}
{Xue}, Y.~Q., {Luo}, B., {Brandt}, W.~N., {et~al.} 2016, \apjs, 224, 15

\end{thebibliography}

\begin{appendix}

\section{X-ray redshifts}

\begin{table*}
\caption{11 sources with X-ray redshift estimates}
\label{tab:zx}
\centering
\begin{tabular}{ccccc}
PID & $z_\mathrm{x}$ & $z_\mathrm{ph}$ & sp & ID(H14) \\[5pt]
19 & 1.92 (1.90-1.94) & 1.980 (1.47-2.6) & L & 105024 \\
32 & 1.90 (1.74-1.96) & 3.163 (0-4.27) & E & 105036 \\
88\tablefootmark{a} & 1.12 (1.1-1.14) & 0.345, 1.360 & L & 59030 \\
95 & 0.86 (0.85-0.87) & 1.478 (0.91-1.78) & EL & 105046 \\
103 & 2.00 (1.98-2.02) & 1.976 (1.92-2.02) & L & 60464 \\
142 & 2.25 (2.15-2.37) & 2.509 (2.03-2.69) & E & 63777 \\
166 & 1.37 (1.32-1.43) & 1.688 (1.55-1.71)\tablefootmark{b} & EL & 9360 \\
213 & 3.17 (3.08-3.31) & 2.167 (1.44-5.74) & E & 105100 \\
215\tablefootmark{c} & 2.66 (2.65-2.67) & 2.513 (2.45-2.61) & L & 15377 \\
252 & 1.52 (1.49-1.56) & 1.907 (1.82-1.95) & E & 19476 \\
338 & 2.02 (2.00-2.07) & 1.018 (0.09-2.58) & E & 105136 \\
\end{tabular}
\tablefoot{Table columns are PID, X-ray redshift $z_\mathrm{x}$
  and the 68\% error interval in parenthesis, photometric redshifts,
  and the 68\% error interval from \citep[][=H14]{hsu14}, X-ray spectral
  feature which drives $z_\mathrm{x}$ estimate: E: Fe K absorption
  edge; L: Fe K line, source ID number of Chandra counterpart in H14.
  \tablefoottext{a}{The $z_\mathrm{x}$ is uncertain as well as
    other redshift estimates. See text for further details.}
  \tablefoottext{b}{Another photometric redshift from
    \citet{straatman16} is $z_\mathrm{ph} = 1.37$.}
  \tablefoottext{c}{There are two Chandra sources in the XMM error
    circle. One of them has a spectroscopic redshift but is fainter in
    X-ray. The other brighter X-ray source only has photmetric
    redshifts shown here and $z_\mathrm{ph}=2.6$ from
    \citet{straatman16}, and seems to be the source of a strong Fe K
    line detected in the XMM-Newton spectrum. See text for further
    details.}
  }
\end{table*}

Table \ref{tab:zx} lists 11 sources for which no spectroscopic
redshifts are available and, instead, redshifts are obtained from X-ray Fe K
band spectral measurements, in addition to the six sources with
redshifts derived by the same technique and published previously
\citep{i12cdfs,vignali15}. Seven sources out of 11 are strongly
absorbed by log \nH\ $\geq 23$ [\psqcm] (see Table \ref{tab:big}) and
the redshift estimates are driven by the Fe K absorption edge. When a
Fe K line, compatible with the estimate by the absorption edge, is
detected, the redshift and its uncertainty are revised. Existing
photometric redshifts were used as a guide and if an X-ray estimate
was better constrained, we adopted the X-ray redshift. Photometric
redshifts of these sources are all available from \citet{hsu14} (H14
hereafter), they are shown in Table \ref{tab:zx} for
reference. Further comments on three sources marked in the Table are
given below.

\smallskip
\noindent PID 88: This X-ray redshift is a tentative assignment, as
there is no agreement between the insecure spectroscopic ($z=0.522$),
the two equally likely photometric redshifts by H14 given in the table
(the probability density function for the photometric redshift shows a
double peaks with comparable heights) and the X-ray estimate. A strong
line feature at 3 keV is identified as a Fe K line to estimate the
redshift but it might be a false detection given the relatively soft
spectrum. Therefore, any of the redshift estimates are highly
uncertain.

\smallskip
\noindent PID 166: The X-ray spectrum of this source is strongly
absorbed with log \nH\ $=23.18$ [\psqcm]. H14 gives a mild constraint
on $z_\mathrm{ph} = 1.688$ while \citet{straatman16} gives
$z_\mathrm{ph} = 1.377$. A combination of the Fe K edge and Fe K line
gives the reasonably constrained $z_\mathrm{x}$ close to the estimate
of \citet{straatman16}.

\smallskip
\noindent PID 215: Two Chandra sources are present at the XMM source
position \citep{luo17}. The fainter source has a spectroscopic
redshift of $z=2.252$ \citep{casey11} while the brighter source (which
is fainter in optical/NIR) has photometric redshift estimates by H14
and $z_\mathrm{ph}=2.6$ by \citet{straatman16}. The XMM spectrum of
this source shows a strong Fe K line feature at 1.75 keV, which is
compatible with the photometric redshift for the brighter Chandra
source. The strong Fe line and the hard
continuum make this source as a likely Compton-thick AGN.

\section{Examples of Fe K band continuum estimate}

In Sect. 5.2, how the Fe K band continuum flux in each spectrum of the
uniform rest-frame intervals is outlined. We describe here
supplemental details on the method and a few examples of its
application. 

First, we pick 10 spectral bins below and above the Fe K bin ($i=13$,
which is excluded from this analysis) so that we have 5 bins on each
side. The lower 5 bins ranges from 4.0 keV to 6.0 keV ($i = $ 8-12)
while the higher from 6.6 keV to 10 keV ($i =$ 14-18). The central
bins on the two bands are at 4.9 keV ($i=10$) and 8.2 keV ($i=16$),
respectively. In each spectrum, we take a median of each band and
adopt the median value as the flux density of the central bin of each band.
\[ f^{\prime}_{10} = median\thinspace (f_8,... f_{12}), \]
\[ f^{\prime}_{16} = median\thinspace (f_{14},... f_{18}), \]
where $f_{i}$ represents flux density at $i$-th channel while
$f^{\prime}_{i}$ is 'estimated' continuum flux density at the given
channel. $f^{\prime }_{10}$ (or $f^{\prime }_{16}$) may coincide with
$f_{10}$ (or $f_{16}$) but it could be a value measured at the other
bin, which holds the median value. Thus we have two continuum
estimates at $i=10$ and $i=16$. Taking a logarithmic mean of the two
gives an estimate of the continuum at $i=13$, because all the bins are
logarithmically equal intervals by design.
\[ log\thinspace f^{\prime}_{13} = {log\thinspace f^{\prime}_{10} + log\thinspace f^{\prime}_{16}\over 2}, \]
where $f^{\prime}_{13}$ is the estimated continuum flux density at
$i=13$, the Fe K band (6.0-6.6 keV).

When, for example, a continuum shape is a simple power-law and data
quality is good, there is no complication in obtaining $f^{\prime}$ at
$i=10$ and $i=16$ and they are usually equal to $f_{10}$ and
$f_{16}$. However, when you have noisy data (which is often the case
for XMM-CDFS sources), data deviate from the underlying continuum shape
and spectral fitting is an usual solution. The above operation using
the median statistic, however, seems to give a reasonable estimate of the two
continuum data points encompassing the Fe K band and the resulted Fe K
band continuum flux, as supported by the intercept of
Fig. \ref{fig:ew-rfe} being very close to 1. Even when an Fe K line is
not found at $i=13$ but at the other interval, the median estimate
remains robust.

\begin{figure}
  \centerline{\includegraphics[width=0.35\textwidth,angle=0]{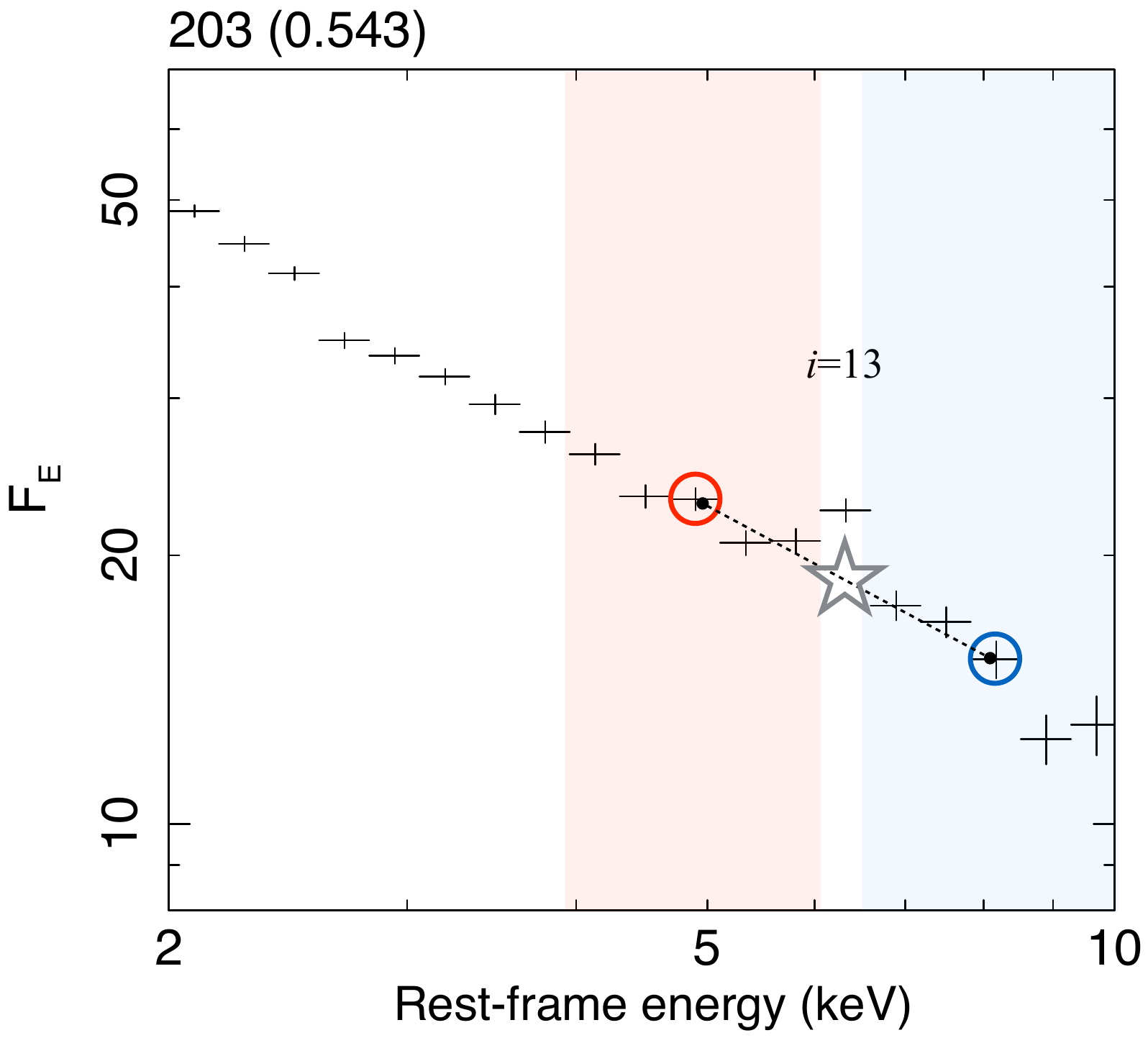}}
  \centerline{\includegraphics[width=0.35\textwidth,angle=0]{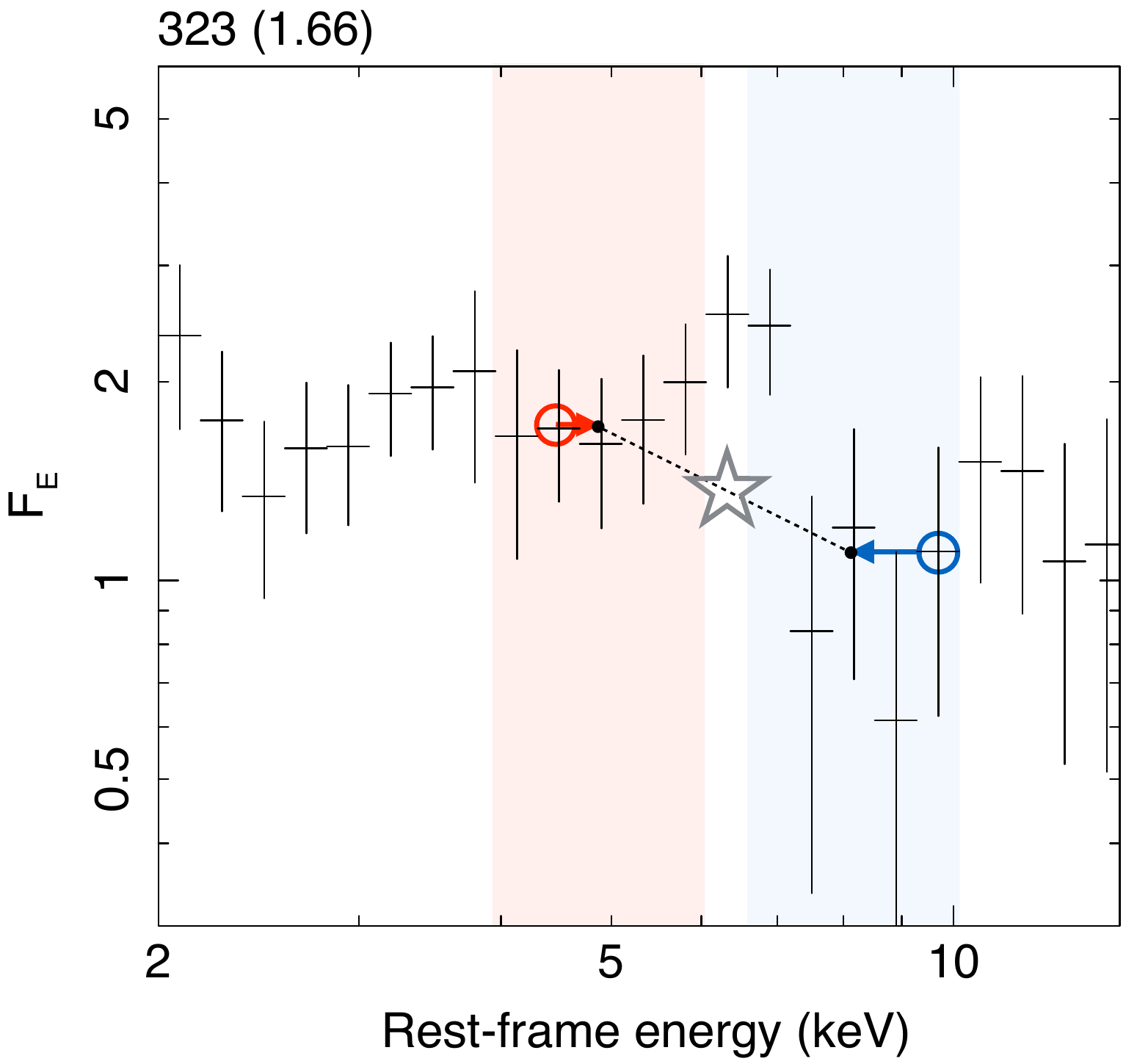}}
\caption{Examples of Fe K continuum estimates for PID 203 and PID
  323. A value inside the parenthesis following the PID in each figure
  is adopted source redshift. The shaded areas indicate the
  lower and higher energy ranges encompassing the Fe K band. The open
  circles indicate where the median values are found in the respective
  energy ranges. The dotted line is an adopted continuum between the
  4.9 keV ($i=10$) and 8.2 keV ($i=16$) with the estimated continuum
  level at $i=13$ (open star).}
\label{fig:ex203323}
\end{figure}

Figure \ref{fig:ex203323} shows examples of applying this method, for
a good quality spectrum (PID 203) and a noisy spectrum (PID 323). For
PID 203, $f^{\prime}_{10} = f_{10}$ and $f^{\prime}_{16} = f_{16}$. On
the other hand, for PID 323, the median values in both lower and higher
energy bands occur at intervals different from the respective central
energy intervals, hence $f^{\prime}_{10} = f_{9}$ and
$f^{\prime}_{16} = f_{18}$ are assigned to estimate $f^{\prime}_{13}$,
as indicated in the figure.

\section{Median/mean stacking for a composite spectrum}

\begin{figure}
  \centerline{\includegraphics[width=0.45\textwidth,angle=0]{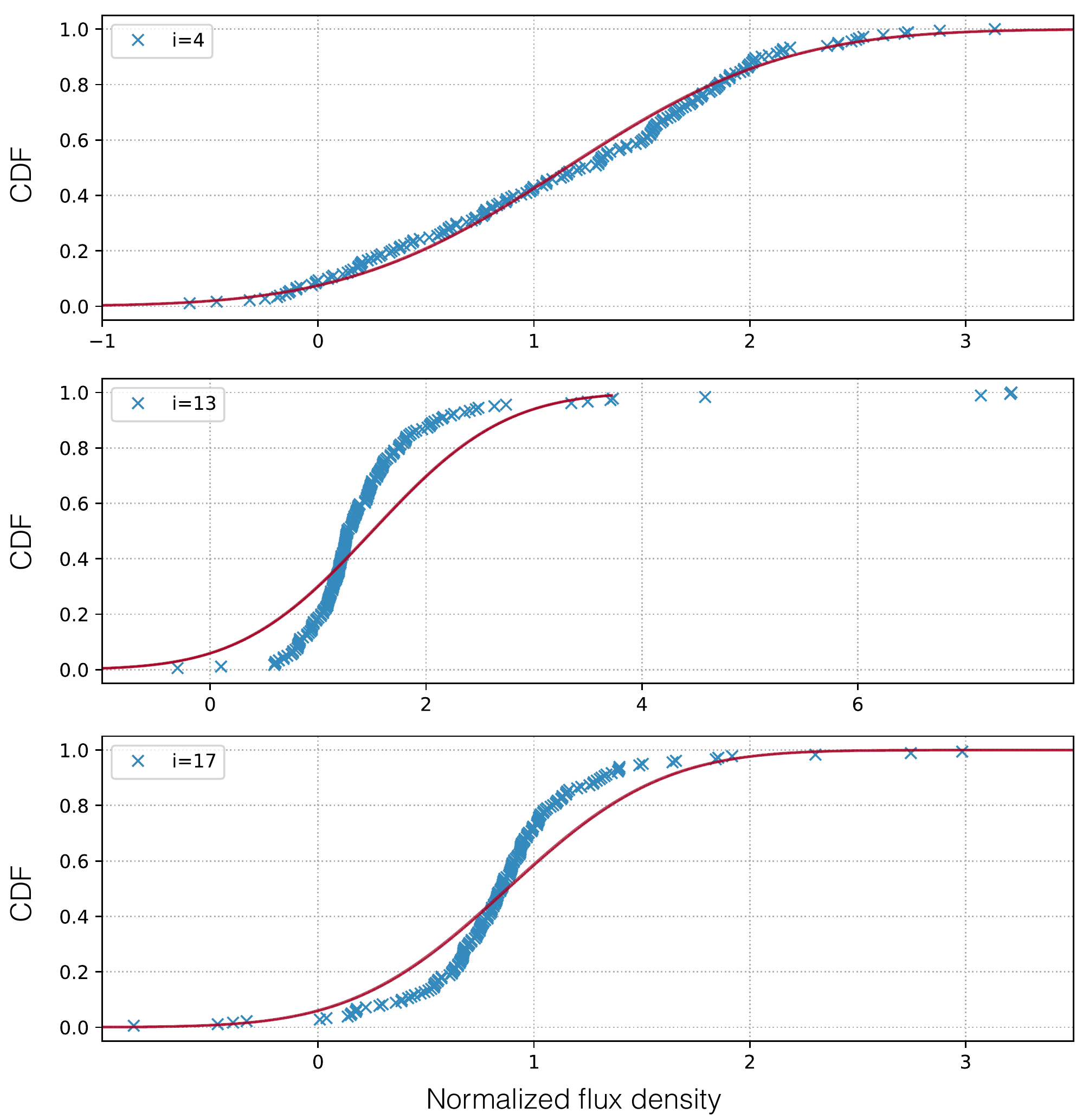}}
\caption{CDF of normalised flux density values (blue crosses) of
  the 180 XMM-CDFS sources in three spectral channels, $i=4$, $i=13$,
  $i=17$, corresponding to the rest frame energies, 3 keV, 6.3 keV (Fe
  K band), and 9 keV, respectively. A theoretical curve of the normal
  distribution which fits the data is overplotted in red in each
  plot.}
\label{fig:ch_dist}
\end{figure}

\begin{figure}
  \centerline{\includegraphics[width=0.35\textwidth,angle=0]{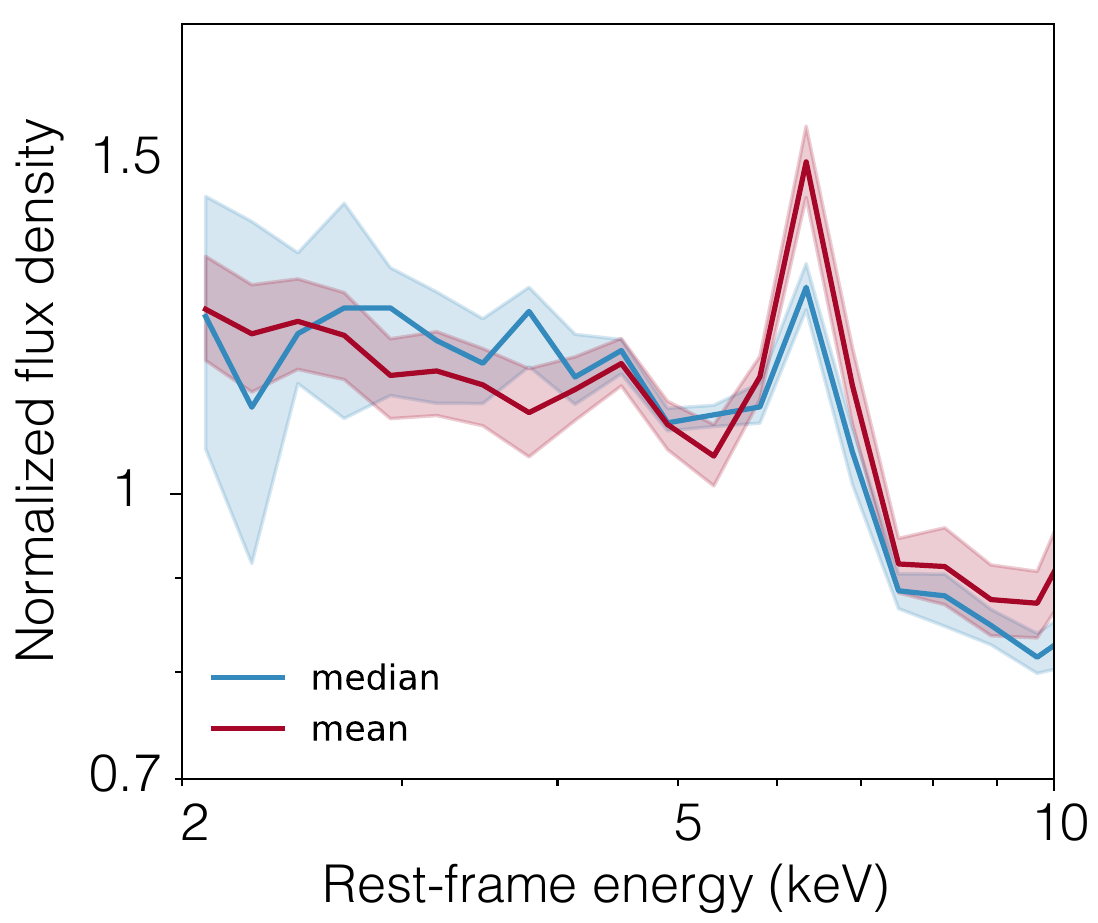}}
\caption{Composite spectra of all the 180 XMM-CDFS sources at $z>0.4$,
  constructed by median (blue) and mean (red) stacking. The composite
  spectrum of Fig. \ref{fig:medsp_all} in the main text is identical
  to that of median stacking.}
\label{fig:meanmedian}
\end{figure}

While previous X-ray spectral stacking work normally use mean
stacking of normalised spectra (or a straight sum), we use median
stacking to construct a composite spectrum, which is a normal practice in
UV/optical/infrared spectroscopy. Median statistic is relatively
robust against outliers and a skewed distribution, compared to mean
statistic. We offer a few examples to illustrate why median stacking
might perform better in obtaining a typical spectral shape than mean
stacking, at least for our dataset. Fig. \ref{fig:ch_dist} shows the
cumulative distribution function (CDF) of three spectral channels of
$i=4$, $i=13$ and $i=17$, corresponding to the rest-frame 3 keV, 6.3
keV and 9 keV, respectively, of the 180 sources to construct the
composite spectra in Fig. \ref{fig:meanmedian}. The CDF for $i=4$
reasonably follows the normal distribution and the mean and median
statistics remain close. The $i=13$ CDF is skewed towards high values
with outliers, as expected from the skewed EW distribution
(Fig. \ref{fig:ewdist}) and the presence of a few very strong Fe K
line sources. This stretches the fitted normal distribution wider and
drags it to higher values. The resulted mean is 1.51, while the median
is 1.27. This difference is translated to a factor of $\sim 2$ larger
EW of the Fe line in the mean stacking than in the median stacking
spectrum, as seen in Fig. \ref{fig:meanmedian}. Judging from the CDF,
the median appears more representative than the mean in this
particular channel. The $i=17$ CDF is closer to a symmetric shape but
outliers make the fitted normal distribution incompatible to the
data. Since our X-ray data of faint sources are noisy, each spectral
channel often contains some outliers. In terms of describing the
distribution, a $t$-distribution, which has an extra parameter for
heavier distribution tails, could be a solution. In fact, a
$t$-distribution gives a good description of the CDF of $i=17$ and of
many other channels ($i=13$ is a notable exception but the central
value, 1.29, is much closer to the median than the mean is). However,
for the simplicity and robustness of median statistic, we opted to use
median stacking to look for typical spectral shape for our dataset.

\section{XMM-CDFS EPIC spectral atlas}

The remaining 182 spectra of Fig. 2 are shown in Fig. \ref{fig:atlas}.

\begin{figure*}
  \centering
  \subfloat[]{\includegraphics[width=0.92\textwidth,angle=0]{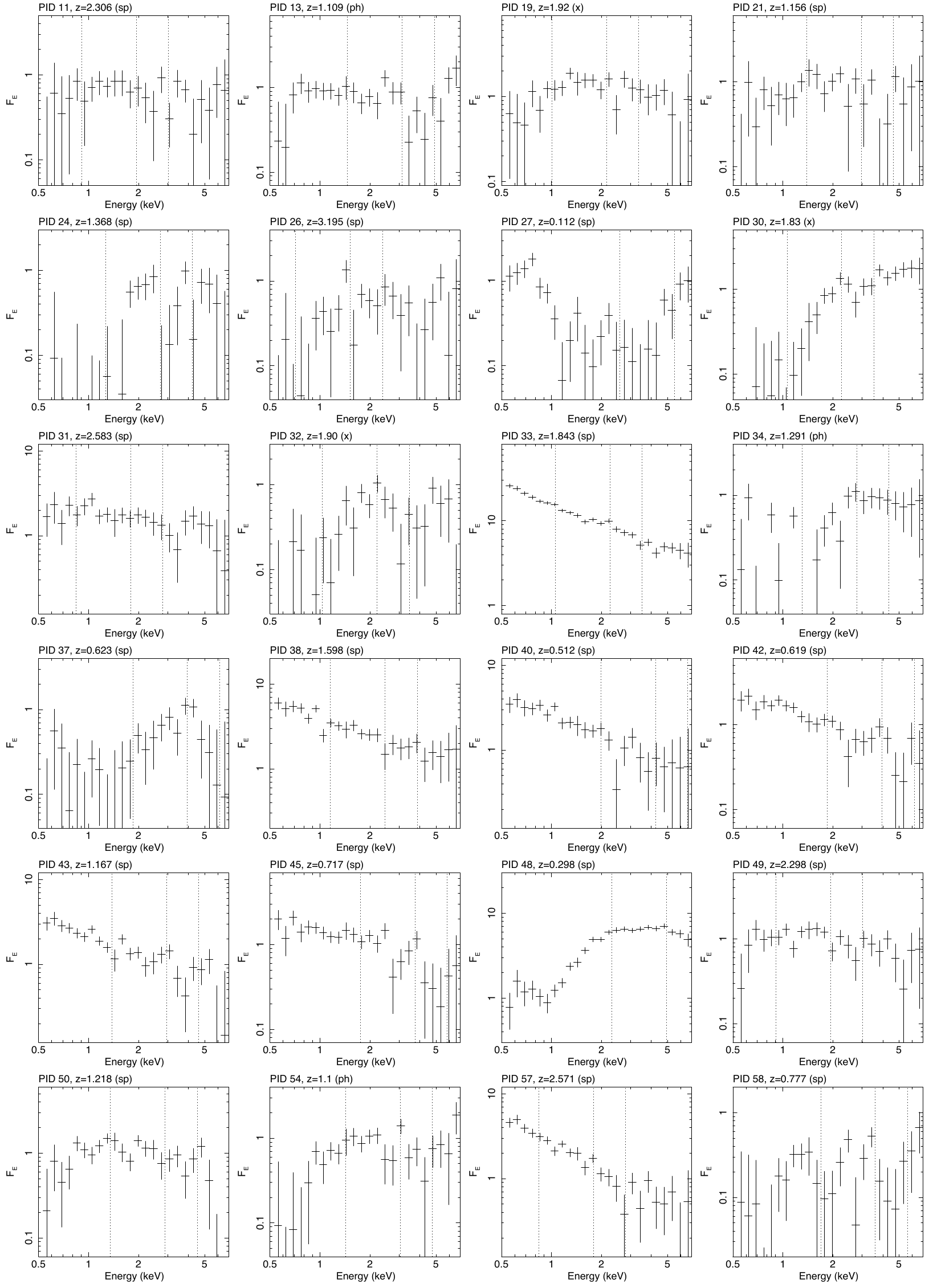}}
  \phantomcaption
\end{figure*}
\begin{figure*}
  \ContinuedFloat
  \centering
  \subfloat[]{\includegraphics[width=0.92\textwidth,angle=0]{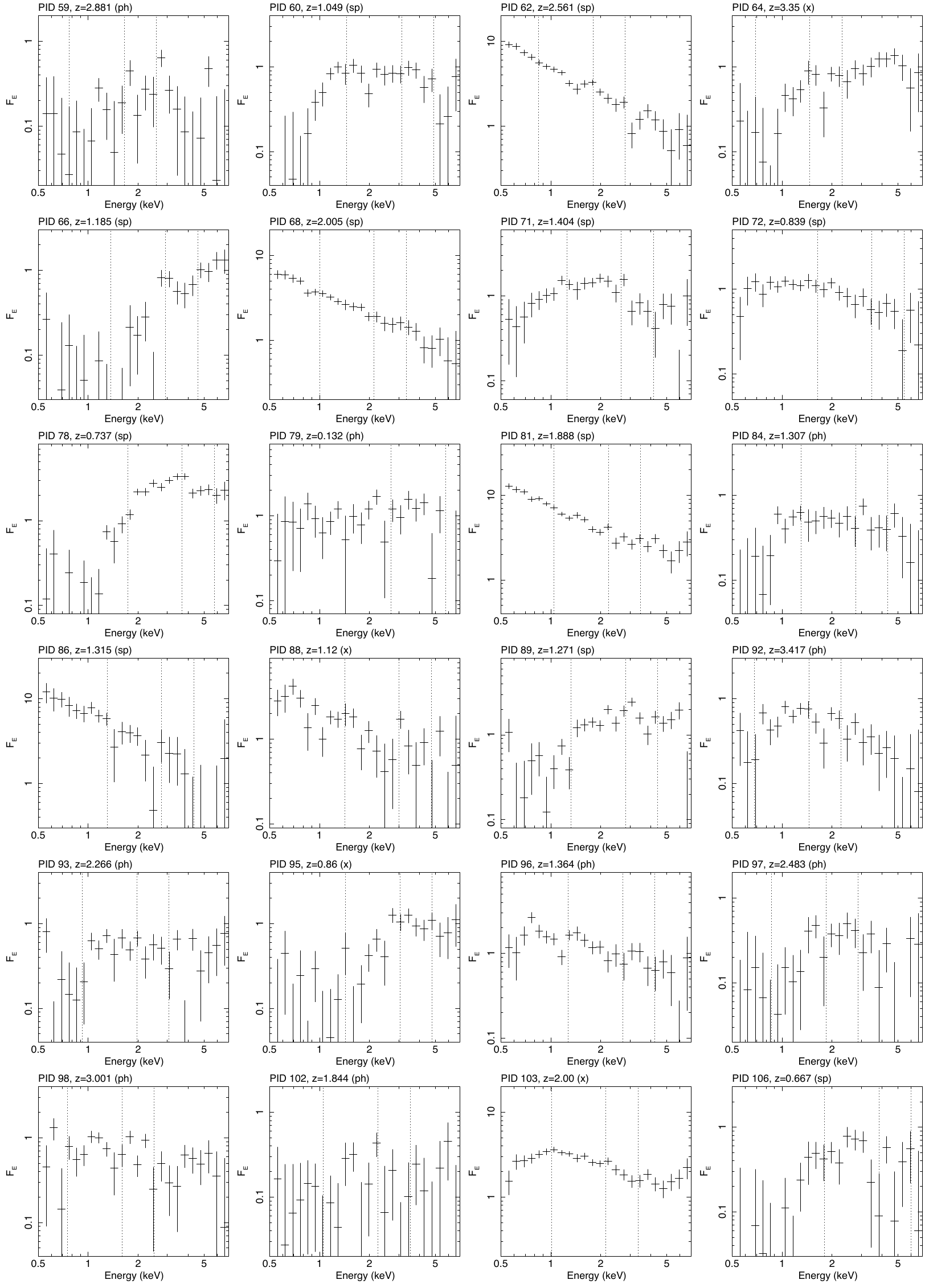}}
    \phantomcaption
\end{figure*}
\begin{figure*}
  \ContinuedFloat
  \centering
  \subfloat[]{\includegraphics[width=0.92\textwidth,angle=0]{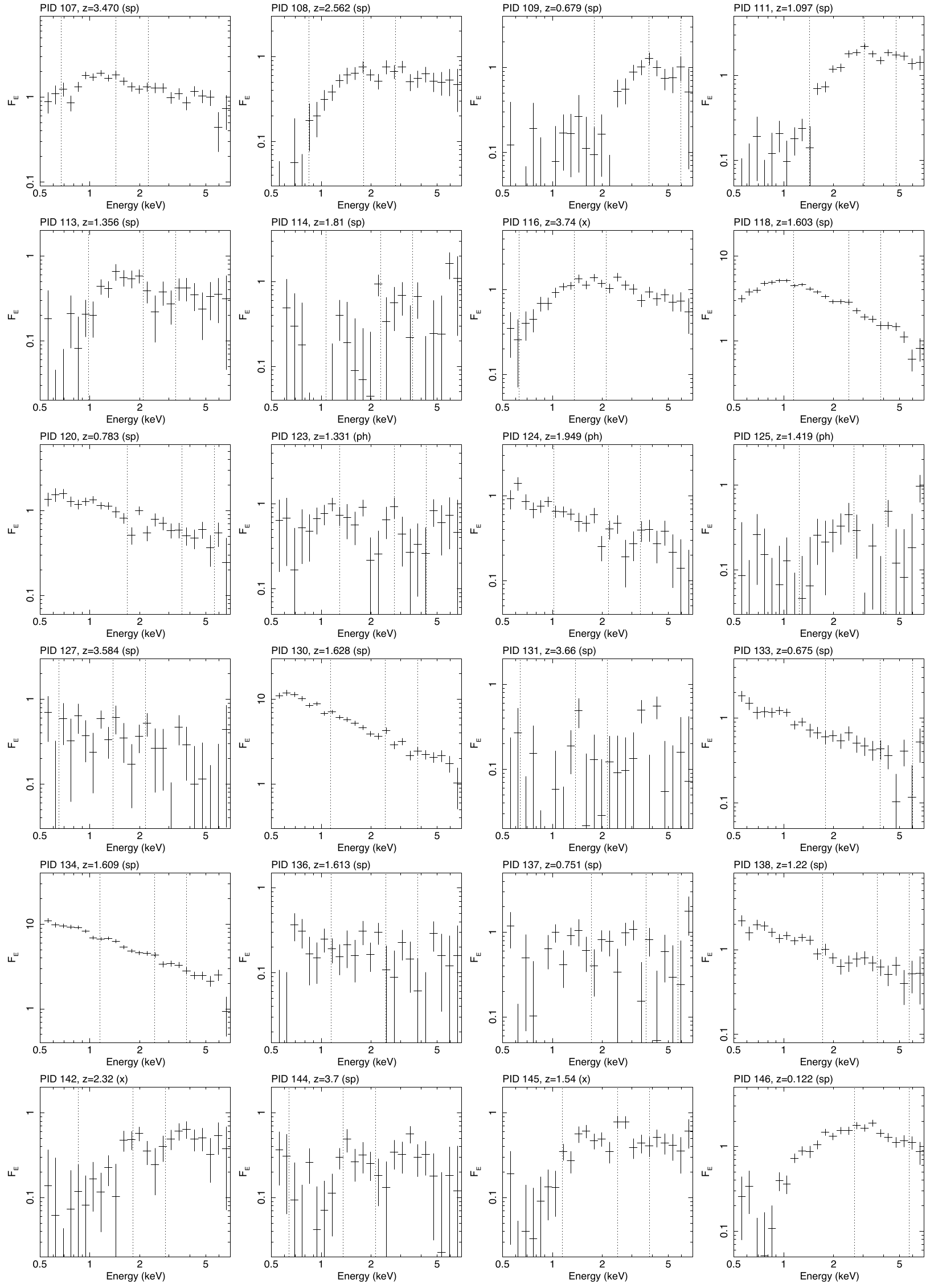}}
  \phantomcaption
\end{figure*}
\begin{figure*}
  \ContinuedFloat
  \centering
  \subfloat[]{\includegraphics[width=0.92\textwidth,angle=0]{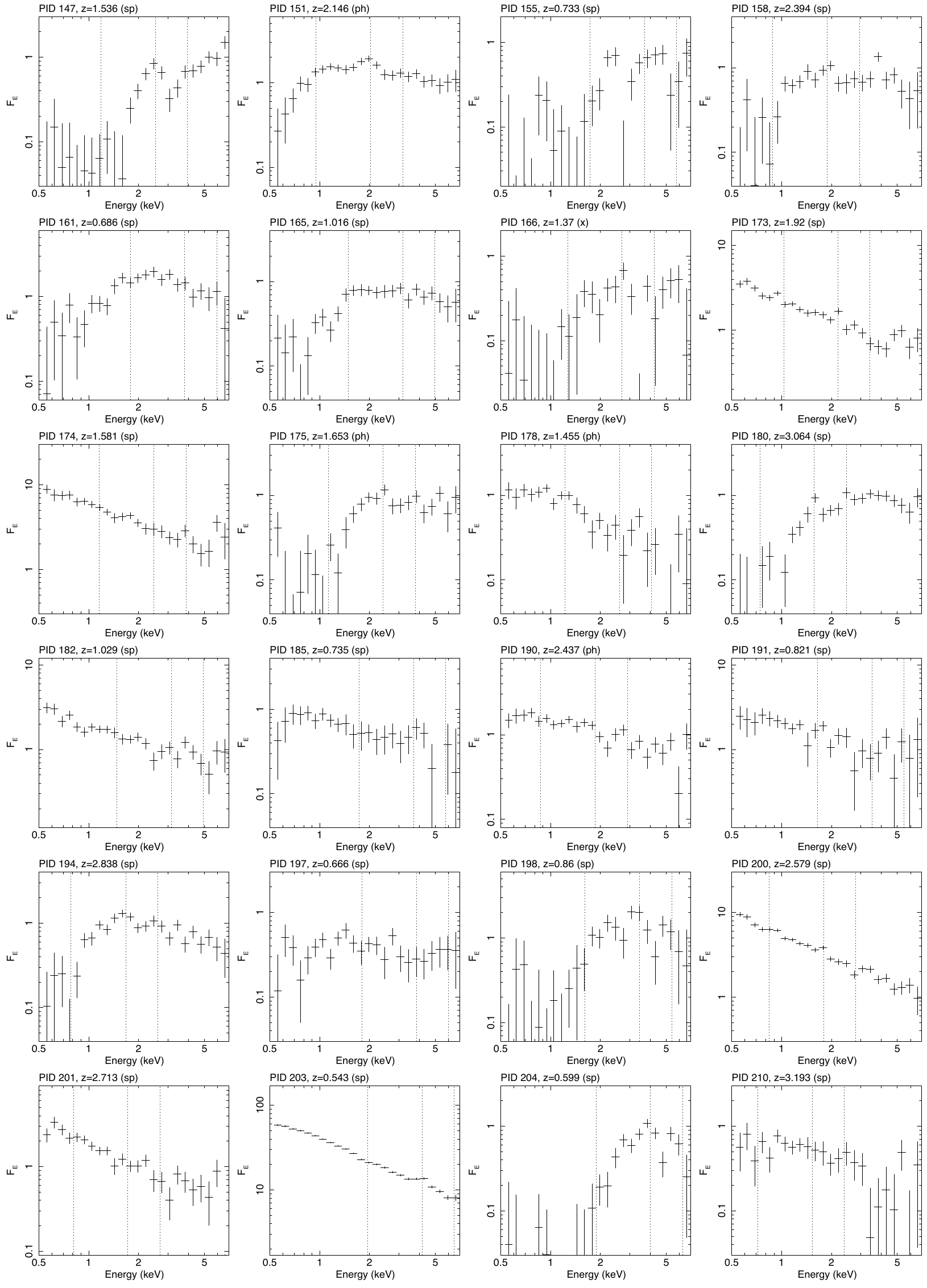}}
  \phantomcaption
\end{figure*}
\begin{figure*}
  \ContinuedFloat
  \centering
  \subfloat[]{\includegraphics[width=0.92\textwidth,angle=0]{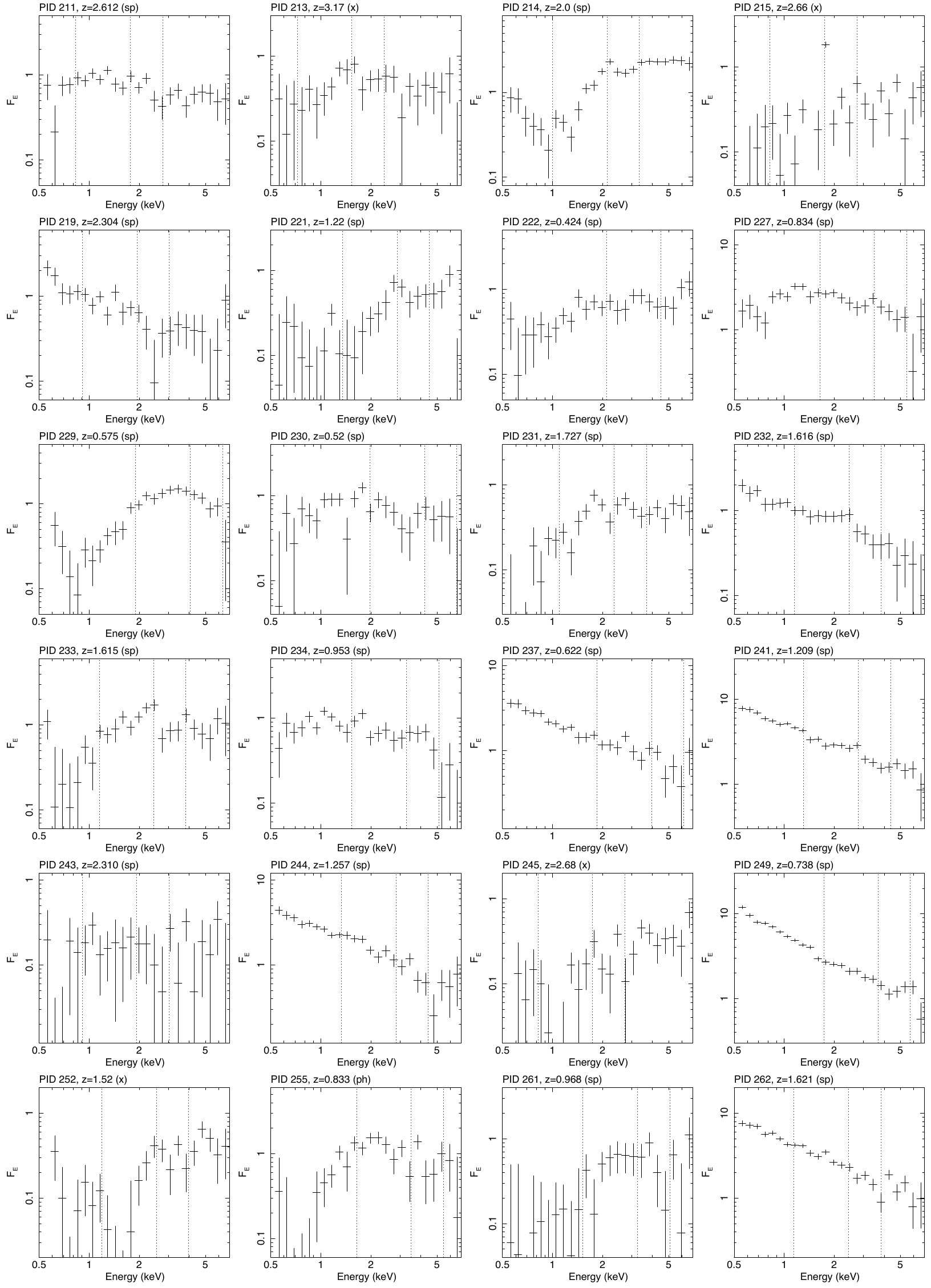}}
  \phantomcaption
\end{figure*}
\begin{figure*}
  \ContinuedFloat
  \centering
  \subfloat[]{\includegraphics[width=0.92\textwidth,angle=0]{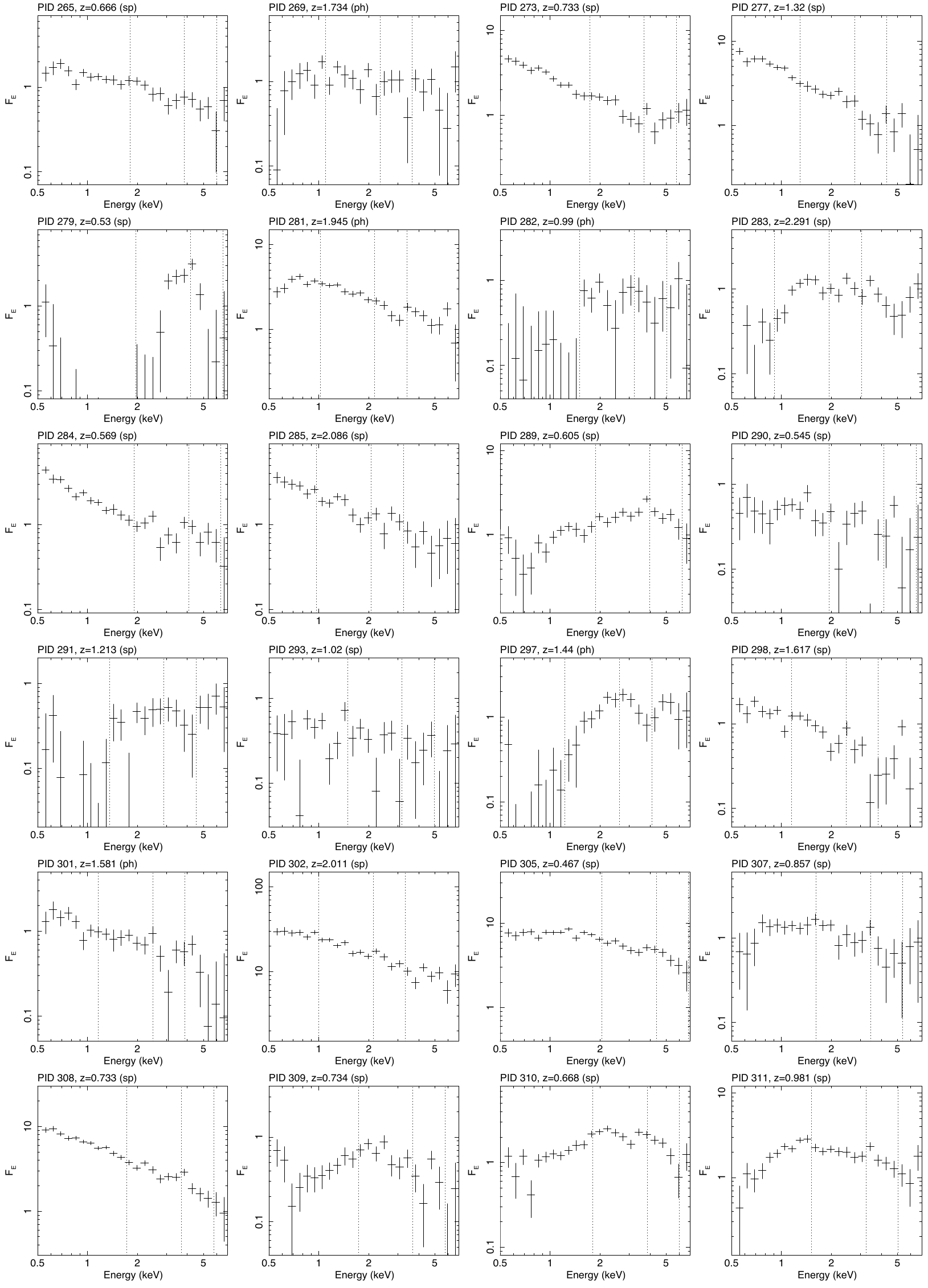}}
  \phantomcaption
\end{figure*}
\begin{figure*}
  \ContinuedFloat
  \centering
  \subfloat[]{\includegraphics[width=0.92\textwidth,angle=0]{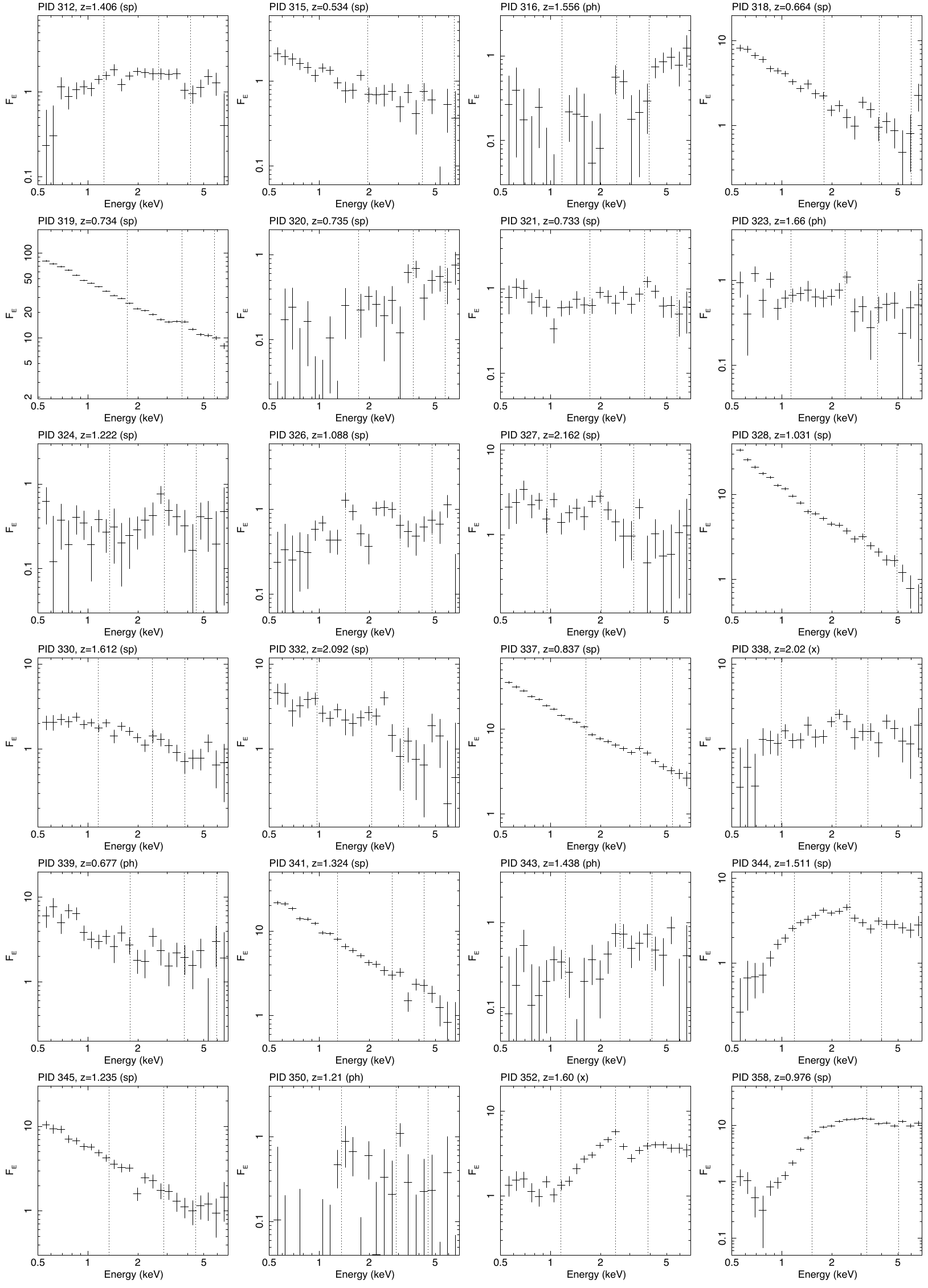}}
  \phantomcaption
\end{figure*}
\begin{figure*}
  \ContinuedFloat
  \centering
  \subfloat[]{\includegraphics[width=0.92\textwidth,angle=0]{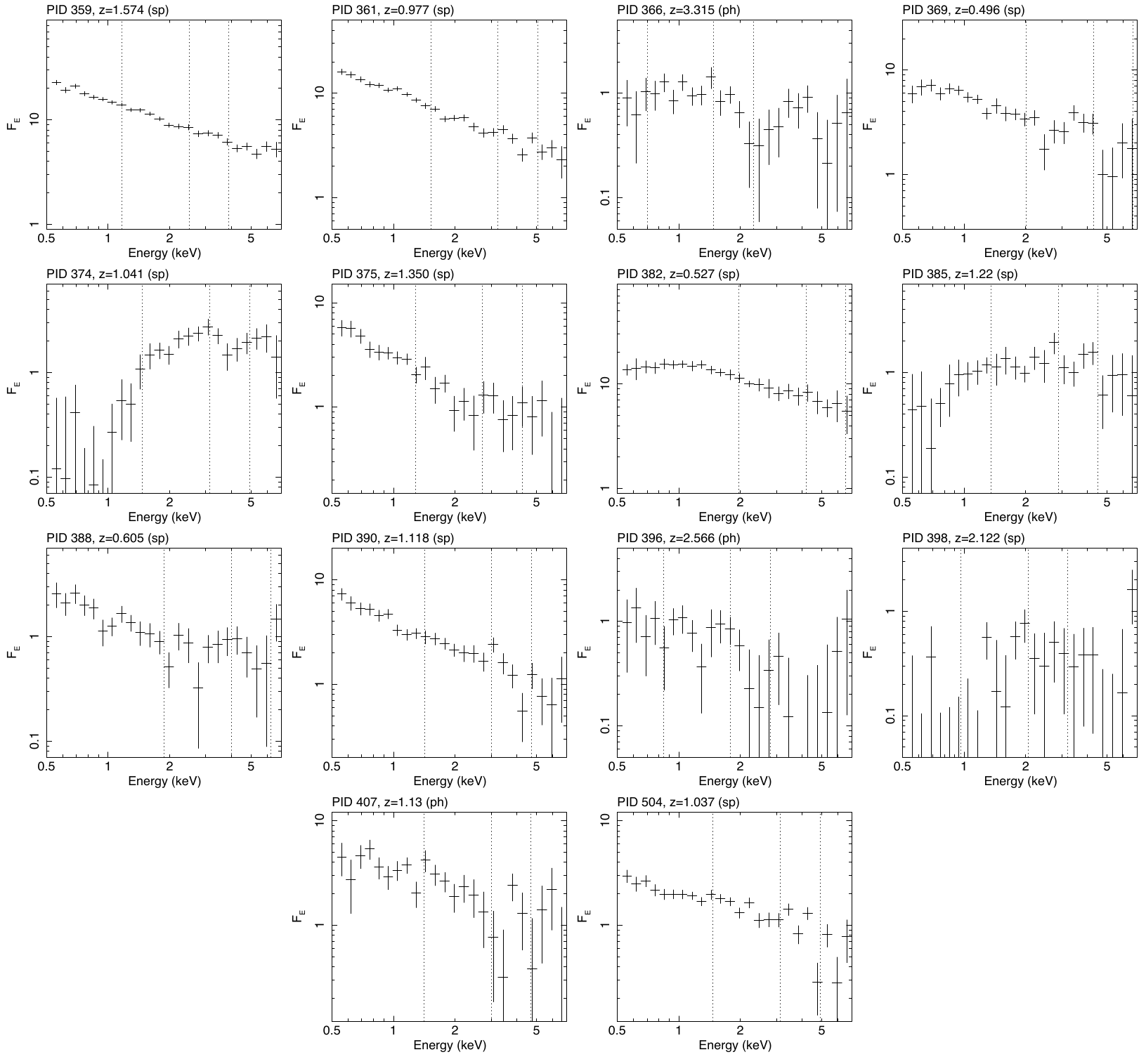}}  \caption{Observed 0.5-7 keV spectra of the remaining 182 XMM-CDFS sources of Fig. 2. Flux density
    is in units of $10^{-15}$ erg~s$^{-1}$~cm$^{-2}$~keV$^{-1}$. The
    energy scale is as observed and the three, vertical dotted lines
    indicate the rest-frame energies of 3 keV, 6.4 keV and 10 keV,
    expected from the adopted redshift.}
 \label{fig:atlas}
\end{figure*}

\end{appendix}



\clearpage
\onecolumn

\setcounter{table}{0}
\begin{longtable}{rlccccccccccc}
  \caption{The 185 XMM-CDFS sources.}\label{tab:big}\\
  \hline\hline
PID & $z$ & class & ref & log$N_{\rm H}$ & log $L$ & log $L^{\prime}$ & $D_{\rm Fe}$ & $E_{\rm Fe}$ & $EW$ & S/M & H/M & $R$(Fe) \\
\hline
\endfirsthead
\caption{continued.}\\
\hline\hline
PID & $z$ & class & ref & log$N_{\rm H}$ & log $L$ & log $L^{\prime}$ & $D_{\rm Fe}$ & $E_{\rm Fe}$ & $EW$ & S/M & H/M & $R$(Fe) \\
\hline
\endhead
\hline
\endfoot
2 & 1.622 & sp & 1 & $<21.9$ & 44.30 & 44.30 & 1 & 6.4 & $0.22^{+0.12}_{-0.12}$ & $4.50\thinspace (0.46)$ & $0.46\thinspace (0.10)$ & $1.50\thinspace (0.33)$ \\
3 & 0.315 & sp & 1 & $22.00^{+0.09}_{-0.06}$ & 42.20 & 42.26 & 0 & 6.4 & $<0.84$ & NA & NA & NA \\
6 & 0.526 & sp & 2 & $22.15^{+0.10}_{-0.18}$ & 42.96 & 43.00 & 0 & 6.4 & $<0.09$ & $2.55\thinspace (0.61)$ & NA & $0.79\thinspace (0.34)$ \\
11 & 2.306 & sp & 3 & $22.67^{+0.14}_{-0.22}$ & 43.88 & 44.04 & 0 & 6.7 & $0.33^{+0.22}_{-0.21}$ & $1.75\thinspace (0.46)$ & $0.72\thinspace (0.20)$ & $0.75\thinspace (0.42)$ \\
13 & 1.109 & ph & 4 & $22.08^{+0.20}_{-0.38}$ & 43.26 & 43.30 & 1 & 6.4 & $0.40^{+0.19}_{-0.19}$ & $1.95\thinspace (0.36)$ & $0.94\thinspace (0.27)$ & $2.26\thinspace (0.48)$ \\
19 & 1.92 & x & 7 & $22.68^{+0.10}_{-0.11}$ & 44.00 & 44.15 & 1 & 6.4 & $0.35^{+0.14}_{-0.13}$ & $1.70\thinspace (0.21)$ & $0.66\thinspace (0.12)$ & $1.47\thinspace (0.26)$ \\
21 & 1.156 & sp & 3 & $22.56^{+0.07}_{-0.11}$ & 43.40 & 43.52 & 0 & 6.4 & $<0.18$ & $2.47\thinspace (0.72)$ & $0.94\thinspace (0.46)$ & $0.64\thinspace (0.46)$ \\
24 & 1.368 & sp & 2 & $23.38^{+0.10}_{-0.15}$ & 43.18 & 43.60 & 0 & 6.4 & $<0.17$ & $0.21\thinspace (0.23)$ & $1.02\thinspace (0.37)$ & $0.10\thinspace (0.50)$ \\
26 & 3.195 & sp & 1 & $23.41^{+0.19}_{-0.21}$ & 43.83 & 44.28 & 0 & 6.4 & $<0.10$ & $0.14\thinspace (0.27)$ & $0.56\thinspace (0.21)$ & $1.86\thinspace (1.35)$ \\
27 & 0.112 & sp & 5 & $<20.78$ & 39.90 & 39.90 & NA & NA & NA & NA & NA & NA \\
30 & 1.83 & x & 6 & $23.70^{+0.06}_{-0.06}$ & 43.72 & 44.48 & 1 & 6.4 & $0.40^{+0.17}_{-0.17}$ & $0.35\thinspace (0.11)$ & $1.02\thinspace (0.14)$ & $1.71\thinspace (0.30)$ \\
31 & 2.583 & sp & 1 & $22.54^{+0.13}_{-0.22}$ & 44.32 & 44.43 & 0 & 6.4 & $<0.07$ & $1.87\thinspace (0.27)$ & $0.51\thinspace (0.11)$ & $0.81\thinspace (0.22)$ \\
32 & 1.90 & x & 7 & $23.26^{+0.12}_{-0.11}$ & 43.54 & 43.91 & 0 & 6.4 & $<0.35$ & $0.64\thinspace (0.22)$ & $0.60\thinspace (0.20)$ & $1.60\thinspace (0.47)$ \\
33 & 1.843 & sp & 3 & $<21.4$ & 44.82 & 44.83 & 1 & 6.4 & $0.07^{+0.04}_{-0.04}$ & $2.63\thinspace (0.09)$ & $0.43\thinspace (0.03)$ & $1.23\thinspace (0.08)$ \\
34 & 1.291 & ph & 4 & $23.58^{+0.10}_{-0.10}$ & 43.36 & 43.93 & 0 & 6.7 & $0.23^{+0.18}_{-0.18}$ & $0.22\thinspace (0.11)$ & $0.55\thinspace (0.15)$ & $1.47\thinspace (0.36)$ \\
37 & 0.623 & sp & 3 & $23.05^{+0.11}_{-0.13}$ & 42.62 & 42.89 & 0 & 6.4 & $<0.08$ & $0.97\thinspace (0.27)$ & NA & $1.48\thinspace (0.46)$ \\
38 & 1.598 & sp & 1 & $21.70^{+0.20}_{-1.00}$ & 44.04 & 44.08 & 0 & 6.4 & $<0.16$ & $2.63\thinspace (0.30)$ & $0.55\thinspace (0.12)$ & $0.83\thinspace (0.26)$ \\
40 & 0.512 & sp & 2 & $<21.8$ & 42.73 & 42.74 & 0 & 6.4 & $<0.18$ & $2.98\thinspace (0.97)$ & NA & $1.17\thinspace (0.59)$ \\
42 & 0.619 & sp & 8 & $<21.8$ & 42.72 & 42.73 & 1 & 6.4 & $0.28^{+0.16}_{-0.16}$ & $2.50\thinspace (0.49)$ & NA & $1.75\thinspace (0.46)$ \\
43 & 1.167 & sp & 2 & $<21.5$ & 43.54 & 43.56 & 1 & 6.5 & $0.30^{+0.18}_{-0.18}$ & $2.42\thinspace (0.32)$ & $0.39\thinspace (0.15)$ & $1.48\thinspace (0.33)$ \\
45 & 0.717 & sp & 2 & $22.46^{+0.18}_{-0.26}$ & 42.93 & 43.04 & 0 & 6.4 & $0.31^{+0.27}_{-0.26}$ & $2.94\thinspace (0.62)$ & NA & $1.55\thinspace (0.55)$ \\
48 & 0.298 & sp & 5 & $22.59^{+0.01}_{-0.02}$ & 43.00 & 43.11 & 1 & 6.4 & $0.17^{+0.06}_{-0.05}$ & NA & NA & NA \\
49 & 2.298 & sp & 5 & $22.56^{+0.12}_{-0.20}$ & 43.99 & 44.11 & 0 & 6.4 & $<0.15$ & $1.95\thinspace (0.24)$ & $0.71\thinspace (0.12)$ & $0.94\thinspace (0.22)$ \\
50 & 1.218 & sp & 9 & $22.34^{+0.11}_{-0.11}$ & 43.51 & 43.59 & 0 & 6.4 & $<0.19$ & $2.08\thinspace (0.30)$ & $0.41\thinspace (0.16)$ & $0.94\thinspace (0.30)$ \\
54 & 1.1 & ph & 4 & $22.61^{+0.09}_{-0.10}$ & 43.30 & 43.43 & 1 & 6.4 & $0.63^{+0.31}_{-0.30}$ & $2.07\thinspace (0.40)$ & $0.96\thinspace (0.31)$ & $1.67\thinspace (0.40)$ \\
57 & 2.571 & sp & 1 & $<21.4$ & 44.36 & 44.36 & 0 & 6.4 & $0.12^{+0.10}_{-0.10}$ & $3.54\thinspace (0.34)$ & $0.39\thinspace (0.08)$ & $1.24\thinspace (0.17)$ \\
58 & 0.777 & sp & 9 & $22.08^{+0.37}_{-2.08}$ & 42.34 & 42.38 & 0 & 6.4 & $<0.35$ & $1.49\thinspace (0.45)$ & NA & $2.02\thinspace (0.77)$ \\
59 & 2.881 & ph & 4 & $23.45^{+0.18}_{-0.15}$ & 43.40 & 43.89 & 1 & 6.7 & $1.32^{+0.67}_{-0.65}$ & $0.88\thinspace (0.44)$ & $1.16\thinspace (0.37)$ & $0.81\thinspace (0.62)$ \\
60 & 1.049 & sp & 8 & $22.48^{+0.06}_{-0.08}$ & 43.20 & 43.30 & 0 & 6.97 & $0.20^{+0.14}_{-0.13}$ & $1.49\thinspace (0.19)$ & $0.39\thinspace (0.15)$ & $0.83\thinspace (0.27)$ \\
62 & 2.561 & sp & 2 & $<21$ & 44.65 & 44.66 & 1 & 6.4 & $0.12^{+0.06}_{-0.06}$ & $3.35\thinspace (0.19)$ & $0.39\thinspace (0.04)$ & $1.21\thinspace (0.11)$ \\
64 & 3.35 & x & 6 & $23.76^{+0.05}_{-0.06}$ & 43.97 & 44.80 & 0 & 6.7 & $0.27^{+0.21}_{-0.21}$ & $0.25\thinspace (0.21)$ & $0.84\thinspace (0.16)$ & $1.64\thinspace (0.56)$ \\
66 & 1.185 & sp & 11 & $23.93^{+0.06}_{-0.07}$ & 43.04 & 44.11 & 0 & 6.5 & $0.22^{+0.17}_{-0.17}$ & $0.12\thinspace (0.14)$ & $1.47\thinspace (0.28)$ & $2.49\thinspace (0.62)$ \\
68 & 2.005 & sp & 2 & $<21.6$ & 44.26 & 44.28 & 0 & 6.4 & $0.09^{+0.07}_{-0.07}$ & $2.93\thinspace (0.21)$ & $0.46\thinspace (0.07)$ & $1.07\thinspace (0.13)$ \\
71 & 1.404 & sp & 2 & $22.48^{+0.06}_{-0.07}$ & 43.69 & 43.79 & 1 & 6.7 & $0.31^{+0.16}_{-0.15}$ & $1.90\thinspace (0.20)$ & $0.36\thinspace (0.10)$ & $1.60\thinspace (0.29)$ \\
72 & 0.839 & sp & 11 & $22.04^{+0.14}_{-0.19}$ & 43.04 & 43.08 & 0 & 6.7 & $0.22^{+0.13}_{-0.14}$ & $2.46\thinspace (0.37)$ & NA & $0.98\thinspace (0.34)$ \\
78 & 0.737 & sp & 11 & $23.04^{+0.04}_{-0.04}$ & 43.43 & 43.70 & 1 & 6.4 & $0.09^{+0.05}_{-0.05}$ & $0.88\thinspace (0.06)$ & NA & $1.15\thinspace (0.12)$ \\
79 & 0.132 & ph & 10 & $21.72^{+0.13}_{-0.14}$ & 41.36 & 41.38 & 0 & 6.4 & $<0.44$ & NA & NA & NA \\
81 & 1.888 & sp & 2 & $<21$ & 44.51 & 44.52 & 0 & 6.4 & $0.09^{+0.06}_{-0.06}$ & $3.09\thinspace (0.15)$ & $0.59\thinspace (0.05)$ & $1.16\thinspace (0.10)$ \\
84 & 1.307 & ph & 4 & $22.49^{+0.09}_{-0.13}$ & 43.23 & 43.34 & 0 & 6.97 & $0.28^{+0.25}_{-0.24}$ & $1.72\thinspace (0.30)$ & $0.49\thinspace (0.18)$ & $0.83\thinspace (0.39)$ \\
86 & 1.315 & sp & 2 & $<21$ & 44.04 & 44.15 & 0 & 6.7 & $1.06^{+0.46}_{-0.45}$ & $3.94\thinspace (1.11)$ & $-0.20\thinspace (0.32)$ & $1.20\thinspace (0.68)$ \\
88 & 1.12 & x & 7 & $<22$ & 43.43 & 43.45 & 1 & 6.4 & $0.72^{+0.28}_{-0.26}$ & $2.72\thinspace (0.68)$ & $0.16\thinspace (0.37)$ & $1.81\thinspace (0.61)$ \\
89 & 1.271 & sp & 2 & $23.08^{+0.07}_{-0.08}$ & 43.70 & 43.98 & 0 & 6.97 & $0.17^{+0.15}_{-0.15}$ & $0.87\thinspace (0.09)$ & $0.72\thinspace (0.09)$ & $1.06\thinspace (0.18)$ \\
92 & 3.417 & ph & 12 & $22.91^{+0.09}_{-0.08}$ & 44.08 & 44.30 & 0 & 6.4 & $<0.35$ & $1.22\thinspace (0.21)$ & $0.50\thinspace (0.10)$ & $1.26\thinspace (0.32)$ \\
93 & 2.266 & ph & 4 & $22.88^{+0.09}_{-0.10}$ & 43.67 & 43.88 & 0 & 6.4 & $0.26^{+0.20}_{-0.20}$ & $1.10\thinspace (0.23)$ & $0.49\thinspace (0.14)$ & $1.27\thinspace (0.28)$ \\
95 & 0.86 & x & 7 & $23.52^{+0.06}_{-0.07}$ & 43.20 & 43.74 & 1 & 6.4 & $0.26^{+0.14}_{-0.13}$ & $0.50\thinspace (0.10)$ & NA & $1.39\thinspace (0.31)$ \\
96 & 1.364 & ph & 4 & $22.15^{+0.13}_{-0.25}$ & 43.60 & 43.65 & 0 & 6.4 & $<0.21$ & $2.27\thinspace (0.30)$ & $0.41\thinspace (0.14)$ & $0.81\thinspace (0.32)$ \\
97 & 2.483 & ph & 4 & $23.38^{+0.11}_{-0.15}$ & 43.36 & 43.80 & 0 & 6.4 & $<0.04$ & $0.41\thinspace (0.22)$ & $0.48\thinspace (0.16)$ & $0.77\thinspace (0.45)$ \\
98 & 3.001 & ph & 12 & $22.62^{+0.10}_{-0.14}$ & 44.08 & 44.20 & 0 & 6.97 & $0.26^{+0.20}_{-0.20}$ & $1.50\thinspace (0.24)$ & $0.36\thinspace (0.10)$ & $0.90\thinspace (0.27)$ \\
102 & 1.844 & ph & 4 & $22.98^{+0.23}_{-0.26}$ & 43.00 & 43.20 & 1 & 6.4 & $0.91^{+0.53}_{-0.46}$ & $1.43\thinspace (0.76)$ & $0.54\thinspace (0.48)$ & $4.58\thinspace (1.52)$ \\
103 & 2.00 & x & 7 & $22.32^{+0.06}_{-0.04}$ & 44.28 & 44.36 & 1 & 6.4 & $0.20^{+0.07}_{-0.07}$ & $2.15\thinspace (0.14)$ & $0.46\thinspace (0.06)$ & $1.26\thinspace (0.13)$ \\
106 & 0.667 & sp & 11 & $22.66^{+0.13}_{-0.20}$ & 42.45 & 42.59 & 0 & 6.7 & $0.83^{+0.50}_{-0.53}$ & $2.71\thinspace (0.91)$ & NA & $-0.30\thinspace (0.40)$ \\
107 & 3.470 & sp & 5 & $22.93^{+0.04}_{-0.03}$ & 44.45 & 44.68 & 1 & 6.4 & $0.14^{+0.08}_{-0.08}$ & $1.14\thinspace (0.08)$ & $0.55\thinspace (0.04)$ & $1.27\thinspace (0.14)$ \\
108 & 2.562 & sp & 11 & $23.36^{+0.04}_{-0.06}$ & 43.81 & 44.23 & 0 & 6.4 & $<0.16$ & $0.52\thinspace (0.10)$ & $0.67\thinspace (0.09)$ & $1.26\thinspace (0.23)$ \\
109 & 0.679 & sp & 14 & $23.51^{+0.09}_{-0.10}$ & 42.88 & 43.40 & 1 & 6.4 & $0.23^{+0.13}_{-0.12}$ & $0.42\thinspace (0.10)$ & NA & $1.75\thinspace (0.30)$ \\
111 & 1.097 & sp & 11 & $23.31^{+0.03}_{-0.02}$ & 43.57 & 43.96 & 0 & 6.7 & $0.09^{+0.06}_{-0.05}$ & $0.53\thinspace (0.04)$ & $0.62\thinspace (0.05)$ & $1.26\thinspace (0.11)$ \\
113 & 1.356 & sp & 23 & $22.68^{+0.06}_{-0.07}$ & 43.23 & 43.38 & 0 & 6.4 & $<0.41$ & $2.22\thinspace (0.45)$ & $0.83\thinspace (0.24)$ & $0.61\thinspace (0.32)$ \\
114 & 1.81 & sp & 1 & $23.20^{+0.25}_{-0.25}$ & 43.20 & 43.54 & 1 & 6.4 & $1.26^{+0.68}_{-0.67}$ & $0.01\thinspace (0.34)$ & $0.52\thinspace (0.30)$ & $3.71\thinspace (1.27)$ \\
116 & 3.74 & x & 6 & $23.41^{+0.04}_{-0.03}$ & 44.30 & 44.76 & 0 & 6.4 & $0.14^{+0.12}_{-0.12}$ & $0.59\thinspace (0.07)$ & $0.69\thinspace (0.05)$ & $1.21\thinspace (0.13)$ \\
118 & 1.603 & sp & 11 & $22.08^{+0.07}_{-0.13}$ & 44.20 & 44.26 & 1 & 6.4 & $0.10^{+0.05}_{-0.05}$ & $2.80\thinspace (0.10)$ & $0.43\thinspace (0.03)$ & $1.24\thinspace (0.10)$ \\
120 & 0.783 & sp & 11 & $<21.8$ & 42.89 & 42.91 & 0 & 6.4 & $<0.11$ & $2.39\thinspace (0.26)$ & NA & $1.18\thinspace (0.25)$ \\
123 & 1.331 & ph & 4 & $<22.2$ & 43.32 & 43.36 & 1 & 6.4 & $1.28^{+0.42}_{-0.44}$ & $2.13\thinspace (0.53)$ & $0.91\thinspace (0.32)$ & $3.73\thinspace (0.90)$ \\
124 & 1.949 & ph & 4 & $<21.8$ & 43.53 & 43.56 & 0 & 6.4 & $<0.15$ & $2.55\thinspace (0.34)$ & $0.61\thinspace (0.13)$ & $0.90\thinspace (0.29)$ \\
125 & 1.419 & ph & 4 & $23.23^{+0.18}_{-0.19}$ & 42.93 & 43.28 & 0 & 6.4 & $<0.29$ & $0.96\thinspace (0.46)$ & $0.70\thinspace (0.41)$ & $1.31\thinspace (0.70)$ \\
127 & 3.584 & sp & 23 & $22.94^{+0.16}_{-0.18}$ & 43.90 & 44.11 & 0 & 6.4 & $<0.30$ & $0.88\thinspace (0.56)$ & $0.46\thinspace (0.24)$ & $0.68\thinspace (0.57)$ \\
130 & 1.628 & sp & 1 & $<21.7$ & 44.40 & 44.40 & 1 & 6.4 & $0.14^{+0.07}_{-0.07}$ & $3.01\thinspace (0.13)$ & $0.47\thinspace (0.04)$ & $1.24\thinspace (0.12)$ \\
131 & 3.66 & sp & 11 & $23.99^{+0.25}_{-0.19}$ & 43.20 & 44.20 & 1 & 6.4 & $0.98^{+0.55}_{-0.53}$ & $0.81\thinspace (0.96)$ & $0.63\thinspace (0.57)$ & $7.70\thinspace (2.33)$ \\
133 & 0.675 & sp & 11 & $<21.8$ & 42.62 & 42.63 & 0 & 6.4 & $<0.23$ & $2.85\thinspace (0.44)$ & NA & $0.99\thinspace (0.32)$ \\
134 & 1.609 & sp & 5 & $<21.4$ & 44.41 & 44.43 & 0 & 6.4 & $<0.07$ & $2.67\thinspace (0.09)$ & $0.47\thinspace (0.03)$ & $1.12\thinspace (0.09)$ \\
136 & 1.613 & sp & 8 & $22.53^{+0.17}_{-0.27}$ & 42.96 & 43.08 & 0 & 6.4 & $<0.26$ & $1.60\thinspace (0.40)$ & $0.46\thinspace (0.22)$ & $0.68\thinspace (0.67)$ \\
137 & 0.751 & sp & 3 & $22.30^{+0.19}_{-0.30}$ & 42.81 & 42.88 & 1 & 6.4 & $0.65^{+0.40}_{-0.40}$ & $2.24\thinspace (0.67)$ & NA & $2.23\thinspace (0.72)$ \\
138 & 1.22 & sp & 8 & $<21.8$ & 43.41 & 43.43 & 1 & 6.97 & $0.27^{+0.15}_{-0.15}$ & $2.39\thinspace (0.23)$ & $0.49\thinspace (0.10)$ & $1.12\thinspace (0.24)$ \\
142 & 2.32 & x & 7 & $23.60^{+0.10}_{-0.09}$ & 43.48 & 44.08 & 0 & 6.4 & $0.25^{+0.19}_{-0.18}$ & $0.48\thinspace (0.18)$ & $0.95\thinspace (0.17)$ & $1.80\thinspace (0.37)$ \\
144 & 3.7 & sp & 11 & $23.82^{+0.11}_{-0.10}$ & 43.58 & 44.45 & 1 & 6.4 & $0.70^{+0.39}_{-0.38}$ & $0.97\thinspace (0.35)$ & $0.56\thinspace (0.18)$ & $2.40\thinspace (0.58)$ \\
145 & 1.54 & x & 4 & $23.00^{+0.06}_{-0.06}$ & 43.30 & 43.54 & 1 & 6.4 & $0.41^{+0.19}_{-0.16}$ & $0.95\thinspace (0.12)$ & $0.56\thinspace (0.09)$ & $1.60\thinspace (0.34)$ \\
146 & 0.122 & sp & 11 & $22.20^{+0.03}_{-0.02}$ & 41.63 & 41.69 & 0 & 6.4 & $<0.23$ & NA & NA & NA \\
147 & 1.536 & sp & 11 & $23.69^{+0.04}_{-0.05}$ & 43.32 & 44.04 & 1 & 6.4 & $0.22^{+0.12}_{-0.12}$ & $0.26\thinspace (0.08)$ & $1.02\thinspace (0.12)$ & $2.36\thinspace (0.35)$ \\
151 & 2.146 & ph & 4 & $22.81^{+0.02}_{-0.03}$ & 44.11 & 44.28 & 0 & 6.4 & $<0.09$ & $1.25\thinspace (0.08)$ & $0.53\thinspace (0.04)$ & $1.19\thinspace (0.12)$ \\
155 & 0.733 & sp & 5 & $23.66^{+0.15}_{-0.21}$ & 42.72 & 43.00 & 0 & 6.4 & $<0.25$ & $0.62\thinspace (0.14)$ & NA & $2.06\thinspace (0.47)$ \\
158 & 2.394 & sp & 11 & $23.23^{+0.05}_{-0.05}$ & 43.89 & 44.26 & 1 & 6.4 & $0.44^{+0.17}_{-0.17}$ & $0.85\thinspace (0.13)$ & $0.81\thinspace (0.10)$ & $1.35\thinspace (0.21)$ \\
161 & 0.686 & sp & 5 & $22.60^{+0.05}_{-0.04}$ & 43.15 & 43.26 & 0 & 6.4 & $<0.14$ & $1.82\thinspace (0.18)$ & NA & $1.20\thinspace (0.21)$ \\
165 & 1.016 & sp & 11 & $22.79^{+0.04}_{-0.03}$ & 43.18 & 43.36 & 0 & 6.4 & $<0.12$ & $1.30\thinspace (0.12)$ & $0.56\thinspace (0.10)$ & $1.17\thinspace (0.19)$ \\
166 & 1.37 & x & 7 & $23.18^{+0.12}_{-0.10}$ & 43.04 & 43.36 & 1 & 6.4 & $0.52^{+0.28}_{-0.27}$ & $0.77\thinspace (0.22)$ & $0.78\thinspace (0.23)$ & $1.75\thinspace (0.47)$ \\
173 & 1.92 & sp & 11 & $<21.5$ & 44.04 & 44.08 & 1 & 6.4 & $0.20^{+0.10}_{-0.10}$ & $2.54\thinspace (0.14)$ & $0.40\thinspace (0.04)$ & $1.25\thinspace (0.12)$ \\
174 & 1.581 & sp & 2 & $<21.6$ & 44.26 & 44.28 & 0 & 6.97 & $0.14^{+0.10}_{-0.10}$ & $2.80\thinspace (0.22)$ & $0.52\thinspace (0.07)$ & $1.09\thinspace (0.19)$ \\
175 & 1.653 & ph & 4 & $23.34^{+0.06}_{-0.04}$ & 43.59 & 44.00 & 0 & 6.4 & $0.12^{+0.12}_{-0.11}$ & $0.57\thinspace (0.08)$ & $0.68\thinspace (0.08)$ & $1.36\thinspace (0.24)$ \\
178 & 1.455 & ph & 4 & $22.08^{+0.28}_{-0.78}$ & 43.32 & 43.38 & 0 & 6.4 & $<0.14$ & $3.34\thinspace (0.56)$ & $0.07\thinspace (0.16)$ & $0.86\thinspace (0.40)$ \\
180 & 3.064 & sp & 11 & $23.75^{+0.03}_{-0.03}$ & 43.91 & 44.73 & 1 & 6.4 & $0.39^{+0.15}_{-0.15}$ & $0.26\thinspace (0.09)$ & $1.06\thinspace (0.10)$ & $1.70\thinspace (0.26)$ \\
182 & 1.029 & sp & 5 & $21.90^{+0.14}_{-0.22}$ & 43.38 & 43.41 & 0 & 6.4 & $<0.09$ & $2.21\thinspace (0.20)$ & $0.54\thinspace (0.11)$ & $1.03\thinspace (0.20)$ \\
185 & 0.735 & sp & 15 & $<22$ & 42.67 & 42.69 & 1 & 6.4 & $0.56^{+0.26}_{-0.26}$ & $2.35\thinspace (0.55)$ & NA & $2.46\thinspace (0.79)$ \\
190 & 2.437 & ph & 4 & $22.18^{+0.08}_{-0.14}$ & 44.15 & 44.20 & 0 & 6.4 & $<0.25$ & $2.29\thinspace (0.18)$ & $0.55\thinspace (0.06)$ & $1.14\thinspace (0.15)$ \\
191 & 0.821 & sp & 3 & $21.78^{+0.26}_{-0.78}$ & 43.18 & 43.18 & 0 & 6.4 & $<0.14$ & $2.43\thinspace (0.47)$ & NA & $0.77\thinspace (0.42)$ \\
194 & 2.838 & sp & 3 & $23.26^{+0.02}_{-0.03}$ & 44.08 & 44.36 & 0 & 6.4 & $<0.04$ & $0.66\thinspace (0.08)$ & $0.57\thinspace (0.06)$ & $1.37\thinspace (0.15)$ \\
197 & 0.666 & sp & 2 & $<22$ & 42.41 & 42.43 & 0 & 6.4 & $<0.39$ & $2.36\thinspace (0.44)$ & NA & $0.71\thinspace (0.34)$ \\
198 & 0.86 & sp & 2 & $23.04^{+0.06}_{-0.07}$ & 43.32 & 43.58 & 1 & 6.4 & $0.32^{+0.15}_{-0.14}$ & $0.86\thinspace (0.15)$ & NA & $1.82\thinspace (0.39)$ \\
200 & 2.579 & sp & 11 & $<21$ & 44.70 & 44.72 & 1 & 6.4 & $0.14^{+0.05}_{-0.04}$ & $2.94\thinspace (0.10)$ & $0.45\thinspace (0.03)$ & $1.10\thinspace (0.07)$ \\
201 & 2.713 & sp & 3 & $<21.4$ & 44.26 & 44.26 & 0 & 6.4 & $<0.22$ & $3.19\thinspace (0.29)$ & $0.43\thinspace (0.07)$ & $1.09\thinspace (0.15)$ \\
203 & 0.543 & sp & 11 & $20.74^{+0.05}_{-0.06}$ & 43.98 & 43.98 & 1 & 6.4 & $0.09^{+0.02}_{-0.02}$ & $2.89\thinspace (0.05)$ & NA & $1.20\thinspace (0.04)$ \\
204 & 0.599 & sp & 11 & $23.32^{+0.06}_{-0.04}$ & 42.65 & 43.04 & 1 & 6.4 & $0.19^{+0.10}_{-0.09}$ & $0.51\thinspace (0.08)$ & NA & $1.57\thinspace (0.21)$ \\
210 & 3.193 & sp & 11 & $22.38^{+0.16}_{-0.23}$ & 43.93 & 44.00 & 0 & 6.4 & $<0.22$ & $1.98\thinspace (0.32)$ & $0.59\thinspace (0.13)$ & $1.08\thinspace (0.44)$ \\
211 & 2.612 & sp & 5 & $22.64^{+0.06}_{-0.07}$ & 43.99 & 44.11 & 1 & 6.4 & $0.20^{+0.12}_{-0.11}$ & $1.68\thinspace (0.17)$ & $0.51\thinspace (0.07)$ & $1.44\thinspace (0.23)$ \\
213 & 3.17 & x & 7 & $23.15^{+0.11}_{-0.15}$ & 43.89 & 44.18 & 0 & 6.4 & $0.35^{+0.24}_{-0.24}$ & $0.68\thinspace (0.21)$ & $0.55\thinspace (0.13)$ & $2.15\thinspace (0.58)$ \\
214 & 2.0 & sp & 17 & $23.68^{+0.03}_{-0.02}$ & 44.00 & 44.68 & 0 & 6.4 & $0.09^{+0.05}_{-0.05}$ & $0.43\thinspace (0.04)$ & $0.89\thinspace (0.06)$ & $1.63\thinspace (0.13)$ \\
215 & 2.66 & x & 7 & $23.69^{+0.11}_{-0.11}$ & 43.72 & 44.30 & 1 & 6.4 & $3.07^{+0.35}_{-0.35}$ & $0.45\thinspace (0.18)$ & $0.60\thinspace (0.13)$ & $7.41\thinspace (0.58)$ \\
219 & 2.304 & sp & 2 & $<22.4$ & 43.83 & 43.83 & 0 & 6.4 & $<0.28$ & $3.26\thinspace (0.56)$ & $0.67\thinspace (0.18)$ & $1.79\thinspace (0.36)$ \\
221 & 1.22 & sp & 8 & $23.59^{+0.11}_{-0.10}$ & 43.08 & 43.64 & 0 & 6.4 & $0.27^{+0.19}_{-0.18}$ & $0.49\thinspace (0.12)$ & $0.77\thinspace (0.17)$ & $2.74\thinspace (0.55)$ \\
222 & 0.424 & sp & 11 & $22.48^{+0.10}_{-0.12}$ & 42.38 & 42.48 & 0 & 6.4 & $0.18^{+0.15}_{-0.14}$ & $1.51\thinspace (0.23)$ & NA & $1.05\thinspace (0.29)$ \\
227 & 0.834 & sp & 3 & $21.85^{+0.15}_{-0.25}$ & 43.51 & 43.53 & 1 & 6.4 & $0.22^{+0.11}_{-0.11}$ & $2.28\thinspace (0.21)$ & NA & $1.31\thinspace (0.21)$ \\
229 & 0.575 & sp & 11 & $22.81^{+0.04}_{-0.04}$ & 42.89 & 43.08 & 0 & 6.4 & $0.11^{+0.07}_{-0.07}$ & $1.23\thinspace (0.10)$ & NA & $1.32\thinspace (0.17)$ \\
230 & 0.52 & sp & 3 & $21.95^{+0.23}_{-0.47}$ & 42.60 & 42.63 & 0 & 6.7 & $0.40^{+0.28}_{-0.26}$ & $1.95\thinspace (0.41)$ & NA & $1.91\thinspace (0.57)$ \\
231 & 1.727 & sp & 2 & $23.15^{+0.05}_{-0.07}$ & 43.48 & 43.79 & 0 & 6.4 & $0.26^{+0.19}_{-0.18}$ & $0.92\thinspace (0.13)$ & $0.61\thinspace (0.10)$ & $1.35\thinspace (0.31)$ \\
232 & 1.616 & sp & 11 & $21.80^{+0.20}_{-0.39}$ & 43.61 & 43.64 & 0 & 6.4 & $0.20^{+0.14}_{-0.14}$ & $2.36\thinspace (0.24)$ & $0.38\thinspace (0.08)$ & $1.48\thinspace (0.28)$ \\
233 & 1.615 & sp & 2 & $22.96^{+0.05}_{-0.06}$ & 43.72 & 43.95 & 1 & 6.4 & $0.24^{+0.13}_{-0.13}$ & $0.97\thinspace (0.12)$ & $0.61\thinspace (0.10)$ & $1.81\thinspace (0.30)$ \\
234 & 0.953 & sp & 24 & $21.74^{+0.25}_{-0.63}$ & 43.11 & 43.11 & 0 & 6.4 & $<0.25$ & $2.19\thinspace (0.26)$ & NA & $1.19\thinspace (0.28)$ \\
237 & 0.622 & sp & 11 & $<21.8$ & 42.93 & 42.94 & 1 & 6.4 & $0.32^{+0.13}_{-0.13}$ & $2.62\thinspace (0.31)$ & NA & $1.53\thinspace (0.28)$ \\
241 & 1.209 & sp & 14 & $<21.4$ & 43.94 & 43.95 & 1 & 6.4 & $0.15^{+0.06}_{-0.06}$ & $2.81\thinspace (0.14)$ & $0.54\thinspace (0.06)$ & $1.32\thinspace (0.13)$ \\
243 & 2.310 & sp & 19 & $22.69^{+0.22}_{-0.29}$ & 43.28 & 43.43 & 0 & 6.4 & $0.60^{+0.49}_{-0.45}$ & $3.05\thinspace (1.36)$ & $0.95\thinspace (0.53)$ & $1.50\thinspace (0.66)$ \\
244 & 1.257 & sp & 15 & $22.08^{+0.15}_{-0.30}$ & 43.66 & 43.72 & 0 & 6.4 & $<0.05$ & $3.02\thinspace (0.25)$ & $0.34\thinspace (0.09)$ & $0.92\thinspace (0.17)$ \\
245 & 2.68 & x & 6 & $23.88^{+0.07}_{-0.09}$ & 43.28 & 44.23 & 0 & 6.4 & $<0.60$ & $0.47\thinspace (0.28)$ & $1.03\thinspace (0.27)$ & $1.96\thinspace (0.80)$ \\
249 & 0.738 & sp & 20 & $<21$ & 43.32 & 43.32 & 1 & 6.4 & $0.13^{+0.07}_{-0.07}$ & $3.14\thinspace (0.18)$ & NA & $1.15\thinspace (0.13)$ \\
252 & 1.52 & x & 7 & $23.83^{+0.08}_{-0.08}$ & 43.00 & 43.91 & 1 & 6.4 & $0.43^{+0.25}_{-0.23}$ & $0.27\thinspace (0.15)$ & $0.95\thinspace (0.20)$ & $1.89\thinspace (0.56)$ \\
255 & 0.833 & ph & 4 & $22.61^{+0.05}_{-0.09}$ & 43.15 & 43.28 & 0 & 6.4 & $<0.08$ & $2.06\thinspace (0.33)$ & NA & $0.59\thinspace (0.30)$ \\
261 & 0.968 & sp & 2 & $23.18^{+0.12}_{-0.14}$ & 42.99 & 43.32 & 0 & 6.4 & $<0.12$ & $0.80\thinspace (0.24)$ & NA & $1.38\thinspace (0.49)$ \\
262 & 1.621 & sp & 3 & $<21.5$ & 44.18 & 44.18 & 0 & 6.4 & $0.14^{+0.08}_{-0.08}$ & $3.10\thinspace (0.16)$ & $0.48\thinspace (0.05)$ & $1.25\thinspace (0.14)$ \\
265 & 0.666 & sp & 5 & $21.90^{+0.14}_{-0.20}$ & 42.85 & 42.88 & 1 & 6.97 & $0.23^{+0.15}_{-0.14}$ & $2.48\thinspace (0.26)$ & NA & $1.20\thinspace (0.23)$ \\
269 & 1.734 & ph & 4 & $22.34^{+0.14}_{-0.14}$ & 43.79 & 43.86 & 0 & 6.4 & $<0.16$ & $1.93\thinspace (0.30)$ & $0.65\thinspace (0.15)$ & $1.30\thinspace (0.45)$ \\
273 & 0.733 & sp & 11 & $<21.2$ & 43.11 & 43.11 & 1 & 6.7 & $0.29^{+0.15}_{-0.15}$ & $2.94\thinspace (0.29)$ & NA & $1.27\thinspace (0.23)$ \\
277 & 1.32 & sp & 9 & $<21.4$ & 43.90 & 43.91 & 0 & 6.4 & $0.12^{+0.10}_{-0.11}$ & $3.24\thinspace (0.29)$ & $0.45\thinspace (0.11)$ & $1.25\thinspace (0.22)$ \\
279 & 0.53 & sp & 2 & $23.48^{+0.10}_{-0.08}$ & 43.08 & 43.57 & 0 & 6.4 & $<0.21$ & $-0.06\thinspace (0.12)$ & NA & $3.50\thinspace (0.57)$ \\
281 & 1.945 & ph & 4 & $21.95^{+0.07}_{-0.12}$ & 44.28 & 44.30 & 0 & 6.4 & $<0.03$ & $2.76\thinspace (0.14)$ & $0.61\thinspace (0.05)$ & $0.96\thinspace (0.10)$ \\
282 & 0.99 & ph & 4 & $22.98^{+0.11}_{-0.10}$ & 43.08 & 43.34 & 0 & 6.4 & $<0.69$ & $0.77\thinspace (0.25)$ & NA & $2.11\thinspace (0.72)$ \\
283 & 2.291 & sp & 11 & $23.08^{+0.03}_{-0.04}$ & 43.98 & 44.28 & 0 & 6.4 & $0.12^{+0.10}_{-0.10}$ & $0.95\thinspace (0.10)$ & $0.66\thinspace (0.08)$ & $1.09\thinspace (0.15)$ \\
284 & 0.569 & sp & 11 & $21.76^{+0.24}_{-0.40}$ & 42.73 & 42.76 & 0 & 6.4 & $<0.23$ & $2.19\thinspace (0.26)$ & NA & $1.28\thinspace (0.28)$ \\
285 & 2.086 & sp & 3 & $<21.8$ & 44.08 & 44.08 & 1 & 6.97 & $0.49^{+0.22}_{-0.22}$ & $2.91\thinspace (0.28)$ & $0.50\thinspace (0.09)$ & $1.03\thinspace (0.18)$ \\
289 & 0.605 & sp & 11 & $22.84^{+0.04}_{-0.05}$ & 43.11 & 43.26 & 0 & 6.4 & $0.08^{+0.06}_{-0.06}$ & $1.17\thinspace (0.08)$ & NA & $1.46\thinspace (0.16)$ \\
290 & 0.545 & sp & 11 & $<22.2$ & 42.26 & 42.28 & 1 & 6.97 & $1.17^{+0.59}_{-0.59}$ & $3.08\thinspace (1.08)$ & NA & $1.43\thinspace (0.54)$ \\
291 & 1.213 & sp & 16 & $23.28^{+0.12}_{-0.13}$ & 43.04 & 43.41 & 0 & 6.97 & $0.40^{+0.31}_{-0.30}$ & $0.74\thinspace (0.22)$ & $0.87\thinspace (0.26)$ & $1.47\thinspace (0.46)$ \\
293 & 1.02 & sp & 11 & $22.18^{+0.16}_{-0.48}$ & 42.85 & 42.90 & 0 & 6.4 & $<0.45$ & $2.03\thinspace (0.52)$ & $0.50\thinspace (0.35)$ & $0.60\thinspace (0.65)$ \\
297 & 1.44 & ph & 10 & $23.30^{+0.06}_{-0.07}$ & 43.69 & 44.08 & 0 & 6.7 & $0.17^{+0.13}_{-0.13}$ & $0.51\thinspace (0.09)$ & $0.57\thinspace (0.10)$ & $1.34\thinspace (0.27)$ \\
298 & 1.617 & sp & 11 & $22.04^{+0.26}_{-0.56}$ & 43.59 & 43.64 & 0 & 6.4 & $0.31^{+0.21}_{-0.21}$ & $3.19\thinspace (0.40)$ & $0.50\thinspace (0.12)$ & $1.76\thinspace (0.39)$ \\
301 & 1.581 & ph & 4 & $22.08^{+0.15}_{-0.23}$ & 43.53 & 43.58 & 1 & 6.4 & $0.36^{+0.20}_{-0.19}$ & $2.38\thinspace (0.33)$ & $0.55\thinspace (0.14)$ & $1.67\thinspace (0.37)$ \\
302 & 2.011 & sp & 1 & $21.84^{+0.11}_{-0.15}$ & 45.15 & 45.18 & 1 & 6.4 & $0.16^{+0.05}_{-0.05}$ & $2.44\thinspace (0.10)$ & $0.45\thinspace (0.03)$ & $1.23\thinspace (0.09)$ \\
305 & 0.467 & sp & 5 & $21.95^{+0.09}_{-0.09}$ & 43.32 & 43.34 & 0 & 6.4 & $<0.05$ & $2.22\thinspace (0.13)$ & NA & $1.22\thinspace (0.13)$ \\
307 & 0.857 & sp & 15 & $22.00^{+0.17}_{-0.29}$ & 43.18 & 43.20 & 0 & 6.4 & $<0.12$ & $2.35\thinspace (0.38)$ & NA & $1.69\thinspace (0.38)$ \\
308 & 0.733 & sp & 11 & $<21$ & 43.49 & 43.49 & 1 & 6.4 & $0.19^{+0.06}_{-0.07}$ & $2.84\thinspace (0.17)$ & NA & $1.26\thinspace (0.14)$ \\
309 & 0.734 & sp & 11 & $22.71^{+0.11}_{-0.13}$ & 42.77 & 42.85 & 0 & 6.4 & $<0.40$ & $2.41\thinspace (0.36)$ & NA & $2.04\thinspace (0.49)$ \\
310 & 0.668 & sp & 11 & $22.57^{+0.03}_{-0.04}$ & 43.26 & 43.38 & 1 & 6.7 & $0.12^{+0.06}_{-0.07}$ & $1.71\thinspace (0.11)$ & NA & $1.14\thinspace (0.14)$ \\
311 & 0.981 & sp & 2 & $22.18^{+0.05}_{-0.10}$ & 43.58 & 43.63 & 1 & 6.4 & $0.24^{+0.08}_{-0.08}$ & $1.99\thinspace (0.14)$ & NA & $1.55\thinspace (0.18)$ \\
312 & 1.406 & sp & 2 & $22.63^{+0.05}_{-0.04}$ & 43.79 & 43.92 & 0 & 6.97 & $0.15^{+0.10}_{-0.10}$ & $1.42\thinspace (0.12)$ & $0.59\thinspace (0.07)$ & $1.10\thinspace (0.18)$ \\
315 & 0.534 & sp & 11 & $<22$ & 42.54 & 42.56 & 0 & 6.4 & $<0.18$ & $2.62\thinspace (0.53)$ & NA & $1.05\thinspace (0.36)$ \\
316 & 1.556 & ph & 4 & $24.23^{+0.08}_{-0.09}$ & 42.82 & 44.58 & 0 & 6.4 & $0.54^{+0.41}_{-0.46}$ & $0.49\thinspace (0.33)$ & $1.94\thinspace (0.63)$ & $7.14\thinspace (3.08)$ \\
318 & 0.664 & sp & 5 & $<21.2$ & 43.08 & 43.11 & 0 & 6.4 & $<0.27$ & $2.90\thinspace (0.41)$ & NA & $1.11\thinspace (0.31)$ \\
319 & 0.734 & sp & 11 & $<21$ & 44.18 & 44.18 & 1 & 6.4 & $0.07^{+0.02}_{-0.02}$ & $2.91\thinspace (0.04)$ & NA & $1.22\thinspace (0.04)$ \\
320 & 0.735 & sp & 11 & $23.28^{+0.15}_{-0.13}$ & 42.61 & 42.97 & 1 & 6.5 & $0.59^{+0.25}_{-0.22}$ & $0.56\thinspace (0.15)$ & NA & $2.08\thinspace (0.48)$ \\
321 & 0.733 & sp & 11 & $22.59^{+0.05}_{-0.07}$ & 42.93 & 43.04 & 0 & 6.4 & $0.15^{+0.12}_{-0.10}$ & $1.36\thinspace (0.15)$ & NA & $1.37\thinspace (0.28)$ \\
323 & 1.66 & ph & 21 & $22.41^{+0.11}_{-0.13}$ & 43.52 & 43.60 & 0 & 6.4 & $0.23^{+0.20}_{-0.19}$ & $1.72\thinspace (0.24)$ & $0.55\thinspace (0.13)$ & $1.84\thinspace (0.42)$ \\
324 & 1.222 & sp & 11 & $22.64^{+0.15}_{-0.18}$ & 43.11 & 43.23 & 1 & 6.4 & $0.58^{+0.26}_{-0.26}$ & $0.99\thinspace (0.23)$ & $0.54\thinspace (0.21)$ & $2.63\thinspace (0.58)$ \\
326 & 1.088 & sp & 3 & $22.63^{+0.07}_{-0.09}$ & 43.20 & 43.34 & 0 & 6.4 & $<0.08$ & $1.54\thinspace (0.23)$ & $0.62\thinspace (0.18)$ & $0.82\thinspace (0.31)$ \\
327 & 2.162 & sp & 2 & $22.26^{+0.15}_{-0.31}$ & 44.18 & 44.28 & 0 & 6.4 & $0.25^{+0.14}_{-0.15}$ & $1.70\thinspace (0.24)$ & $0.44\thinspace (0.11)$ & $1.49\thinspace (0.29)$ \\
328 & 1.031 & sp & 11 & $<21$ & 43.94 & 43.94 & 1 & 6.4 & $0.15^{+0.06}_{-0.05}$ & $3.98\thinspace (0.18)$ & $0.28\thinspace (0.05)$ & $1.25\thinspace (0.11)$ \\
330 & 1.612 & sp & 2 & $21.94^{+0.13}_{-0.18}$ & 43.88 & 43.91 & 0 & 6.4 & $0.16^{+0.11}_{-0.11}$ & $2.52\thinspace (0.22)$ & $0.56\thinspace (0.08)$ & $1.13\thinspace (0.20)$ \\
332 & 2.092 & sp & 2 & $22.15^{+0.19}_{-0.30}$ & 44.28 & 44.34 & 1 & 6.97 & $0.49^{+0.24}_{-0.22}$ & $2.10\thinspace (0.32)$ & $0.39\thinspace (0.11)$ & $1.50\thinspace (0.32)$ \\
337 & 0.837 & sp & 11 & $<21$ & 43.98 & 43.98 & 1 & 6.4 & $0.13^{+0.04}_{-0.04}$ & $3.11\thinspace (0.09)$ & NA & $1.21\thinspace (0.07)$ \\
338 & 2.02 & x & 7 & $22.86^{+0.07}_{-0.07}$ & 44.04 & 44.26 & 1 & 6.4 & $0.33^{+0.11}_{-0.11}$ & $1.10\thinspace (0.14)$ & $0.66\thinspace (0.09)$ & $1.61\thinspace (0.20)$ \\
339 & 0.677 & ph & 4 & $<22$ & 43.30 & 43.32 & 0 & 6.4 & $<0.28$ & $2.63\thinspace (0.60)$ & NA & $1.33\thinspace (0.47)$ \\
341 & 1.324 & sp & 11 & $<21$ & 44.20 & 44.20 & 0 & 6.97 & $0.11^{+0.09}_{-0.08}$ & $4.12\thinspace (0.25)$ & $0.39\thinspace (0.07)$ & $1.00\thinspace (0.14)$ \\
343 & 1.438 & ph & 12 & $23.57^{+0.14}_{-0.16}$ & 43.11 & 43.67 & 0 & 6.4 & $<0.43$ & $0.50\thinspace (0.16)$ & $0.66\thinspace (0.18)$ & $1.60\thinspace (0.51)$ \\
344 & 1.511 & sp & 15 & $22.84^{+0.02}_{-0.02}$ & 44.20 & 44.41 & 1 & 6.4 & $0.12^{+0.06}_{-0.06}$ & $1.26\thinspace (0.06)$ & $0.59\thinspace (0.04)$ & $1.32\thinspace (0.12)$ \\
345 & 1.235 & sp & 2 & $<21$ & 43.87 & 43.87 & 0 & 6.4 & $<0.15$ & $3.62\thinspace (0.36)$ & $0.48\thinspace (0.10)$ & $1.19\thinspace (0.27)$ \\
350 & 1.21 & ph & 3 & $22.99^{+0.28}_{-0.29}$ & 42.91 & 43.15 & 0 & 6.7 & $1.02^{+0.77}_{-0.68}$ & $1.24\thinspace (0.72)$ & $0.02\thinspace (0.54)$ & $0.96\thinspace (1.73)$ \\
352 & 1.60 & x & 6 & $23.34^{+0.02}_{-0.02}$ & 44.26 & 44.65 & 1 & 6.4 & $0.21^{+0.06}_{-0.05}$ & $0.73\thinspace (0.04)$ & $0.68\thinspace (0.04)$ & $1.58\thinspace (0.13)$ \\
358 & 0.976 & sp & 3 & $22.98^{+0.02}_{-0.01}$ & 44.36 & 44.61 & 1 & 6.97 & $0.05^{+0.03}_{-0.03}$ & $0.96\thinspace (0.03)$ & NA & $1.13\thinspace (0.06)$ \\
359 & 1.574 & sp & 3 & $21.70^{+0.08}_{-0.10}$ & 44.72 & 44.72 & 0 & 6.4 & $<0.05$ & $2.56\thinspace (0.07)$ & $0.50\thinspace (0.03)$ & $1.09\thinspace (0.07)$ \\
361 & 0.977 & sp & 11 & $<21$ & 44.04 & 44.04 & 1 & 6.7 & $0.11^{+0.06}_{-0.06}$ & $2.92\thinspace (0.14)$ & NA & $0.94\thinspace (0.10)$ \\
366 & 3.315 & ph & 4 & $22.52^{+0.11}_{-0.14}$ & 44.23 & 44.34 & 0 & 6.4 & $<0.24$ & $1.56\thinspace (0.23)$ & $0.28\thinspace (0.09)$ & $1.45\thinspace (0.51)$ \\
369 & 0.496 & sp & 3 & $<21.9$ & 43.08 & 43.11 & 0 & 6.4 & $0.19^{+0.15}_{-0.14}$ & $2.19\thinspace (0.34)$ & NA & $1.45\thinspace (0.32)$ \\
374 & 1.041 & sp & 2 & $23.02^{+0.04}_{-0.03}$ & 43.59 & 43.84 & 1 & 6.5 & $0.14^{+0.09}_{-0.08}$ & $0.93\thinspace (0.12)$ & $0.66\thinspace (0.12)$ & $1.53\thinspace (0.28)$ \\
375 & 1.350 & sp & 2 & $<21$ & 43.78 & 43.79 & 0 & 6.4 & $<0.37$ & $3.23\thinspace (0.59)$ & $0.52\thinspace (0.21)$ & $2.04\thinspace (0.53)$ \\
382 & 0.527 & sp & 2 & $21.72^{+0.10}_{-0.12}$ & 43.63 & 43.65 & 1 & 6.4 & $0.07^{+0.04}_{-0.04}$ & $2.31\thinspace (0.22)$ & NA & $1.18\thinspace (0.20)$ \\
385 & 1.22 & sp & 2 & $22.61^{+0.07}_{-0.08}$ & 43.53 & 43.65 & 0 & 6.4 & $0.17^{+0.14}_{-0.13}$ & $1.31\thinspace (0.20)$ & $0.57\thinspace (0.15)$ & $1.56\thinspace (0.44)$ \\
388 & 0.605 & sp & 2 & $<22$ & 42.72 & 42.73 & 0 & 6.7 & $0.27^{+0.23}_{-0.21}$ & $1.77\thinspace (0.34)$ & NA & $1.46\thinspace (0.47)$ \\
390 & 1.118 & sp & 2 & $<21.7$ & 43.74 & 43.75 & 1 & 6.4 & $0.29^{+0.14}_{-0.13}$ & $2.67\thinspace (0.27)$ & $0.39\thinspace (0.11)$ & $1.35\thinspace (0.25)$ \\
396 & 2.566 & ph & 4 & $<22$ & 43.90 & 43.90 & 0 & 6.4 & $<0.18$ & $2.00\thinspace (0.73)$ & $0.08\thinspace (0.24)$ & $2.16\thinspace (0.81)$ \\
398 & 2.122 & sp & 2 & $23.34^{+0.19}_{-0.19}$ & 43.53 & 43.89 & 1 & 6.4 & $1.53^{+0.56}_{-0.54}$ & $0.24\thinspace (0.28)$ & $0.51\thinspace (0.26)$ & $3.35\thinspace (0.97)$ \\
407 & 1.13 & ph & 22 & $<22$ & 43.71 & 43.75 & 0 & 6.4 & $0.45^{+0.43}_{-0.39}$ & $3.16\thinspace (0.71)$ & $0.72\thinspace (0.33)$ & $0.95\thinspace (0.51)$ \\
504 & 1.037 & sp & 11 & $21.98^{+0.10}_{-0.14}$ & 43.46 & 43.49 & 1 & 6.7 & $0.17^{+0.11}_{-0.10}$ & $2.43\thinspace (0.18)$ & $0.37\thinspace (0.08)$ & $0.93\thinspace (0.17)$ \\
\end{longtable}
\tablebib{ 1:~\citet{treister09}; 2: \citet{silverman10}; 3:
  \citet{cooper12}; 4: \citet{hsu14}; 5: \citet{balestra10}; 6:
  \citet{i12cdfs}; 7: This work; 8: \citet{vanzella08}; 9:
  \citet{ravikumar07}; 10: \citet{rafferty10}; 11: \citet{szokoly04};
  12: \citet{luo10}; 13: \citet{taylor09}; 14: \citet{popesso09}; 15:
  \citet{lefvre13}; 16: \citet{kurk13}; 17: \citet{georgantopoulos13};
  18: \citet{mignoli05}; 19: \citet{morris05}; 20: \citet{mignoli04};
  21: \citet{straatman16}; 22: \citet{cardamone10}; 23:
  \citet{pentericci18}; 24: \citet{urrutia19}.}  \tablefoot{
  Absorption column density, \nH, is in units of \psqcm. Rest-frame
  2-10 keV luminosity, $L$, is in units of \ergps. $L^{\prime}$ is
  absorption corrected value in the same units. $D_{\rm Fe}$ is the Fe
  K line detection flag, 1: detection with the $>90$\% significance;
  0: less significant or no detection. Fe K line energy, $E$, and its
  equivalent width, $EW$, are in units of keV and measured in the
  galaxy rest-frame. S/M and H/M are the rest-frame X-ray colours (see
  text). $R({\rm Fe})$ is the ratio of the Fe K band interval ($i$=13)
  and the estimated underlying continuum. S/M, H/M and $R({\rm Fe})$
  are not available for the five sources with $z<0.4$. H/M are not
  available for lowz ($z<1$) sources. The online version of this table
  includes Fe K line detection flag with the 68\% detection threshold,
  the signal-to-noise ratio indicator of each spectrum in the
  rest-frame 3-10 keV band, Chandra source identification numbers of
  \citet{luo17} and \citet{xue16} and the radio source identification
  numbers of \citet{bonzini13} when a source is classified as
  radio-loud AGN.\\
  \tablefoottext{a}{The XMM detected emission of
  PID~301 is composed of two X-ray sources resolved by Chandra. The
  redshift quoted here is the photometric redshift for the absorbed,
  hard source. The softer part of the spectrum is dominated by the
  softer source with spectroscopic redshift $z=0.566$.}  }

\end{document}